\documentclass[aps,prd,twocolumn,showpacs,notitlepage,eqsecnum,
superscriptaddress]{revtex4-1}

\usepackage[centertags]{amsmath}
\usepackage{amssymb}
\usepackage{latexsym}
\usepackage{enumerate}
\usepackage{graphicx}
\usepackage{mathrsfs}
\usepackage{hyperref}
\usepackage{stmaryrd}
\usepackage{import}
\usepackage{tensor}
\usepackage{color}
\usepackage{bm}
\usepackage{multirow}
\usepackage{breakurl}
\usepackage{float}
\usepackage{placeins}

\newcommand{\bi}{\begin{itemize}}
\newcommand{\ei}{\end{itemize}}
\newcommand{\be}{\begin{equation}}
\newcommand{\ee}{\end{equation}}

\renewcommand{\l}{\left(}
\renewcommand{\r}{\right)}
\renewcommand{\a}{\alpha}

\newcommand{\g}{\gamma}
\newcommand{\G}{\Gamma}

\newcommand{\D}{\Delta}
\newcommand{\e}{\epsilon}

\newcommand{\La}{\Lambda}
\newcommand{\la}{\lambda}
\renewcommand{\O}{\Omega}
\renewcommand{\o}{\omega}

\newcommand{\Th}{\Theta}
\newcommand{\q}{\quad}
\newcommand{\qq}{\qquad}

\newcommand{\vp}{\varphi}

\newcommand{\pa}{\partial}

\allowdisplaybreaks

\begin{document}

\title{High-order post-Newtonian expansion of the redshift invariant for eccentric-orbit non-spinning 
extreme-mass-ratio inspirals}

\author{Christopher Munna}
\affiliation{MIT Kavli Institute, Massachusetts Institute of Technology, Cambridge, MA 02139, USA }
\affiliation{Department of Physics and Astronomy, University of North Carolina, Chapel Hill, North Carolina 27599, USA}
\author{Charles R. Evans}
\affiliation{Department of Physics and Astronomy, University of North Carolina, Chapel Hill, North Carolina 27599, USA}

\begin{abstract}
We calculate the eccentricity dependence of the high-order post-Newtonian (PN) series for the generalized redshift 
invariant $\langle u^t \rangle_\tau$ for eccentric-orbit extreme-mass-ratio inspirals on a Schwarzschild background.  
These results are calculated within first-order black hole perturbation theory (BHPT) using Regge-Wheeler-Zerilli (RWZ) 
gauge.  Our \textsc{Mathematica} code is based on a familiar procedure, using PN expansion of the 
Mano-Suzuki-Takasugi (MST) analytic function formalism for $l$ modes up to a certain maximum and then using a 
direct general-$l$ PN expansion of the RWZ equation for arbitrarily high $l$.  We calculate dual expansions in PN 
order and in powers of eccentricity, reaching 10PN relative order and $e^{20}$.  Detailed knowledge of the 
eccentricity expansion at each PN order allows us to find within the eccentricity dependence numerous closed-form 
expressions and multiple infinite series with known coefficients.  We find leading logarithm sequences in the PN 
expansion of the redshift invariant that reflect a similar behavior in the PN expansion of the energy flux to 
infinity.  A set of flux terms and special functions that appear in the energy flux, like the Peters-Mathews flux 
itself, are shown to reappear in the redshift PN expansion. 
\end{abstract}


\maketitle

\section{Introduction}
\label{sec:intro}

Using a recently developed method and \textsc{Mathematica} code \cite{Munn20,Munn20c}, we calculated previously 
high-order post-Newtonian (PN) expansions of the energy and angular momentum radiated to infinity by non-spinning 
eccentric-orbit extreme-mass-ratio inspirals (EMRIs) in first-order black hole perturbation theory (BHPT) (see also 
\cite{MunnETC20}).  The resulting expansions, in both PN order and eccentricity $e$, were taken to high PN order 
(19PN) and $e^{10}$ and to somewhat lower PN order (10PN) and higher order ($e^{20}$) in 
eccentricity.  The detailed behavior in eccentricity allowed us to find numerous closed-form expressions and 
infinite series in $e$ with identifiable coefficient sequences.  In the process we found a set of leading-logarithm 
connections between low-order multipole moments of the orbital motion and arbitrarily high PN order sequences in the 
fluxes \cite{MunnEvan19a,MunnEvan20a,Munn20c}.  Since then, fluxes at the horizon have also been found, to 18PN 
(relative to the leading horizon flux) and $e^{10}$ as well as to 10PN and $e^{20}$ \cite{Munn20c,MunnEvan20b}.  
Taken together these expansions are useful since fluxes are the most significant contributors to EMRI orbital phase 
evolution \cite{HindFlan08}.  High-order PN expansions of the fluxes and ultimately waveform amplitudes associated 
with Kerr EMRIs could make important early-phase baseline contributions to more comprehensive efforts to develop 
``fast'' waveform models for the LISA mission \cite{KatzETC21}.

These deep PN expansions in first-order BHPT in the dissipative sector can also be extended to perturbations of the 
metric and of local, conservative, gauge-invariant quantities.  The first such local quantity to be examined for its 
connections between BHPT and PN theory was Detweiler's redshift invariant \cite{Detw05} for circular orbits, $u^t$, 
which was initially calculated through 3PN order \cite{Detw08}.  Ultimately, Kavanagh, Ottewill, and Wardell 
\cite{KavaOtteWard15} used analytic expansion methods to compute this term to 21.5PN for circular orbits.  The 
redshift invariant was generalized to eccentric orbits by Barack and Sago \cite{BaraSago11}, who defined it in that
case as the average of $u^t$ taken in proper time over one radial libration, $\langle u^t \rangle_\tau$.  Its behavior 
was calculated to 3PN order in \cite{AkcaETC15} using results from the full PN theory (see \cite{Blan14} for review 
of status of PN theory).  The redshift is one of multiple gauge-invariants that can be calculated in both BHPT and 
PN theory and compared.  Others that have been identified, either for circular or eccentric orbits, include the 
first-order in the mass ratio effects on apsidal advance of eccentric orbits \cite{BaraSago11}, location of the 
innermost stable circular orbit \cite{BaraSago09}, spin-precession invariant $\psi$ (correction to geodetic precession) 
\cite{DolaETC14a,BiniDamo14b,KavaOtteWard15}, tidal invariants \cite{DolaETC14b,KavaOtteWard15}, and octupole 
invariants \cite{NolaETC15}.  Conservative-sector invariants calculated in BHPT may supply calibration of 
effective-one-body (EOB) potentials (see, e.g., 
\cite{BaraDamoSago10, LetiBlanWhit12, BiniDamo14c, BiniDamoGera15,Leti15,HoppKavaOtte16, KavaETC17, 
BiniDamoGera18, BiniDamoGera19, BiniDamoGera20a, BiniDamoGera20b}), which is important since EOB allows rapid 
evaluation of the dynamics of merging binaries and covers broad regions of parameter space.  Recent work has also 
shown that the redshift invariant, in particular, can be directly translated to the local sector of post-Minkowskian 
(PM) dynamics, allowing derivation of higher-order PM scattering mechanics 
\cite{BiniDamoGera19,BiniDamoGera20a,BiniDamoGera20b}.  This paper turns the use of our recently developed code to 
the task of uncovering the higher-order (10PN and $e^{20}$) behavior of the redshift invariant and 
in the process we show intriguing physical connections between the conservative and dissipative sectors.

The present method derives from work of \cite{BiniDamoGera16a,HoppKavaOtte16,BiniDamoGera16c}.  Mode functions for 
$l \ge 2$ are computed in the Regge-Wheeler-Zerilli (RWZ) gauge \cite{ReggWhee57,Zeri70}.  For $l$ modes up to a 
certain order, PN expansions of the mode functions are found using the Mano-Suzuki-Takasugi (MST) formalism 
\cite{ManoSuzuTaka96a}, as shown in previous applications \cite{BiniDamo13, BiniDamo14a, BiniDamo14b, BiniDamo14c, 
KavaOtteWard15, HoppKavaOtte16}.  Modes of the metric perturbation are derived from the mode functions and the 
modes of the redshift invariant $\langle u^t \rangle^{l}_\tau$ are found through projection of the metric perturbation 
on the four velocity.  Finite local value of the redshift invariant is then obtained by directly applying mode-sum 
regularization to the scalar quantity.  Mode-sum regularization requires knowledge of all $l$ modes.  Beyond the 
range in $l$ covered by the MST expansion, we use a direct PN expansion ansatz for general-$l$ solutions of the 
Regge-Wheeler (RW) equation \cite{BiniDamo13}.  The modes of $\langle u^t \rangle^{l}_\tau$ derived at low $l$ by 
MST and general $l$ by the ansatz are augmented by direct solution of the $l=0,1$ modes to complete the regularization.

The structure of this paper is as follows.  In Sec.~\ref{sec:MSTreview} we briefly outline the problem setup and 
the MST formalism, with a focus on how the mode functions in the RWZ gauge can be PN expanded.  That section 
includes discussion of the metric perturbations and how they are likewise PN expanded.  The metric perturbations 
evaluated at the location of the small body are needed to find the regularized (conservative sector) self-force.  
As mentioned, for conservative sector quantities the $l$-mode expansion of the metric must be made for all $l$.  The 
MST formalism is used to find $l$-modes up to a modest $l$ related to the sought-after PN order.  
Sec.~\ref{sec:genLexps} details the separate procedure used to obtain general, higher $l$-modes.  In 
Sec.~\ref{sec:l01andReg} we briefly recall the final two, non-radiative modes that are not covered by the RWZ 
formalism and discuss the mode-sum regularization procedure, which is specialized here for extracting the 
redshift invariant.  Secs.~\ref{sec:ut} is then the heart of the paper, outlining the expected form of the 
eccentric-orbit PN expansion of the redshift and displaying our results for the numerous non-log and log parts of 
the eccentricity dependence up to 10PN order.  (We show results in this paper up to 8.5PN with the remainder being 
posted at \cite{BHPTK18}.)  The redshift invariant is shown for two different compactness parameters.  This section 
then summarizes the results, including a discussion of the uncovered connection between the redshift PN expansion 
and the PN expansion of the energy flux to infinity.  We also compare our PN expansion numerically to self-force 
results published previously for compact orbits.  Sec.~\ref{sec:ConsConc} concludes with summary and outlook.

Throughout this paper we primarily choose units such that $c = G = 1$, though in making PN expansions we reintroduce 
$\eta = 1/c$ as a PN (slow motion) parameter for bookkeeping purposes.  Our metric signature is $(-+++)$.  Our 
notation for the RWZ formalism follows that found in \cite{ForsEvanHopp16, MunnETC20}, which in part derives from 
notational changes for tensor spherical harmonics and perturbation amplitudes introduced by Martel and Poisson 
\cite{MartPois05}.  For the MST formalism, we largely make use of the discussion and notation found in the review 
by Sasaki and Tagoshi \cite{SasaTago03}.

\section{Brief review of RWZ and MST formalisms}
\label{sec:MSTreview}

We briefly outline the setup of the problem of calculating conservative sector perturbations for bound EMRI motion 
on a Schwarzschild background.  We further summarize the MST analytic function expansions, the use of which are 
required for modes with small $l$ in the PN expansion.  This process is more extensively detailed in \cite{Munn20} 
and is based on earlier work in \cite{BiniDamo13,BiniDamo14a,KavaOtteWard15,BiniDamoGera16a,HoppKavaOtte16}.  

\subsection{Bound orbits on a Schwarzschild background}
\label{sec:orbits}

The secondary is treated as a point mass $\mu$ in bound geodesic orbit about a Schwarzschild black hole of mass $M$ 
with $\varepsilon = \mu/M \ll 1$.  The line element in Schwarzschild coordinates $x^{\mu} = \{t,r,\theta, \varphi \}$ is
\be
ds^2 = -f dt^2 + f^{-1} dr^2
+ r^2 \left( d\theta^2 + \sin^2\theta \, d\varphi^2 \right) ,
\ee
with $f = 1 - 2M/r$.  For motion $x^\a = x_p^\a(\tau)$ confined to the equatorial plane, the four-velocity is 
\be
\label{eqn:four_velocity}
u^\a(\tau) = \frac{dx_p^{\alpha}(\tau)}{d\tau} 
= \l \frac{{\mathcal{E}}}{f_{p}}, u^r, 0, \frac{{\mathcal{L}}}{r_p^2} \r ,
\ee
where $\mathcal{E}$ and $\mathcal{L}$ are the conserved specific energy and angular momentum, respectively.  The 
radial proper velocity $u^r$ is then found from the normalization of $u^\mu$.  Orbital motion is conveniently 
described by an alternative (Darwin) parameter set $\{ \chi, p, e \}$ \cite{Darw59,CutlKennPois94,BaraSago10} with
\begin{align}
\label{eqn:defeandp}
{\mathcal{E}}^2 &= \frac{(p-2)^2-4e^2}{p(p-3-e^2)} ,
\q
{\mathcal{L}}^2 = \frac{p^2 M^2}{p-3-e^2} ,  
\notag 
\\
& \qq \q  r_p \l \chi \r = \frac{pM}{1+ e \cos \chi} .
\end{align}
One radial libration occurs with each $2 \pi$ advance in $\chi$.  The dimensionless quantity $1/p$ serves as a 
PN compactness parameter.  Integrals can be written down from separate ordinary differential equations (ODEs) for 
the development of $\vp$, $t$, and $\tau$ in terms of $\chi$ \cite{HoppETC15,ForsEvanHopp16}.  Each integrand can 
be expanded as a PN series (e.g., in $1/p$) and the integrals can be solved order by order in powers of $1/p$.  
Definite integrals yield the fundamental frequencies $\O_r$ and $\O_\vp$.  The radial period is given by
\begin{align}
\label{eqn:O_r}
T_r = \int_{0}^{2 \pi}  \frac{r_p \l \chi \r^2}{M (p - 2 - 2 e \cos \chi)}
 \left[\frac{(p-2)^2 -4 e^2}{p -6 -2 e \cos \chi} \right]^{1/2}    d \chi ,
 \notag
\end{align}
with $\Omega_r = 2 \pi/T_r$.  The azimuthal frequency is given by
\be
\label{eqn:O_phi}
\O_\varphi = \frac{4}{T_r} \left(\frac{p}{p - 6 - 2 e}\right)^{1/2} \, 
K\left(-\frac{4 e}{p - 6 - 2 e}  \right) ,
\ee
where $K(m)$ is the complete elliptic integral of the first kind \cite{GradETC07}.  Each frequency is PN expanded.  
Once the azimuthal frequency known, the usual PN compactness parameter $y = (M \O_\vp)^{2/3}$ can obtained as a 
power series in $1/p$, or vice versa.  Eccentric motion also leads to expansions in powers of Darwin eccentricity 
$e$.

\subsection{The RWZ master equation}
\label{sec:TDmasterEq}

Bound motion acts as a periodic source for the first-order gravitational perturbations.  On a Schwarzschild 
background these can encoded by a pair (even and odd parity) of RWZ-gauge master functions 
\cite{ReggWhee57,Zeri70,MartPois05}.  The master equations in the frequency domain (FD) take the form
\be
\label{eqn:masterInhomogFD}
\left[\frac{d^2}{dr_*^2} +\omega^2 -V_l (r) \right]
X_{lmn}(r) = Z_{lmn} (r) .
\ee
Here $r_{*} = r + 2 M \ln | r/2 M - 1 |$ is the tortoise coordinate, the frequency spectrum is discrete 
$\o \equiv \o_{mn} = m \O_\vp + n \O_r$, and the source functions are given by
\begin{align}
Z_{lmn}(r) &= \frac{1}{T_r} \int_0^{2 \pi} \big( G_{lm}(t) \, \delta[r - r_p(t)] 
\notag \\
& \qq \qq 
+ F_{lm}(t) \, \delta'[r - r_p(t)] \big) e^{i \o t} dt .
\end{align}
The functions $G_{lm}(t)$ and $F_{lm}(t)$ \cite{HoppEvan10} follow from the point-particle stress-energy tensor.
Both the source term and the potential $V_{l}(r)$ are ($l+m$) parity-dependent.

The homogeneous form of this equation yields two independent solutions: $X_{lmn}^{\rm in} = X_{lmn}^{-}$, with 
causal (downgoing wave) behavior at the horizon, and $X_{lmn}^{\rm up} = X_{lmn}^{+},$ with causal (outgoing wave) 
behavior at infinity.  The odd-parity homogeneous solutions can be determined directly using the MST formalism 
\cite{ManoSuzuTaka96a}, which we outline below.  The corresponding even-parity solutions are derived from the 
odd-parity solutions using one form of the Detweiler-Chandrasekar transformation 
\cite{Chan75,ChanDetw75,Chan83,Bern07,ForsEvanHopp16}. 

\begin{widetext}
\subsection{The MST homogeneous solutions and the source integration}

The MST solution for $X_{lmn}^+$ can be expressed \cite{KavaOtteWard15} as
\begin{align}
\label{eqn:XupMST}
X^{+}_{lmn} &= e^{iz} z^{\nu+1} \left(1- \frac{\e}{z}\right)^{-i \e} \sum_{j=-\infty}^{\infty} a_j (-2 i z)^{j} 
\frac{ \G(j + \nu + 1 - i \e) \G(j + \nu - 1 - i \e) }{\G(j + \nu + 3 + i \e) \G(j + \nu + 1 + i \e) } \times \notag \\
& \hspace{27em} U(j + \nu + 1 - i \e, 2 j + 2 \nu + 2, -2 i z) , 
\end{align}
where $U$ is the irregular confluent hypergeometric function, $\e = 2 M \o \eta^3$, $z = r \o \eta$, and 
$\eta = 1/c$ (which serves as a 0.5PN expansion parameter).  In this equation, $\nu$ is the renormalized angular 
momentum, which is an eigenvalue chosen to make the series coefficients $a_j$ converge in both limits as 
$j \rightarrow \pm\infty$.  Both $\nu$ and $a_j$ are determined through a continued fraction method 
\cite{ManoSuzuTaka96a,SasaTago03}, which leads to series in $\e$ for both (which are then PN series).  

Similarly, the solution for $X_{lmn}^-$ is given by
\begin{align}
X^{-}_{lmn} &= e^{-iz} \left(\frac{z}{\e} - 1\right)^{-i \e} \left(\frac{\e}{z}\right)^{i \e + 1}  
\sum_{j= -\infty}^{\infty} a_j \frac{\G(j + \nu - 1 - i\e) \G(-j - \nu - 2 - i \e)}{\G(1 - 2i\e)} \times \notag \\
& \hspace{20em} {}_2F_1(j + \nu - 1 - i\e, -j - \nu - 2 - i\e; 1 - 2 i \e; 1 - z/\e) ,
\label{eqn:XinMST}
\end{align}
which is expressed in terms of the ordinary (Gauss) hypergeometric function.  The $\nu$ and $a_j$ appearing here are 
the same as those found in solving for the up ($+$) solution \eqref{eqn:XupMST}.

The process of expanding these homogeneous solutions by collecting on powers of $\eta$ is fully described in 
\cite{Munn20}, based on the methods initially presented in \cite{BiniDamo13} and \cite{KavaOtteWard15}.  The 
homogeneous solutions are normalized initially by making the choice $a_0 = 1$ in solving the recurrence relation 
for $a_j$.  However, it proves useful to remove $z$-independent factors from these solutions to reduce their size and 
complexity, as described in \cite{Munn20, KavaOtteWard15}.  This step temporarily rescales the solutions, which are 
then used to form a Green function to find the inner and outer solutions that reflect the behavior of the source.  
Integration with the Green function yields normalization coefficients on both sides of the source region 
\begin{align}
\label{eqn:ClmnIntChi}
C_{lmn}^{\pm} = \frac{1}{W_{lmn} T_r} \int_0^{2 \pi} \left(\frac{dt}{d\chi} \right) 
 \bigg[  \frac{1}{f_p}G_{lm}(\chi) X^{\mp}_{lmn}  
+\left( \frac{2 M}{r_p^2 f_p^2}  X^{\mp}_{lmn} - \frac{1}{f_p} \frac{d X^{\mp}_{lmn}}{dr} \right)  F_{lm}(\chi) \bigg] 
e^{i \o t(\chi)} d\chi , 
\end{align}
where a subscript $p$ denotes functions that are evaluated along the worldline of the particle and $W_{lmn}$ is 
the Wronskian.  In the dissipative sector, it is necessary to rescale these coefficients in order that their complex 
square yields the fluxes \cite{Munn20}.  To find local conservative quantities, the time domain (TD) extended 
solutions are constructed from the combinations $C_{lmn}^+ X_{lmn}^+$ and $C_{lmn}^- X_{lmn}^-$ 
\cite{HoppEvan10, Munn20}, which automatically produce the proper normalization.  In the present application, these 
time domain functions are then PN expanded.

\subsection{The metric perturbations}
\label{sec:metPerExps}

The first-order generalized redshift invariant is a quantity that depends upon the metric perturbation and its
regularized behavior evaluated along the particle worldline.  The general metric perturbation expressions all involve 
products of the normalization coefficients $C^{\pm}_{lmn}$ with some linear functional of the homogeneous solutions 
$X^{\pm}_{lmn}$, prior to summing to transfer from the FD to TD \cite{HoppEvan10,HoppKavaOtte16}.  As discussed 
in \cite{HoppKavaOtte16}, singular and discontinuous parts of the FD metric perturbations cancel on the particle's 
worldline in summing over $m$, leaving the $l$-dependent functions being $C^0$.  The even-parity TD amplitudes are
\begin{align}
K^{lm,\pm}(t,r) &= f \pa_r \Psi_{lm}^{e,\pm} + A(r) \Psi_{lm}^{e, \pm},  \notag \\
h_{rr}^{lm,\pm}(t,r) &= \frac{\La}{f^2} \left[ \frac{\la + 1}{r} \Psi_{lm}^{e, \pm} - K^{lm,\pm} \right] 
+ \frac{r}{f} \, \pa_r K^{lm,\pm} ,    
\notag \\
h_{tr}^{lm,\pm}(t,r)   &=  r \pa_t \pa_r \Psi^{e, \pm}_{lm} + r B(r) \pa_t \Psi^{e,\pm}_{lm} ,
\notag \\
h_{tt}^{lm,\pm}(t,r)  &= f^2 h_{rr}^{lm,\pm} , 
\end{align}
where $\Psi_{lm}^e$ is the solution to the TD Zerilli-Moncrief equation \cite{HoppEvan10}.  The $+$ and $-$ 
superscripts correspond to whether the solution is constructed from $C_{lmn}^+ X_{lmn}^+$ or $C_{lmn}^- X_{lmn}^-$ 
(see below).  The expressions above use the following definitions
\begin{align}
\la &= \frac{1}{2}(l+2)(l-1), \qq  \qq    \La = \la + \frac{3 M}{r},    \notag \\
A(r) &= \frac{1}{r \La} \left[ \la(\la+1) + \frac{3 M}{r} \left( \la + \frac{2 M}{r} \right) \right],  \qq \qq
B(r) = \frac{1}{r f \La} \left[ \la \left(1 - \frac{ 3M}{r} \right) - \frac{3 M^2}{r^2} \right].  
\end{align}
The $l$-mode decomposition of the full even-parity metric perturbation $p_{\mu\nu}$ can then be written as
\begin{align}
p_{rr}^{l}(t,r,\theta,\vp) &= \sum_{m=-l}^l  \left[ h_{rr}^{lm, +}(t,r)  \Theta[r - r_p(t)] + h_{rr}^{lm, -}(t,r) 
\Theta[r_p(t) - r]   \right] Y_{lm}(\theta,\vp) ,    
\notag \\
p_{tr}^{l}(t,r,\theta,\vp)   &=  \sum_{m=-l}^l \left[ h_{tr}^{lm, +}(t,r)  \Theta[r - r_p(t)] + h_{tr}^{lm, -}(t,r) 
\Theta[r_p(t) - r]   \right] Y_{lm}(\theta,\vp),   
\notag \\
p_{tt}^{l}(t,r,\theta,\vp)  &= f^2 \,  p_{rr}^{l}, 
\notag \\
p_{AB}^{l}(t,r,\theta,\vp)  &= \sum_{m=-l}^l r^2 \O_{AB} \left[ K^{lm,+}(t,r)  \Theta[r - r_p(t)] +  
K^{lm,-}(t,r)  \Theta[r_p(t) - r]  \right]  Y_{lm}(\theta,\vp)  .
\end{align}

Likewise we can express the odd-parity TD amplitudes as follows
\begin{equation}
h_{t}^{lm,\pm}(t,r)  = \frac{f}{2} \pa_r (r \Psi_{lm}^{o, \pm} ) ,   \qquad
h_{r}^{lm}(t,r,\theta,\vp) = \frac{r}{2 f} \pa_t \Psi_{lm}^{o, \pm} ,  
\end{equation}
with the $l$-mode decomposition of the full odd-parity metric perturbation given by
\begin{align}
p_{tB}^{l}(t,r,\theta,\vp)  &= \sum_{m=-l}^l  \left[ h_{t}^{lm, +}(t,r)  \Theta[r - r_p(t)] + h_{t}^{lm, -}(t,r)  
\Theta[r_p(t) - r]   \right] X_B^{lm}(\theta,\vp) ,    
\notag \\
p_{rB}^{l}(t,r,\theta,\vp)  &= \sum_{m=-l}^l  \left[ h_{r}^{lm, +}(t,r)  \Theta[r - r_p(t)] + h_{r}^{lm, -}(t,r)  
\Theta[r_p(t) - r]   \right] X_B^{lm}(\theta,\vp) .
\end{align}

These reconstructions of the metric are valid for all $t$ and for $r > 2 M$.  The redshift invariant, however, 
merely requires the behavior along the trajectory $r = r_p(t)$ itself.  To parameterize the background motion, and 
therefore the self-force, it is computationally convenient to use $\chi$ instead of $t$.  Accordingly, we modify 
the notation so that quantities are thought to be functions of $\chi$ (e.g., $r = r_p(\chi), f = f_p(\chi), 
t = t_p(\chi),$ etc).  Expressing everything, including derivatives, in terms of $\chi$, we find the following 
local behavior for the $l$-modes of the metric perturbation
\begin{align}
\label{eqn:MPchi}
p_{rr}^{l}(\chi) &=   \left(\frac{d \chi}{d r}\right) \sum_{mn} \frac{Y_{lm}(\pi/2,0)}{f} C^{\pm}_{lmn} 
e^{i m \vp -i \o t} 
\Bigg\{\left[ \left(\frac{d r}{d \chi}\right)\frac{\La(\la + 1)}{f r} - \left(\frac{d r}{d \chi}\right) 
\frac{\La}{f} A(\chi) + r \left(\frac{d A(\chi)}{d \chi}\right) \right] X^{\pm}_{lmn}(\chi) \notag \\ &
+ (r A(\chi) - \La ) \left(\frac{d X^{\pm}_{lmn}(\chi)}{d \chi} \right) + r \frac{d}{d \chi} 
\left[ f \left(\frac{d \chi}{d r}\right) \left(\frac{d X^{\pm}_{lmn}(\chi)}{d \chi} \right)   \right] \Bigg\}, 
\notag \\
p_{tr}^{l}(\chi)  &=  \sum_{mn} Y_{lm}(\pi/2,0) C^{\pm}_{lmn} e^{i m \vp -i \o t} (-i \o) 
\left[ r \left(\frac{d \chi}{d r}\right)  \left(\frac{d X^{\pm}_{lmn}(\chi)}{d \chi} \right) 
+ r B(\chi) X^{\pm}_{lmn} (\chi)  \right] ,     \notag \\
p_{tt}^{l}(\chi) &= f^2 p_{rr}^{l,\pm}, \notag \\
p_{AB}^{l}(\chi) &= r^2 \O_{AB}  \sum_{mn} Y_{lm}(\pi/2,0)  C^{\pm}_{lmn} e^{i m \vp -i \o t}
\left[ f \left(\frac{d \chi}{d r}\right) \pa_{\chi} X_{lmn}^{e,\pm} + A(\chi) X_{lmn}^{e, \pm} \right],  \notag \\ 
p_{tB}^{l}(\chi) &=   \left( \frac{f}{2} \right) \sum_{mn} X^{lm}_B(\pi/2,0) C^{\pm}_{lmn} e^{i m \vp -i \o t}
\left(\frac{d \chi}{d r}\right)   \frac{d}{d \chi} (r X^{\pm}_{lmn}) ,  \notag \\ 
p_{rB}^{l}(\chi) &=  \left( \frac{r}{2f} \right) 
\sum_{mn} X^{lm}_B(\pi/2,0)  C^{\pm}_{lmn} e^{i m \vp -i \o t}  (-i \o) X^{\pm}_{lmn} .
\end{align}
Since the $l$-modes are $C^0$ at $r = r_p(\chi)$, the same result emerges in using either the $+$ or $-$ side 
mode functions.  

\section{General-$l$ expansions}
\label{sec:genLexps}

The MST formalism, as briefly summarized in Sec.~\ref{sec:MSTreview}, provides mode functions for specific $l$.  
We used that procedure in several previous papers \cite{Munn20,ForsEvanHopp16,MunnEvan19a,MunnETC20, 
MunnEvan20a,MunnEvan20b} that dealt with gravitational wave fluxes, 
taking advantage of the fact that the PN expansions of higher $l$ fluxes begin at successively higher PN order.  To 
determine the redshift invariant or other conservative quantities, $l$ modes of the local behavior of 
the metric perturbation are needed.  This introduces a difficulty not encountered with the fluxes---the PN expansions 
of higher $l$ contributions, $p_{\mu \nu}^{l}(\chi)$, do not begin with successively higher PN order.  Thus, to 
obtain correct PN coefficients in the expansion of the metric $p_{\mu \nu}^{l}(\chi)$, a sum over \textit{all} $l$ 
must be made.  This necessitates finding analytic expansions for arbitrary $l$.

\subsection{The homogeneous solutions and normalization constants}

To generate expansions for general $l$, we might try directly expanding the odd-parity MST solutions 
\eqref{eqn:XupMST} and \eqref{eqn:XinMST} while leaving $l$ arbitrary.  However, the $\G$ functions in the summations 
make such an approach apparently intractable.  An alternative method utilizes an ansatz 
\cite{BiniDamo13,KavaOtteWard15} for the homogeneous solutions of the RW equation
\begin{align}
\label{eqn:RWZans}
X_{lmn}^- &= \left(\frac{\e}{z} \right)^{-\nu - 1} (1 + A_2 \eta^2 + A_4 \eta^4 + \cdots + A_{2l} \eta^{2l} 
+ \mathcal{O}(\eta^{2l+1}) ) , \notag \\
X_{lmn}^+ &= (z)^{-\nu} (1 + B_2 \eta^2 + B_4 \eta^4 + \cdots + B_{2l} \eta^{2l} + \mathcal{O}(\eta^{2l+1}) )  ,
\end{align}
as a general-$l$ PN expansion with undetermined coefficients.  Here $A_i$ and $B_i$ are functions of $z, \e, l$.  
The original ansatz \cite{BiniDamo13} employed different prefactors, namely $r^{l+1}$ and $r^{-l}$.  This was 
modified \cite{KavaOtteWard15} to use $\nu$ in the exponents, which removes logarithmic terms from the $A_i$ and 
$B_i$ coefficients.  The PN expansion of $\nu$ itself is found using the continued fraction method (but for general 
$l$) and then the expansions are plugged into the homogeneous RW equation 
\be
\left[ \left(1 - \frac{\e}{z} \right) \frac{\pa}{\pa z} \left(\left(1 - \frac{\e}{z} \right) 
\frac{\pa}{\pa z} \right) + \eta^2 
+ \left(1 - \frac{\e}{z} \right) \left(\frac{l (l + 1)}{z^2} - \frac{3 \e}{z^3} \right)\eta^2 \right] 
X^{\pm}_{lmn} = 0 .
\ee
The ODE is then solved order-by-order.  For even parity, Zerilli equation solutions are derived from the RW solutions 
via the Detweiler-Chandrasekhar transformation \cite{Munn20}.  The ansatz \eqref{eqn:RWZans} does not fully 
incorporate the boundary conditions, which makes it break down at PN orders at and above $\mathcal{O}(\eta^{2l})$.  
If a target PN order $P$ is set, the ansatz will be useless for $l \le P$ and portions of the solution for those 
values of $l$ must be determined separately with the MST formalism.

Proceeding in this way, we obtain a general-$l$ PN expansion for $\nu$, the first few terms of which are
\begin{align}
\nu =&\,\,  l+\frac{24+13 l+28 l^2+30 l^3+15 l^4}{6 l+10 l^2-20 l^3-40 l^4-16 l^5} \e^2+ (51840+102816 l
-850608 l^2-1855326 l^3-675625 l^4+733273 l^5 \notag \\ &
+1217380 l^6+1397512 l^7+1355518 l^8+1520455 l^9 +1678310 l^{10}+1096830 l^{11}-8295 l^{12}
-605640 l^{13}  - \notag \\ &  
456120 l^{14} -147840 l^{15} -18480 l^{16}) \e^4/[8 (l-1) l^3 (1+l)^3 (2+l) (2 l-3) (2 l-1)^3 (1+2 l)^3 
(3+2 l)^3 (5+2 l)]  \notag \\& +\mathcal{O}(\e^6) ,
\end{align}
and obtain the general-$l$ expansions for the mode functions, which are again truncated after the first few terms 
\begin{align}
(z^l) X^+_{lmn} &= X_{\rm up}^{\rm ser} 
= 1 + \left[\frac{\e \left(-3+2 l+l^2\right)}{(1+l)(2z)}+\frac{z^2}{-2+4 l}\right] \eta^2    \\ &
+ \left[\frac{\e^2 l \left(12-29 l+4 l^2+11 l^3+2 l^4\right)}{4 (3+2 l) \left(-1+l+2 l^2\right) z^2}
+\frac{\e \left(4-l+8 l^2+l^3\right) z}{4 l \left(-1+l+2 l^2\right)} 
+\frac{(1+l) z^4}{8 (-3+2 l) \left(-1+l+2 l^2\right) }\right] \eta^4 +\mathcal{O}(\eta^6),  \notag \\
\left(\frac{\e}{z}\right)^{l+1} X^-_{lmn} &= X_{\rm in}^{\rm ser}  =
1- \left[\frac{\e}{2z} \left(-\frac{4}{l}+l\right)+\frac{z^2}{6+4 l}\right] \eta^2  \\ &
+ \left[\frac{\e^2 (-3+l) (-2+l) (1+l) (2+l)}{l (-1+2 l) 4 z^2}
+\frac{\e (-12+(-7+l) l (2+l)) z}{l (1+l) (3+2 l)4}+\frac{z^4}{(15+16 l+4 l^2)8}\right] \eta^4 
+ \mathcal{O}(\eta^6) .
\notag
\end{align}
Here we defined $X^{\rm ser}$ as the normalized PN series that begin at $\mathcal{O}(1)$.  It is useful to 
factor out leading terms $(z^{-l})$ and $(\e/z)^{-l-1}$ at each step of the calculation so that PN orders do not 
depend on $l$.  Eventually, all $l$-dependent powers of $\eta$ will cancel in the metric perturbation due to their 
corresponding presence in the Wronskian. 

The next few steps in the general-$l$ procedure are identical to the specific-$l$ case \cite{Munn20,HoppKavaOtte16}.  
The Wronskian and source terms are expanded and then the $C_{lmn}^\pm$ normalization coefficients are computed using 
\eqref{eqn:ClmnIntChi}.  The general-$l$ expansions are significantly lengthier than their specific-$l$ counterparts, 
making this step a bottleneck in the calculation.  Of course, in applications to the orbital phase evolution in 
EMRI waveforms, the accuracy requirements on the conservative part of the self-force are relaxed relative to those 
on dissipative terms by a factor of the mass ratio \cite{HindFlan08}.

\subsection{Sums of spherical harmonics over $m$}

The construction of the full metric perturbation involves summation over all three mode indices $l,m,n$.  The 
summation over $n$ is straightforward, as only finite $n$ are needed to reach any particular order in the expansion 
over eccentricity $e$.  The summation over $l$ will range from $l=0$ to $l=\infty$, but the form of the summands 
will involve products and quotients of polynomials in $l$.  Infinite sums over these expressions are still trivial 
to execute in \textsc{Mathematica}.  This leaves the more difficult task of summing $m$ modes from $-l$ to $l$ for 
general $l$.  In the process of constructing the $l$-modes of the metric perturbation \eqref{eqn:MPchi}, we find 
the following two classes of sums 
\begin{align}
&\sum_{m=-l}^{l} m^{N} | Y_{lm}(\pi/2,0) |^2  \q \text{(even parity),}    \\
&\sum_{m=-l}^{l} m^{N} | \pa_{\theta}Y_{lm}(\pi/2,0) |^2 \q \text{(odd parity),}
\end{align}
where $N$ is any positive integer.  Sums of these types occur because one spherical harmonic factor explicitly 
appears in \eqref{eqn:MPchi} while a second spherical harmonic implicitly resides in the calculation of 
$C^{\pm}_{lmn}$.  Powers of $m$ come from powers of $\epsilon$ and PN expansion of the Fourier kernel.  
Closed-form expressions must be derived for both of these sums.

The evaluation of the first (even-parity) summation starts by using the spherical harmonic addition theorem, reduced 
to the following form for this case
\be
\label{eqn:addtheorem}
\sum_{m=-l}^l e^{im\vp} | Y_{lm}(\theta,0) |^2 
= \left(\frac{2 l + 1}{4 \pi} \right) P_l(\cos^2\theta + \sin^2\theta \cos\vp)  .
\ee
Then, the even-parity sum can be derived by differentiating multiple times
\be
\label{eqn:legendreEven}
 \sum_{m=-l}^l m^N | Y_{lm}(\pi/2,0) |^2 
 = \frac{\pa^N}{\pa \vp^N} \left[\sum_{m=-l}^l (-i)^N e^{im\vp} | Y_{lm}(\pi/2,0) |^2 \right]_{\vp=0}
 = (-i)^N \left(\frac{2 l + 1}{4 \pi} \right) \frac{\pa^N}{\pa \vp^N} \left[P_l(\cos\vp) \right]_{\vp=0}.
\ee
When $N$ is odd, the LHS is real while the RHS is imaginary.  Thus, sums for odd $N$ must vanish.  Equivalently, 
we can set $z = i \vp$ and make the Taylor expansion of $P_l(\cos (-i z))$ in $z$.  Except for the added factor of 
$(2 l + 1)/4 \pi$, the coefficient of the $z^N/(N!)$ term in the expansion will correspond to the desired sum 
over $m^N$. 

The odd-parity summation requires more effort but it can be derived by taking a pair of $\theta$ derivatives
of the even-parity addition formula
\be
\frac{\pa^2}{\pa\theta^2} \left(\sum_{m=-l}^l e^{m z} | Y_{lm}(\theta,0) |^2 \right)
= 2 \sum_{m=-l}^l e^{m z} | \pa_{\theta}Y_{lm}(\theta,0) |^2 
+ 2 \sum_{m=-l}^l e^{m z} \left(\pa^2_{\theta}Y_{lm}(\theta,0) \right) Y_{lm}(\theta,0) .
\ee
Rearranging and fixing the polar angle, we find
\be
\sum_{m=-l}^{l} e^{mz} | \pa_{\theta}Y_{lm}(\pi/2,0) |^2 
= \frac{1}{2}\frac{\pa^2}{\pa\theta^2} \left[\sum_{m=-l}^l e^{m z} | Y_{lm}(\theta,0) |^2 \right]_{\theta=\pi/2}
- \sum_{m=-l}^l e^{m z} \left[\pa^2_{\theta}Y_{lm}(\pi/2,0) \right] Y_{lm}(\pi/2,0) . 
\ee
The first portion can be easily written in terms of $P_l$ and the second term can be reduced using the spherical 
harmonic differential equation itself.  We then find
\begin{gather}
\label{eqn:legendreOdd}
\sum_{m=-l}^{l} e^{mz} | \pa_{\theta}Y_{lm}(\pi/2,0) |^2 
=\left(\frac{2 l + 1}{8 \pi} \right)\frac{\pa^2}{\pa\theta^2} 
\bigg(P_l(\cos^2\theta + \sin^2\theta \cos(-iz)) \bigg)_{\theta=\pi/2}  
\notag \\
+\, l (l+1) \left(\frac{2 l + 1}{4 \pi} \right) P_l(\cos(-iz)) 
- \left(\frac{2 l + 1}{4 \pi} \right) \pa^2_zP_l(\cos(-iz)),
\end{gather}
with the $N$th term in the Taylor series in $z$ giving the desired odd-parity summation over $m^N$.

An alternative means of evaluating the two classes of summations involves expressing them in terms of the Gauss 
${}_2F_1$ hypergeometric functions.  Indeed, it can be shown \cite{NakaSagoSasa03} that 
\be
\sum_{m=-l}^l e^{m z} | Y_{lm}(\pi/2,0) |^2 
= \left(\frac{2 l + 1}{4 \pi} \right) e^{l z} \, {}_2F_1(1/2,-l,1,1-e^{-2z}) ,
\ee
which is readily Taylor expanded in $z$ with the $N$th power term directly providing the even-parity result.  The 
approach based around expanding Legendre functions in \eqref{eqn:legendreEven} for calculating the even-parity sums 
is slightly faster than this second, alternative route with hypergeometric functions, so we retain use of the former 
in our \textsc{Mathematica} code.
  
The same paper gives the following odd-parity summation (except for an errant factor of 1/4)
\be
\label{eqn:oddParSumNSS}
\sum_{m=-l}^l e^{m z} | \pa_\theta Y_{lm}(\pi/2,0) |^2 = 
\left(\frac{2 l + 1}{\pi^2} \right) e^{(l-1) z} \frac{\G(3/2) \G(l + 1/2)}{\G(l)} \, {}_2F_1(3/2,-l+1,-l+1/2,e^{-2z}) .
\ee
This expression is not immediately easily expanded in $z$.  Instead, we can apply the hypergeometric identity
\begin{align}
{}_2F_1(a,b,c,z) =&  \frac{\G (c-a-b) \G (c)}{\G (c-a) \G (c-b)}{}_2F_1(a,b;a+b+1-c;1-z) \notag \\& 
 +\frac{\G (a+b-c) \G (c)}{\G (a) \G (b)} 
 (1-z)^{c-a-b}  {}_2F_1(c-a,c-b;c+1-a-b;1-z) ,
\end{align}
to make headway.  When substituted in \eqref{eqn:oddParSumNSS}, the second term on the right hand side in the 
identity vanishes for all $l$ of interest here since there is a Gamma function in the denominator $\G(b) = \G(-l+1)$ 
with negative argument.  Hence, the hypergeometric function itself in \eqref{eqn:oddParSumNSS} can be replaced 
with the following,
\be 
\frac{\G (-2) \G (1/2-l)}{\G (-1/2) \G (-l-1)} {}_2F_1(3/2,1-l;3;1-e^{-2z}) 
= (-1)^{l} \frac{\G(l+2)}{4 \sqrt{\pi}} \G (1/2-l){}_2F_1(3/2,1-l;3;1-e^{-2z}) ,
\ee
where we have canceled the formally diverging terms using
\begin{align}
\frac{\G (-2)}{\G (-l-1)} &= \left(\prod_{k=-l-1}^{-3} k \right) = \frac{1}{2}(-1)^{l+1} \G(l+2) .
\end{align}
Including the rest of the factors from \eqref{eqn:oddParSumNSS} and noting
\begin{align}
\G (-l+1/2) \G(l + 1/2) = \pi \sec l\pi = \pi (-1)^l ,
\end{align}
we arrive at 
\be
\sum_{m=-l}^{l} e^{mz} | \pa_{\theta}Y_{lm}(\pi/2,0) |^2 = 
\left(\frac{2 l + 1}{8 \pi} \right) e^{(l-1) z} l (l+1) \,{}_2F_1(3/2,-l+1;3;1-e^{-2z}) .
\ee
Taylor expanding the right hand side this expression in $z$ and plucking off the $m^N$ term provides the desired 
odd-parity sums and turns out to be much faster to execute in \textsc{Mathematica} than its Legendre function 
alternative.

With a means of handling the sums over $m$, the general-$l$ PN expansions for the mode functions can be inserted in 
\eqref{eqn:ClmnIntChi} to obtain the PN expansion of $C^{\pm}_{lmn}$ and in \eqref{eqn:MPchi} to obtain the $l$-modes 
of the metric perturbation.  The calculation proceeds much the same way as in the specific-$l$ case, though the 
general-$l$ expansions are found to be orders of magnitude larger and more cumbersome to manipulate.  

\section{Additional considerations in the conservative sector}
\label{sec:l01andReg}

Our previous papers \cite{Munn20,MunnETC20,MunnEvan20a} utilizing this code have focused on the dissipative sector.  
In the present effort, leading to a PN expansion of the redshift invariant, there are additional considerations 
that arise exclusively in the conservative sector.  The first of these is the computation of the low-order modes 
($l = 0,1$) and the second is mode-sum regularization. 

\subsection{Non-radiative modes}
\label{sec:l0l1}

The $l=0$ and $l=1$ modes are not addressed by the RWZ master equation and the metric perturbations for these 
modes must be found directly \cite{Zeri70,DetwPois04,BaraLous05,BaraSago07,SagoBaraDetw08}.  We follow the 
presentation found in \cite{HoppKavaOtte16}.  The $l=m=0$ monopole mode is even parity and was found by Zerilli to be
\be
p_{tt}^{0} = 2 \mu \left[ \frac{\mathcal{E}}{r} - \frac{f}{\mathcal{E} f_p r_p} 
\left(2 \mathcal{E}^2 - f_p \left(1 + \frac{\mathcal{L}^2}{r_p^2}\right)\right) \right] 
\Theta[r - r_p(t)], \qq  p_{rr}^0 = \frac{2 \mu \mathcal{E}}{f^2 r} \theta[r - r_p(t)] .
\ee
However, in this particular gauge the metric perturbation is not asymptotically flat, which can be seen by 
inspecting the $p_{tt}$ component.  Recovering asymptotic flatness is effected by introducing a gauge transformation 
\cite{SagoBaraDetw08,HoppKavaOtte16} involving just the $\xi^0$ component of the gauge generator.  This affects 
only $p_{tt}$ (for the $l=0$ mode) and leaves 
\be
p_{tt}^{0} = 2 \mu \frac{\mathcal{E}}{r} \Theta[r - r_p(t)]  
+ \frac{2 \mu f}{\mathcal{E} f_p r_p} \left[2 \mathcal{E}^2 - 
f_p \left(1 + \frac{\mathcal{L}^2}{r_p^2}\right)\right] \Theta[r_p(t) - r] .
\ee

For $l=1$, both even-parity ($m=1$) and odd-parity ($m=0$) contributions are present.  Gauge freedom allows the 
odd-parity mode to appear in a single metric component, 
\be
p_{t\vp}^{1} = -2 \mu \mathcal{L} 
\sin^2\theta \left(\frac{1}{r} \Theta[r - r_p(t)]  + \frac{r^2}{r_p^3} \Theta[r_p(t) - r] \right) ,
\ee
which is in a form suitable for our first-order perturbation calculations \cite{HoppKavaOtte16}.  The even-parity 
$l=m=1$ dipole mode is more complicated and expressions can be found in \cite{Zeri70,DetwPois04,HoppKavaOtte16}.  
However, this multipole part of the metric perturbation is understood to be a pure-gauge mode and its contribution 
to the redshift invariant (and presumably all other gauge-invariant quantities) vanishes locally.  

\subsection{Mode-sum regularization}
\label{sec:regularization}

The last major hurdle in the computation of local conservative quantities is that of regularization.  The 
retarded-time metric perturbation emerges from \eqref{eqn:MPchi} after summing over $l$, which diverges on the 
worldline of the particle.  A local gauge invariant quantity (like the redshift invariant) computed from the 
retarded field would itself diverge.  What is needed instead is to extract the \textit{effective} (regular) 
metric perturbation experienced by the particle, which is found through regularization.  

Regularization can be approached by using the splitting prescription of Detweiler and Whiting \cite{DetwWhit03}, 
which decomposes the retarded metric perturbation into regular and singular fields
\be
p_{\mu \nu}(x) = p_{\mu \nu}^{S}(x) + p_{\mu \nu}^{R}(x) .
\ee
As its name implies, the singular field is divergent at the location of the particle, sharing this aspect with the 
retarded field.  The singular and retarded fields satisfy the same inhomogeneous field equation but with different 
boundary conditions.  A consequence is that the regular field $p_{\mu \nu}^{R}$ is a solution to the homogeneous 
field equation.  Because of symmetry, the singular field makes no contribution to the self-force, leaving those 
effects to the regular part.  In this way, the regular metric perturbation added to the background Schwarzschild 
metric can be thought of as a smooth effective metric in which the point particle executes (perturbed) geodesic 
motion.  

The regular-singular decomposition can be incorporated as part of mode-sum regularization \cite{Bara01,BaraOri03b}.  
This approach takes advantage of the fact that while the retarded and singular fields are divergent on the worldline, 
their individual $l$-modes are finite.  Thus, if the $l$-modes of the singular field can be determined, they can be 
subtracted from the $l$-modes of the retarded field, allowing a convergent sum to be formed for the regular metric 
perturbation.  Like the retarded metric, the singular metric is gauge dependent.  The singular metric, or 
alternatively the self-force itself, has $l$ dependence that can be represented as an expansion in which each term 
has dependence that is polynomial in $l$ or reciprocal of a product of polynomials in $l$.  Most work has focused on 
expanding the singular field in Lorenz gauge and there the $l$-independent coefficients (i.e., regularization 
parameters) of these terms have been calculated for multiple orders \cite{HeffOtteWard12a}.  Only the first 
regularization parameter is needed in the case of the metric itself to achieve a convergent result, while the first 
two parameters are needed to regularize the self-force.  Because our approach is analytic, only the regularization 
parameters that are essential for convergence are needed.

However, because our focus in this paper is on a gauge-invariant scalar quantity (i.e., the redshift invariant), we 
can instead directly regularize the redshift invariant, avoiding the complications of components and gauge.  As 
noted by Detweiler \cite{Detw08}, the regularization scheme becomes gauge invariant when working with gauge 
invariant quantities.  Furthermore, we avoid the whole usual issue of whether to regularize the metric components 
in a tensor spherical harmonic basis or by treating each component in an expansion over scalar harmonics 
\cite{WardWarb15}.  We are thus able to extract the finite result using a single regularization parameter found 
by using Lorenz gauge.  

\section{The generalized redshift invariant}
\label{sec:ut}

\subsection{Background and implementation}

For an eccentric orbit, the redshift invariant is the average of $u^t = dt/d\tau$ integrated over proper time 
$\tau$ for one radial libration period \cite{BaraSago11, AkcaETC15, HoppKavaOtte16}.  This 
quantity is equivalent to the coordinate-time period, $T_r$, divided by the proper-time period, $\mathcal{T}_r$, and 
generalizes Detweiler's original redshift invariant, which was defined as the instantaneous value of $u^t$ for 
circular orbits \cite{Detw08,BlanETC10}.  All of the necessary tools to calculate the redshift invariant have been 
summarized in the previous sections.  

As mentioned before, this particular gauge-invariant quantity encodes important details of the conservative 
motion of the system.  The first-order conservative dynamics contribute at $\mathcal{O}(\varepsilon^0)$ in the 
cumulative EMRI phase, a level needed for the creation of accurate waveform templates in the LISA mission, making the 
redshift invariant especially valuable.  In addition, there is an exact correspondence between the PN expansion of 
$\langle u^t \rangle_\tau$ and the expansion of the $Q(1/r, p_r ;\nu)$ EOB potential, which governs the deviation from 
geodesic behavior in the EOB Hamiltonian \cite{Leti15,DamoJaraScha15,BiniDamoGera16a,HoppKavaOtte16,BiniDamoGera16c}.  
The transformation between these quantities is outlined in \cite{Leti15}.

Given our first-order self-force calculation, we seek the first-order correction to the ratio $T_r/\mathcal{T}_r$.  
To achieve a gauge-invariant result, we make the assumption that the (observable) radial libration frequency is held 
fixed in going from the background geodesic to the first-order perturbed orbit.  The result is that all of the 
necessary gauge-invariant information is contained within the first-order correction to $\mathcal{T}_r$ alone 
\cite{BaraSago11, AkcaETC15, HoppKavaOtte16}.  Thus, we can express $\langle u^t \rangle_\tau$ as
\be
 \left<u^t\right>_\tau = \frac{T_r}{\mathcal{T}_r + \D \mathcal{T}_r} =
 \frac{T_r}{\mathcal{T}_r} -  \D \mathcal{T}_r \frac{T_r}{\mathcal{T}_r^2} 
=  \left<u^t\right>_\tau^0 +  \left<u^t\right>_\tau^1  .
\ee
The first term, $T_r/\mathcal{T}_r$, is the geodesic value of the redshift invariant, which can be trivially 
calculated using the Darwin parameterization of the background orbit.  The second term is the conservative self-force 
correction, scaling as $\mu/M$ and requiring the calculation of the first-order piece of the proper time radial 
period $\D \mathcal{T}_r $.  This correction was found \cite{BaraSago11, AkcaETC15} to be given by a projection of 
the regular part of the metric perturbation
\be
\D \mathcal{T}_r = - \mathcal{T}_r \left< \frac{1}{2} p_{\mu \nu}^R u^{\mu} u^{\nu}\right>_\tau 
= - \mathcal{T}_r \left< \frac{1}{2} p_{\mu \nu} u^{\mu} u^{\nu} - H^{S} \right>_\tau .
\ee
Here the average is taken over a $\tau$ period and in the second equality the projection is made on the 
retard-time metric perturbation.  The term $H^{S}$ is the projection of the singular metric that must be subtracted 
off.  To obtain finite results, this subtraction is done in an $l$-mode by $l$-mode fashion using the leading order 
regularization parameter \cite{BaraSago11, HeffOtteWard12a, HoppKavaOtte16}
\begin{align}
\label{eqn:H0par}
H^{S} = \sum_{l} H_{[0]} = \sum_l \frac{2 \mu}{\pi \sqrt{\mathcal{L}^2 + r^2}} 
\mathcal{K}\left(\frac{\mathcal{L}^2}{\mathcal{L}^2 + r^2} \right) ,  
\end{align}
where $\mathcal{K}$ is the complete elliptic integral of the first kind.  Like the rest of our quantities, $H_{[0]}$
can be PN expanded in $1/p$ and expanded in $e$.  The series expansion of $H_{[0]}$ is trivial to calculate, with 
the leading few terms being
\begin{align}
H_{[0]} = \left(\frac{\mu}{M}\right) 
\left[ 
\frac{1}{p} \left( 1+ e \cos\chi \right) - \frac{1}{4 p^2} \left( 1 + e \cos\chi \right)^3 
+ \frac{1}{64 p^3} \left( \left( 1+ e \cos\chi \right)^3 
\left( 9 \left(1+ e \cos\chi \right)^2 - 16 (3 + e^2) \right) \right)
+ \cdots 
\right] .
\notag
\end{align}

The regularized redshift invariant is constructed from the individual $l$-dependent differences 
\be 
\left< \frac{1}{2} p_{\mu \nu}^{l} u^{\mu} u^{\nu}  \right>_\tau - \left< H_{[0]} \right>_\tau ,
\ee
which are then summed from $l=0$ to $l=\infty$.  The $l$-modes of the retarded-time metric perturbation, 
$p_{\mu \nu}^{l}$, are calculated in three different blocks.  The modes $l=0$ and $l=1$ are expanded using the 
non-radiative solutions in Sec.~\ref{sec:l0l1}, while the modes from $l=2$ to the integer part of the PN order 
minus 1 (which in this paper for 10PN means $l=9$) are expanded using specific-$l$ MST solutions, and lastly the 
remaining modes from the desired PN order to infinity are expanded using the general-$l$ ansatz from 
Sec.~\ref{sec:genLexps}.  Once $(1/2) p_{\mu \nu}^{l} u^{\mu} u^{\nu}$ is assembled (and regularized) for both 
specific and general $l$, the summation over $l$ is computationally efficient.

This procedure was first implemented in \cite{BiniDamoGera16a}, where the redshift invariant was expanded to 
6.5PN and $e^2$ in eccentricity and to 4PN and $e^4$.  Shortly thereafter, the expansion was taken 
\cite{HoppKavaOtte16} to 4PN through $e^{10}$.  Those efforts were followed by \cite{BiniDamoGera16c}, who extended 
the result to 4PN and $e^{20}$, as well as 9.5PN through $e^4$.  More recently, these latter authors improved the 
eccentric knowledge to 9.5PN and $e^{8}$ \cite{BiniDamoGera20b}, as that level was needed to complete a novel 
transcription of the redshift invariant to the scattering angle for hyperbolic orbits, which can be used to compute 
the full post-Minkowskian dynamics to high order.

This paper extends the PN and eccentricity expansion further by taking the redshift invariant to 10PN and $e^{20}$.  
More importantly, we have further analyzed each eccentricity function (in keeping with work in \cite{ForsEvanHopp16, 
MunnEvan19a, MunnETC20, MunnEvan20a}) to find those that can be manipulated either into closed-form expressions or 
into known infinite series.  By known we mean cases where an infinite series is derived from sums over Fourier spectra 
of low-order multipole moments, as we showed occurs in the dissipative sector for gravitational wave fluxes radiated 
to infinity.  Other sums over Fourier spectra of low-order multipoles were shown \cite{MunnEvan19a,MunnEvan20a} 
to yield sequences of closed-form expressions in the PN expansion of the fluxes to infinity.  Surprisingly, a number 
of these special functions, both closed form and infinite series, reappear in parts of the PN expansion of the 
(conservative) redshift invariant.  In the case of non-closed-form functions, we present resummations that rely on 
factoring out powers of $1 - e^2$ (often referred to as eccentricity singular factors) that improve the convergence 
of the remaining series as $e\rightarrow 1$ \cite{HoppKavaOtte16, ForsEvanHopp16, MunnEvan19a}.  In what follows, we 
present the redshift invariant in two different PN series, using first the compactness parameter $1/p$ and then the 
parameter $y$.  

\subsection{Redshift invariant as an expansion in $1/p$}

In terms of the compactness parameter $1/p$, circular-orbit studies \cite{KavaOtteWard15} lead us to expect the 
following form of the PN expansion of the redshift invariant 
\begin{align}
 \left<u^t\right>_\tau^1 
=& \, \, \left(\frac{\mu}{M}\right) \frac{1}{p} \bigg[ \mathcal{U}_0 + \frac{ \mathcal{U}_1}{p} 
+ \frac{ \mathcal{U}_2}{p^2} + \frac{\mathcal{U}_{3} }{p^3}
+ \Big( \mathcal{U}_{4} + \mathcal{U}_{4L} \log p \Big) \frac{1}{p^4}
+ \Big( \mathcal{U}_{5} + \mathcal{U}_{5L} \log p \Big) \frac{1}{p^5}
+ \frac{ \mathcal{U}_{11/2}}{p^{11/2}}  
\notag \\&
+ \Big( \mathcal{U}_{6} + \mathcal{U}_{6L} \log p \Big) \frac{1}{p^6}   
+ \frac{ \mathcal{U}_{13/2}}{p^{13/2}}     
+ \Big( \mathcal{U}_{7} + \mathcal{U}_{7L} \log p 
+ \mathcal{U}_{7L2} \log^2 p \Big) \frac{1}{p^7}
+ \frac{\mathcal{U}_{15/2}}{p^{15/2}}  
\notag \\&
+ \Big( \mathcal{U}_{8} + \mathcal{U}_{8L} \log p 
+ \mathcal{U}_{8L2} \log^2 p \Big) \frac{1}{p^8}  
+ \Big( \mathcal{U}_{17/2} + \mathcal{U}_{17/2L} \log p \Big) \frac{1}{p^{17/2}}  
+ \Big( \mathcal{U}_{9} + \mathcal{U}_{9L} \log p + \mathcal{U}_{9L2} \log^2 p \Big) \frac{1}{p^9}  
\notag \\&
+ \Big( \mathcal{U}_{19/2} + \mathcal{U}_{19/2L} \log p \Big) \frac{1}{p^{19/2}}  
+ \Big( \mathcal{U}_{10} + \mathcal{U}_{10L} \log p 
+ \mathcal{U}_{10L2} \log^2 p + \mathcal{U}_{10L3} \log^3 p \Big) \frac{1}{p^{10}}  
+ \cdots \bigg]  ,  
\end{align}
where each one of the $\mathcal{U}_k$ is a function of eccentricity $e$ (which if 
appropriately scaled is sometimes called an enhancement function).  Our perturbation results when sorted on $p$ 
dependence allow the $\mathcal{U}_k(e)$ functions to be read off.  Due to the increasing complexity of the expansion 
with PN order, we limit our presentation in this paper to 8.5PN order.  The full results to 10PN will be available on 
the Black Hole Perturbation Toolkit \cite{BHPTK18} website.  

We find that the first few leading terms (0PN, 1PN, 2PN, and 3PN) all have simple closed-form expressions
\begin{align}
\mathcal{U}_0 &= -(1 - e^2) , \notag \\
\mathcal{U}_1 &= -2 \left(1 - e^2 \right)^2, \notag \\
\mathcal{U}_2 &= \left(1-e^2\right)^2 \left(9-5 e^2 \right) +\left(1-e^2 \right)^{3/2} \left(-14+9 e^2 \right), 
\notag \\
\mathcal{U}_3 &= \left(1-e^2 \right)^2 \left(28-8 e^2-4 e^4 \right)+ \left(1-e^2 \right)^{3/2} 
\left[-\frac{205}{3}+\frac{41 \pi^2}{32}+e^2 \left(-\frac{241}{6}+\frac{41 \pi ^2}{64}\right)+\frac{27 e^4}{2}\right] .
\end{align}
The functions $\mathcal{U}_0(e)$ and $\mathcal{U}_1(e)$ were previously known \cite{HoppKavaOtte16}.  Those authors 
also gave $\mathcal{U}_2(e)$ and $\mathcal{U}_3(e)$ as power series in eccentricity through $e^{10}$, but our 
calculation shows these terms to be in fact closed in form. 

At 4PN order, a $\log(p)$ term makes its first appearance.  The 4PN non-log function contains combinations of 
transcendentals similar to the 3PN flux at infinity.  Intriguingly, the 4PN log term, $\mathcal{U}_{4L}(e)$, is 
exactly proportional to the Peters-Mathews quadrupole flux term, $\mathcal{L}_0(e)$ \cite{PeteMath63}
\begin{align}
\label{eqn:U4L}
\mathcal{U}_{4L} = \frac{64}{5} \left(1-e^2\right)^{3/2} \left( 1+\frac{73 e^2}{24}+\frac{37 e^4}{96} \right) 
= \frac{64}{5} \left(1-e^2\right)^{5} \mathcal{L}_0(e) .
\end{align}
As we recall, the Peters-Mathews flux is found to be a sum over the Fourier power spectrum $g(n,e)$ of the Newtonian 
mass quadrupole \cite{PeteMath63, BlanScha93} (see also \cite{MunnEvan19a}),
\be
\mathcal{L}_{0}(e)=\sum_{n=1}^\infty g(n,e)=\frac{1}{(1-e^2)^{7/2}}\bigg(1+\frac{73}{24}e^2 +\frac{37}{96}e^4\bigg) . 
\ee

The similarities with the 3PN flux (and the fact that the 3PN log flux term shows up in the 3PN non-log flux) led 
us to seek a compact expression for the 4PN non-log term, $\mathcal{U}_{4}(e)$, resembling that of $\mathcal{L}_3(e)$.  
The procedure is described in Sec.~IV of \cite{MunnEvan19a}.  We found the following segregation of terms 
\begin{align}
\label{eqn:U4}
\mathcal{U}_{4} =& \left(1-e^2\right)^{3/2} \bigg[-\frac{1963}{45}-\frac{21182 e^2}{45}+\frac{1469 e^4}{9}
-\frac{129 e^6}{16} +\sqrt{1-e^2} \bigg( -\frac{1508}{45}+\frac{5281 e^2}{90}-\frac{159 e^4}{2}+5 e^6 \bigg) \bigg]  
\\
\notag
- &2 \left[\g_E +\log \left(\frac{8 \left(1-e^2\right)^{3/2}}{1+\sqrt{1-e^2}}\right) \right] \mathcal{U}_{4L}(e) 
+ \left(1-e^2\right)^{3/2} \bigg(\frac{677 \pi ^2}{512} +\frac{17879 \pi ^2}{1536}e^2 
+\frac{29665 \pi ^2}{12288} e^4\bigg)  
- \frac{128}{5} \left(1-e^2\right)^{5} \La_0(e) .
\end{align}
As is evident, the 4PN non-log redshift invariant separates into a set of closed terms, including one with 
$\mathcal{U}_{4L}(e)$, along with an added term containing a function denoted by $\La_0(e)$.  The $\La_0(e)$ function 
turns out to be an infinite series.  It is, in fact, the first in a sequence of functions that we previously 
identified \cite{MunnEvan19a}.  In that paper, we showed that the functions $\La_1(e)$, $\La_2(e)$, etc 
appeared in the leading-logarithm sequence in the energy flux at infinity.  The first such term that showed up in 
the flux was $\La_1(e)$, also known as $\chi(e)$ \cite{ArunETC08a,Blan14} which appears in the 3PN energy flux. 
The sequence of functions $\La_k(e)$ that we defined contained a first element, $\La_0(e)$, which is given by
\be
\La_0(e) =\sum_{n=1}^{\infty} \log\Big(\frac{n}{2}\Big) \, g(n,e) .
\ee
Even though this particular function made no appearance in the energy flux, it does interestingly now appear in 
the redshift at 4PN order.

At 5PN order there are log and non-log terms.  The 5PN log term in the redshift is found to be another closed-form 
function
\begin{align}
\label{eqn:U5L}
\mathcal{U}_{5L} = \left(1-e^2\right)^{3/2} \left(-\frac{956}{105}-\frac{2026 e^2}{21}-\frac{211 e^4}{10}
+\frac{2393 e^6}{420}\right) .
\end{align}
Just as with the 4PN case, the 5PN log term is coupled into the 5PN non-log term.  Following the procedure used 
above, we can seek a compact expression for the 5PN non-log redshift.  The result is similar 
\begin{align}
\label{eqn:U5}
\mathcal{U}_{5} =& \,\, \left(1-e^2\right)^{3/2} \bigg[ \left(\frac{711289}{1575}-\frac{166903 e^2}{252}
-\frac{5861983 e^4}{4200} +\frac{4691137 e^6}{7200}-\frac{5625 e^8}{64}\right) 
+\sqrt{1-e^2} \bigg(\frac{61433}{1050}   \notag \\&
+\frac{366389 e^2}{6300} +\frac{1825589 e^4}{3150}-\frac{897 e^6}{4}+26 e^8 \bigg) \bigg]   
-2 \left[\g_E +\log \left(\frac{8 \left(1-e^2\right)^{3/2}}{1+\sqrt{1-e^2}}\right) \right] \mathcal{U}_{5L}  
+ \mathcal{U}^{\chi}_{5}      
\\
\notag
& + \pi^2 \left(1-e^2\right)^{3/2} \bigg[ -\frac{64771}{768}-\frac{122659 e^2}{768}+\frac{106757 e^4}{2048}
+\frac{9679 e^6}{3072} +\sqrt{1-e^2} \left(\frac{369}{64}-\frac{369 e^2}{128}-\frac{369 e^4}{128}\right) \bigg] ,
\end{align}
with a set of new closed form terms and with a single remaining infinite series, denoted by $\mathcal{U}^{\chi}_5$, 
that soaks up the appearance of $\log(2)$, $\log(3)$, etc transcendentals.  This 
$\chi$-like function has not (yet) been determined in terms of multipoles and thus it is only known in our present 
calculation as an expansion to $e^{20}$:
\begin{align}
\mathcal{U}^{\chi}_{5} =& \,\, (1-e^2)^{3/2} \bigg[ \bigg(\frac{248 \log (2)}{7}-\frac{243 \log (3)}{7}\bigg)
+\bigg(-\frac{39380 \log (2)}{21} + \frac{80919 \log (3)}{70}\bigg) e^2  + \notag \\&
\bigg(\frac{1159803 \log (2)}{35}-\frac{3082941 \log (3)}{320}-\frac{9765625 \log (5)}{1344}\bigg) e^4
+\bigg(-\frac{611462239 \log (2)}{1890}+\frac{4054941 \log (3)}{896}   \notag \\&
+\frac{3299921875 \log (5)}{24192}\bigg) e^6  +\bigg(\frac{32367232 \log (2)}{15}
+\frac{73001048877 \log (3)}{114688} -\frac{3402892421875 \log (5)}{3096576}  -  \notag \\&
\frac{96889010407 \log (7)}{442368}\bigg) e^8
- \bigg(\frac{63126936562 \log (2)}{4725} + \frac{34356489334353 \log (3)}{5734400}  
- \frac{32100152734375 \log (5)}{6193152}     \notag \\&
- \frac{28419552326641 \log (7)}{7372800}\bigg) e^{10}
+\bigg(\frac{5276536312963 \log (2)}{60750}+\frac{10364402788528311 \log (3)}{458752000}   \notag \\&
-\frac{7174481733359375 \log (5)}{445906944}-\frac{322609178993859793 \log (7)}{10616832000}\bigg) e^{12}
+\bigg(-\frac{5186372399556326 \log (2)}{10418625}  +  \notag \\&
\frac{1135462692217658751 \log (3)}{44957696000}
+\frac{1026818179405390625 \log (5)}{43698880512}  
+\frac{9149100203955857957 \log (7)}{63700992000}\bigg) e^{14}  +  \notag \\&
\bigg(\frac{748264799632979347 \log (2)}{333396000}
-\frac{609082161701573645199 \log (3)}{822083584000}   
+\frac{13203573865427998046875 \log (5)}{89495307288576}    \notag \\&
-\frac{14949272565618057218377 \log (7)}{32614907904000}
-\frac{81402749386839761113321 \log (11)}{2237382682214400}\bigg) e^{16}    \notag \\&
+\bigg(-\frac{883299584070658267147 \log (2)}{108020304000}
+\frac{48265448696337440140569 \log (3)}{11509170176000}     \notag \\&
-\frac{7667417741966072528515625 \log (5)}{4832746593583104}
+\frac{222715444298861595602129 \log (7)}{211344603217920}    \notag \\&
+\frac{236352943152570442830239113 \log (11)}{362455994518732800}\bigg) e^{18}
+\bigg(\frac{290351111127874858038001 \log (2)}{10802030400000}    \notag \\&
-\frac{974062330634519749930801293 \log (3)}{73658689126400000}
+\frac{423462131605851394906953125 \log (5)}{51549296998219776}   \notag \\&
-\frac{3843873591325549271865416797 \log (7)}{2113446032179200000}
-\frac{1568842386573481329215281289749 \log (11)}{289964795614986240000}   \notag \\&
-\frac{91733330193268616658399616009 \log (13)}{289964795614986240000}\bigg) e^{20}  + \cdots \bigg] . 
\end{align}

The first half-integer PN term arises at 5.5PN order \cite{ShahFrieWhit14}.  For circular orbits, this term is 
known to be associated with the effect of the 1.5PN energy flux tail showing up in the redshift factor 
\cite{BlanFayeWhit14a}.  In our calculations for eccentric orbits this function emerged as being exactly proportional 
to the eccentricity dependence of the 1.5PN energy flux enhancement function $\vp(e)$ 
\cite{BlanScha93,ArunETC08a,Blan14,ForsEvanHopp16,MunnEvan19a}
\be
\label{eqn:U5p5}
\mathcal{U}_{11/2} = - \frac{13696}{525} \pi  \left(1-e^2\right)^{13/2}  \vp(e) ,
\ee
a result that appears in \cite{BiniDamoGera20a} in slightly different notation.  It is well known that the tail 
enhancement function is found \cite{BlanScha93} as an infinite series over the Newtonian quadrupole moment power 
spectrum
\be
\varphi(e) = \sum_{n=1}^\infty \frac{n}{2} \, g(n,e) ,
\ee
and is therefore easily calculable to arbitrary order in $e^2$.  In \cite{MunnEvan19a} we found this function 
to be the first element, $\Theta_0(e)$, in an infinite sequence of functions, $\Theta_k(e)$, that appear in the 
subleading-log (or 3PN-log) terms in the energy flux.  Furthermore, as shown in \cite{ForsEvanHopp16}, it is 
the combination $\left(1 - e^2 \right)^5 \varphi(e)$ that is an infinite series in $e^2$ with diminishing 
coefficients and which limits on a finite number as $e \rightarrow 1$.

At 6PN order, the $\log(p)$ term is found to also be a closed-form function in $e^2$, but with the wrinkle that 
a lower-order log term, $\mathcal{U}_{4L}$, reappears
\begin{align}
\label{eqn:U6L}
\mathcal{U}_{6L} = \left(1-e^2\right)^{3/2} \left(-\frac{419576}{2835}-\frac{3187312 e^2}{2835}
-\frac{831494 e^4}{945} -\frac{59098 e^6}{315}-\frac{12889 e^8}{5040}\right)  
+\frac{9}{2} \, (1-e^2)^{3/2} \, \mathcal{U}_{4L} .
\end{align}
As with 5PN, we can make some headway in analyzing the structure of the 6PN non-log term.  We find, for example, 
that the 6PN log term couples into the 6PN non-log.  We also find several closed-form functions multiplying the 
appearance of the $\pi^2$ and $\pi^4$ transcendental combinations.  Beyond that there is an infinite series with 
rational coefficients and another series, $\mathcal{U}^{\chi}_6$, that contains $\log(2)$, $\log(3)$, etc 
transcendentals.  At this point, we have not been able to manipulate the rational-number series into a closed-form 
expression, which was possible in the case of $\mathcal{U}_{4}$ and $\mathcal{U}_{5}$.  Instead, both of these 
series are given as expansions to $e^{20}$
\begin{align}
\label{eqn:U6}
\mathcal{U}_{6} =&  \,\,(1-e^2)^{3/2} \bigg(\frac{17083661}{4050}+\frac{2700577231 e^2}{132300}
+\frac{161896927 e^4}{8400}+\frac{30273005737 e^6}{9525600}-\frac{58376487559 e^8}{152409600}  \notag \\&
-\frac{5527943783 e^{10}}{14515200}-\frac{36259997113 e^{12}}{174182400}
-\frac{67759073909 e^{14}}{487710720}-\frac{12979122851 e^{16}}{123863040}  \notag \\ &
-\frac{37020766301 e^{18}}{445906944}-\frac{433129626457 e^{20}}{6370099200} + \cdots \bigg)
-2 \left[\g_E +\log \left(\frac{8 \left(1-e^2\right)^{3/2}}{1+\sqrt{1-e^2}}\right) \right] \mathcal{U}_{6L}  \notag \\&
+\pi^4 \left(1-e^2\right)^{3/2} \left( \frac{2800873}{262144} + \frac{27872821 e^2}{524288}
+\frac{41197641 e^4}{2097152}-\frac{135909 e^6}{4194304}  \right)   \notag \\&
+ \pi^2 \left(1-e^2\right)^{3/2}  \bigg[-\frac{1231647119}{1769472}-\frac{4365848063 e^2}{884736}
-\frac{8200800977 e^4}{2359296}-\frac{68605901 e^6}{262144}-\frac{38113839e^8}{8388608}  \notag \\&
+\left(1-e^2\right)^{3/2} \left(-\frac{8339}{1024}+\frac{72005 e^2}{1024}+\frac{191331 e^4}{8192}\right) \bigg]
+  \mathcal{U}^{\chi}_{6} , \\
\mathcal{U}^{\chi}_{6} =& \,\, (1-e^2)^{3/2} \bigg[ \bigg(\frac{1215 \log (3)}{7}-\frac{652336 \log (2)}{2835}\bigg)
+\bigg(\frac{30101992 \log (2)}{2835}-\frac{85779 \log (3)}{40}-\frac{9765625 \log (5)}{4536}\bigg)   \notag \\&
\times e^2
+\bigg(-\frac{177664066 \log (2)}{945}-\frac{23175639 \log (3)}{2240}+\frac{3353515625 \log (5)}{36288}\bigg) e^4
+\bigg(\frac{5725580404 \log (2)}{2835}    \notag \\&
+\frac{15974601543 \log (3)}{17920} 
-\frac{3033371509375 \log (5)}{2612736}-\frac{96889010407 \log (7)}{373248}\bigg) e^6
+\bigg(-\frac{887519089375 \log (2)}{40824}    \notag \\&
-\frac{6552704842893 \log (3)}{573440}  
+\frac{693828478128125 \log (5)}{83607552}+\frac{87590207685169 \log (7)}{11943936}\bigg) e^8   \notag \\&
+\bigg(\frac{2392238577122749 \log (2)}{10206000}  
+\frac{3631791410461107 \log (3)}{57344000}
-\frac{162963875003125 \log (5)}{3981312}  -   \notag \\&
\frac{1019942547198706943 \log (7)}{11943936000}\bigg)e^{10}  
+\bigg(-\frac{105745169881200781 \log (2)}{52488000}
+\frac{3877667156883681 \log (3)}{65536000}   +   \notag \\&
\frac{62356409327340625 \log (5)}{573308928}   
+\frac{102210299549702981063 \log (7)}{171992678400}\bigg) e^{12}
+\bigg(\frac{3133171953524932019 \log (2)}{240045120}  -  \notag \\&
\frac{83389586865982335 \log (3)}{22478848}
+\frac{11457047215621220496875 \log (5)}{18877916381184}  
-\frac{6522844667781563185519 \log (7)}{2293235712000}   \notag \\&
-\frac{81402749386839761113321 \log (11)}{471947909529600}\bigg) e^{14}
+\bigg(-\frac{4813254866666992632161 \log (2)}{72013536000}     \notag \\&
+\frac{177380873493979007695959 \log (3)}{5754585088000}
-\frac{25670728576758333505178125 \log (5)}{2416373296791552}   \notag \\&
+\frac{8810523169175992438990487 \log (7)}{880602513408000}   
+\frac{254291295259872635488850159 \log (11)}{60409332419788800}\bigg) e^{16}    \notag \\&
+\bigg(\frac{9192410559014602065623101 \log (2)}{31109847552000}   
-\frac{53375870964426295835250711 \log (3)}{368293445632000}     \notag \\&
+\frac{122391122829784578980588140625 \log (5)}{1565809896320925696}  
-\frac{7642813738393655784243245839 \log (7)}{285315214344192000}   \notag \\&
-\frac{188665985427182853199195707341 \log (11)}{3994413000818688000}   
-\frac{91733330193268616658399616009 \log (13)}{39145247408023142400}\bigg) e^{18}    \notag \\&
+\bigg(-\frac{11551363401726154564820010019 \log  (2)}{9332954265600000}  
+\frac{5974056293971987433032285851 \log (3)}{14731737825280000}    \notag \\&
-\frac{531598234297333443052911578125 \log (5)}{1391831018951933952}   
+\frac{2732223161911355671416537362081 \log (7)}{57063042868838400000}   \notag \\&
+\frac{12865262475131526047951261471976161 \log (11)}{39145247408023142400000}   \notag \\&
+\frac{63598994107503492021893060068543 \log (13)}{1118435640229232640000}\bigg) e^{20} + \cdots 
\bigg] .
\end{align}

At 6.5PN the redshift function is found to be another (presumably) infinite rational-number series, which we 
calculated to $e^{20}$
\begin{align}
\label{eqn:U6p5}
\mathcal{U}_{13/2} =& \pi \, (1-e^2)^{3/2} \bigg(\frac{81077}{3675}+\frac{7082924 e^2}{11025}
+\frac{19545681 e^4}{15680}+\frac{171593203 e^6}{705600}-\frac{24169567 e^8}{4644864}  \notag \\&
-\frac{5517037829 e^{10}}{10160640000}+\frac{1289360091619 e^{12}}{7803371520000}
-\frac{1698130226071 e^{14}}{229419122688000}   
-\frac{377560795098119 e^{16}}{587312954081280000}  \notag \\&
+\frac{6704699069179 e^{18}}{14158437285888000}
+\frac{8520571675445796049 e^{20}}{76115758848933888000000} + \cdots \bigg) .
\end{align}
As a term that is 1PN relative to the 5.5PN redshift function, it is possible that the 6.5PN redshift may represent 
some combination of the 1.5PN and 2.5PN energy flux tail functions (or more properly, of the 0PN and 1PN source 
motion multipole moments).  We will leave the search for that connection to a future paper.  

The 7PN terms represent an added level of complexity, with a first appearance of a $\log^2(p)$ term.  The 
$\log^2(p)$ term is of closed form and is directly proportional to the 3PN log energy flux function, $F(e)$ 
\cite{ArunETC08a,Blan14}.  The log term reveals a structure similar to that of the 4PN non-log function, featuring a 
return appearance of $\mathcal{U}_{7L2}$ and containing another one of the 3PN energy flux functions, in this case 
$\chi(e) = \La_1(e)$ \cite{ArunETC08a,Blan14,MunnEvan19a} itself.  Our expansion of the rational-number series part, 
however, stops short of providing enough information to allow it to be manipulated into a closed form.  Finally, 
the 7PN non-log part is an infinite series with numerous transcendentals and powers of transcendental numbers as 
coefficients.  We present only a few coefficients here, saving the rest for the online repository \cite{BHPTK18}.  
These three functions are
\begin{align}
\label{eqn:U7}
\mathcal{U}_{7} =& \, \, (1-e^2)^{3/2} \bigg[ \frac{12624956532163}{382016250}
-\frac{10327445038 \g_E}{5457375}+\frac{109568 \g_E^2}{525}-\frac{9041721471697 \pi^2}{2477260800} 
-\frac{23851025 \pi ^4}{16777216}   \notag\\&
- \frac{16983588526 \log (2)}{5457375}+\frac{438272}{525} \g_E \log (2)
+\frac{438272 \log ^2(2)}{525}-\frac{2873961 \log (3)}{24640}-\frac{1953125 \log (5)}{19008}  \notag \\&
-\frac{2048 \zeta (3)}{5}  + \bigg(  \frac{40501543520891}{125023500}-\frac{37267116191 \g_E}{1091475}
+\frac{931328 \g_E^2}{315}-\frac{37593336465137 \pi ^2}{990904320}   \notag \\&
-\frac{12464105531 \pi ^4}{251658240}-\frac{196923520603 \log (2)}{5457375}
+\frac{1753088 \g_E \log (2)}{1575}-\frac{6683648 \log ^2(2)}{1575}  \notag \\&
-\frac{55044555129 \log (3)}{784000} 
+\frac{1872072}{175} \g_E \log (3)+\frac{1872072}{175} \log (2) \log (3)+\frac{936036 \log ^2(3)}{175}  \notag \\&
+\frac{2439453125 \log (5)}{114048}-\frac{17408 \zeta (3)}{3} \bigg) e^2   
+ \bigg( \frac{17120008508868443}{27505170000}-\frac{917258418203 \g_E}{10914750}
+\frac{8852752 \g_E^2}{1575}    \notag \\&
-\frac{536485713928139 \pi ^2}{9909043200}    
+\frac{24259399603 \pi ^4}{62914560}-\frac{952475674979 \log (2)}{1212750}
+\frac{393441568 \g_E \log (2)}{1575}   \notag \\&
+\frac{740753296 \log ^2(2)}{1575}  +\frac{66160453137993 \log (3)}{55193600}-\frac{4212162}{35} \g_E \log (3)
-\frac{4212162}{35} \log (2) \log (3)     \notag \\&
-\frac{2106081 \log ^2(3)}{35} -  \frac{1472676171875 \log (5)}{2838528} 
-\frac{678223072849  \log (7)}{6082560}-\frac{165472 \zeta (3)}{15}  \bigg) e^4 + \cdots \bigg]  , 
\\
\label{eqn:U7L}
\mathcal{U}_{7L} =& \, \, (1-e^2)^{3/2} \bigg( \frac{5163722519}{5457375}+\frac{36697636511 e^2}{2182950}
+\frac{833592361883 e^4}{21829500}+\frac{219994202647 e^6}{14553000}    
\notag \\&
+  \frac{346091521507 e^8}{232848000}  
+\frac{5895396959 e^{10}}{9504000}+\frac{41803247 e^{12}}{108000}+\frac{16344589 e^{14}}{60480}
+\frac{15771376787 e^{16}}{77414400}    
\notag \\&
+\frac{178135189 e^{18}}{1105920}  
+\frac{29119129511 e^{20}}{221184000} + \cdots \bigg) 
-4 \left[\g_E +\log \left(\frac{8 \left(1-e^2\right)^{3/2}}{1+\sqrt{1-e^2}}\right) \right] \mathcal{U}_{7L2}  
\notag \\&
- \frac{109568}{525} (1 - e^2)^8 \chi(e), 
\\
\label{eqn:U7L2}
\mathcal{U}_{7L2} =& \, \, (1-e^2)^{3/2} \left(\frac{27392}{525}\right) \left(1 +\frac{85}{6} e^2
+ \frac{5171}{192} e^4 + \frac{1751}{192} e^6 + \frac{297}{1024} e^8 \right) 
=  (1-e^2)^{8} \left(\frac{27392}{525}\right) F(e) .
\end{align}

As might be expected, the 7.5PN term, as the third half-integer term and 2PN correction to the 5.5PN redshift, is 
an infinite series with rational coefficients
\begin{align}
\label{eqn:U7p5}
\mathcal{U}_{15/2} =& \pi \, (1-e^2)^{3/2} \bigg( \frac{82561159}{467775}+\frac{73692788269 e^2}{26195400}
+\frac{336208042337 e^4}{69854400}+\frac{51301033584109 e^6}{22632825600}  \notag \\&
+\frac{117144137627477 e^8}{241416806400}-\frac{5009246358913537 e^{10}}{57940033536000}
-\frac{4421924451335987 e^{12}}{166867296583680}   \notag \\&
-\frac{6030593084685188237 e^{14}}{454249862922240000}
-\frac{639047634512255985157 e^{16}}{74756548869488640000}  \notag \\&
-\frac{213545959135760979200821 e^{18}}{37677300630222274560000}
-\frac{90046530071830147716010793 e^{20}}{22606380378133364736000000}  + \cdots \bigg)  .
\end{align}

The 8PN redshift contributions are similar in structure to features seen at 5PN, 6PN, and 7PN.  Like 7PN, there is an 
8PN $\log^2(p)$ term that is closed in form.  The 8PN log term contains a return appearance of $\mathcal{U}_{8L2}$ 
and it further separates, like the 5PN and 6PN non-log terms, into a $\chi$-like series containing log transcendental 
number coefficients and a remaining rational-number series.  In this case, like 7PN log, the rational-number series 
in the 8PN log term is only known to $e^{20}$.  The 8PN non-log term is a complex series, with only a few coefficients 
shown here, leaving the rest to the online repository \cite{BHPTK18}.  We find
\begin{align}
\label{eqn:U8}
\mathcal{U}_{8} =& \, \, (1-e^2)^{3/2} \bigg[ -\frac{7516581717416867}{34763478750}
-\frac{1526970297506 \g_E}{496621125}-\frac{108064 \g_E^2}{2205} 
-\frac{246847155756529 \pi ^2}{18496880640}    \notag \\&
+\frac{22759807747673 \pi ^4}{6442450944} -\frac{1363551923554 \log (2)}{496621125}
-\frac{3574208 \g_E \log (2)}{3675}   -\frac{2143328 \log ^2(2)}{1575}   \notag \\&
-\frac{2201898578589 \log (3)}{392392000} + \frac{37908}{49} \g_E \log (3)+\frac{37908}{49} \log (2) \log (3)
+\frac{18954 \log ^2(3)}{49}  -\frac{41408 \zeta (3)}{105}    \notag \\&
+\frac{798828125 \log (5)}{741312} + \bigg( -\frac{11055650107673008247}{6883168792500}
+\frac{4072595375711 \g_E}{496621125} - \frac{36209008 \g_E^2}{3675}    \notag \\&
-\frac{45536963437709173 \pi ^2}{277453209600}
+\frac{1900985448773627 \pi ^4}{64424509440}-\frac{6752485219069 \log (2)}{23648625}   
+\frac{77260192 \g_E \log (2)}{1575}  \notag \\&
+\frac{1369273264 \log ^2(2)}{11025} + \frac{46741772758059 \log (3)}{156956800}
-\frac{70898166 \g_E \log (3)}{1225}  -\frac{35449083 \log ^2(3)}{1225}  \notag \\&
-\frac{70898166 \log (2) \log (3)}{1225} -\frac{6266287109375 \log (5)}{72648576}
-\frac{678223072849 \log (7)}{46332000}  +\frac{958624 \zeta (3)}{105}  \bigg) e^2   \notag \\&
+ \bigg( \frac{1505993024318088529}{1835511678000}
+\frac{157108097019259 \g_E}{496621125} - \frac{7047512 \g_E^2}{147}    
-\frac{128983569284008979 \pi ^2}{443925135360}  \notag \\&
+\frac{810412974158849 \pi ^4}{42949672960}+\frac{4268621081358131 \log  (2)}{496621125}  
-\frac{28157926288 \g_E \log (2)}{11025}-\frac{6057603976 \log ^2(2)}{1225}     \notag \\&
-\frac{3927930747653331 \log (3)}{717516800}+\frac{1973769687 \g_E \log (3)}{2800}   
+\frac{1973769687 \log (2) \log (3)}{2800}   \notag \\&
+\frac{1973769687 \log ^2(3)}{5600} - \frac{29898969861484375 \log (5)}{16273281024}
+\frac{3173828125 \g_E \log (5)}{7056} +\frac{428016 \zeta (3)}{7}     \notag \\&
+\frac{3173828125 \log (2) \log (5)}{7056}+\frac{3173828125 \log ^2(5)}{14112}
+\frac{3748926479678051 \log (7)}{1976832000} \bigg) e^4  + \cdots \bigg]  ,  \\
\label{eqn:U8L}
\mathcal{U}_{8L} =& \, \, (1-e^2)^{3/2} \bigg( \frac{769841899153}{496621125}
-\frac{3814689327311 e^2}{993242250}-\frac{144438148878409 e^4}{993242250}
-\frac{9467353875299  e^6}{52972920}   \notag \\&
-\frac{35883020985379 e^8}{977961600}  -\frac{1471341925457 e^{10}}{188348160}
-\frac{139981090287 e^{12}}{32032000} -\frac{33108069083 e^{14}}{11289600}   \notag \\&
-\frac{195156076961 e^{16}}{90316800}
-\frac{51917170997 e^{18}}{30965760}-\frac{1393506257047 e^{20}}{1032192000}  + \cdots \bigg)  \notag \\&
-4 \left[\g_E +\log \left(\frac{8 \left(1-e^2\right)^{3/2}}{1+\sqrt{1-e^2}}\right) \right] \mathcal{U}_{8L2} 
+ \mathcal{U}^{\chi}_{8L},  \\
\mathcal{U}^{\chi}_{8L} =& \,\, (1-e^2)^{3/2} \bigg[\bigg(\frac{4280672 \log (2)}{11025}
-\frac{18954 \log (3)}{49}\bigg) +\bigg(-\frac{97532944 \log (2)}{2205}
+\frac{35449083 \log (3)}{1225}\bigg) e^2   \notag \\&
+\bigg(\frac{13021836344 \log (2)}{11025}-\frac{1973769687 \log (3)}{5600}
-\frac{3173828125 \log (5)}{14112}\bigg) e^4
+\bigg( \frac{20333297889 \log (3)}{78400}   \notag \\&
-\frac{1607044879436 \log (2)}{99225}
+\frac{1754489453125 \log (5)}{254016}\bigg) e^6
+\bigg(\frac{56824736038921 \log (2)}{396900}    \notag \\&
+\frac{453776788877859 \log (3)}{10035200}  
-\frac{2483678962109375 \log (5)}{32514048} 
-\frac{8816899947037 \log (7)}{663552}  \bigg) e^8    \notag \\&
+\bigg(-\frac{852887396476489 \log (2)}{735000}  
-\frac{280310702846358393 \log (3)}{501760000}
+\frac{10223282240234375 \log (5)}{21676032}   + \notag \\&
\frac{18727987261867681 \log (7)}{55296000}\bigg) e^{10}  
+\bigg(\frac{221370828402431906 \log (2)}{22325625}
+\frac{23025946077060861357 \log (3)}{8028160000}    \notag \\&
-\frac{330898321172265625 \log (5)}{173408256}  
-\frac{853513907519888633221 \log (7)}{238878720000}\bigg) e^{12}   \notag \\&
+\bigg(-\frac{8963770141645538357 \log (2)}{121550625}   
+\frac{1232021848498062153183 \log (3)}{786759680000}    \notag \\&
+\frac{1864847696637355859375 \log (5)}{458838245376}
+\frac{10521634442344300732457 \log (7)}{477757440000}\bigg) e^{14}   \notag \\&
+\bigg(\frac{2479395019071424128029 \log (2)}{5834430000}
-\frac{12742371516403267758044721 \log (3)}{100705239040000}   \notag \\&
+\frac{20498612836182160865234375 \log (5)}{939700726530048}
-\frac{7441303652495936331330901 \log (7)}{81537269760000}   \notag \\&
-\frac{128046524785498944231253933 \log (11)}{23492518163251200}\bigg) e^{16}
+\bigg(-\frac{1104365325145934208143569 \log (2)}{567106596000}   \notag \\&
+\frac{191754205895862468592394607 \log (3)}{201410478080000}
-\frac{50803935116307668662066015625 \log (5)}{152231517697867776}  \notag \\&
+\frac{10860283420254440580447366481 \log (7)}{39627113103360000}
+\frac{1744072191687631808797341997021 \log (11)}{13592099794452480000}\bigg) e^{18}  \notag \\&
+\bigg(\frac{2220617736605099271594254231 \log (2)}{283553298000000}
-\frac{1024327182220393415834369745243 \log (3)}{257805411942400000}  \notag \\&
+\frac{10855085329137334241788375390625 \log (5)}{4871408566331768832}
-\frac{9898173809145672448664085120763 \log (7)}{15850845241344000000}  \notag \\&
-\frac{103629515649148597467998465637518081 \log (11)}{76115758848933888000000}   \notag \\&
-\frac{201538126434611150798503956371773 \log (13)}{3044630353957355520000}\bigg) e^{20} + \cdots \bigg]
, \\
\label{eqn:U8L2}
\mathcal{U}_{8L2} =& \, \, (1-e^2)^{3/2} \bigg( -\frac{27016}{2205}-\frac{9052252 e^2}{3675}
-\frac{1761878 e^4}{147}-\frac{20257439 e^6}{2205}-\frac{698783 e^8}{784}+\frac{491013 e^{10}}{39200} \bigg) .
\end{align}

Finally, we provide in this paper one further term in the redshift invariant (8.5PN) and save the remaining results 
to 10PN and $e^{20}$ for release at the Black Hole Perturbation Toolkit \cite{BHPTK18}.  The 8.5PN redshift has both a 
$\log(p)$ and non-log term, each of which demonstrates additional connections to the energy flux at infinity.  The 
8.5PN log term turns out to be proportional to a function, $\Th_1(e)$, that we identified previously 
\cite{MunnEvan19a} as contributing to the 4.5PN energy flux.  This function is given by the following sum over the 
Newtonian mass quadrupole spectrum
\be
\Th_1(e) =\sum_{n=1}^{\infty} \left(\frac{n}{2} \right)^3  \, g(n,e) ,
\ee
and can be evaluated exactly to any desired order in $e^2$.  Thus, the 8.5PN log redshift is part of a 
leading-log sequence in the redshift that is analogous to the leading-log series in the energy and angular 
momentum fluxes.  We find explicitly 
\begin{align}
\label{eqn:U8p5L}
\mathcal{U}_{17/2L} =& \,\, - \pi \left(\frac{11723776}{55125} \right) (1-e^2)^{19/2} \, \Th_1(e)   
\notag \\
=& \,\, \pi (1-e^2)^{3/2} \bigg(-\frac{11723776}{55125}-\frac{179108156 e^2}{33075}-\frac{3476454503 e^4}{165375}
-\frac{30371758363 e^6}{1587600}-\frac{29489429729 e^8}{7620480}  
\notag \\&
-\frac{6377226117523 e^{10}}{76204800000}+\frac{51655953119 e^{12}}{130636800000}
-\frac{56934823428673  e^{14}}{477956505600000}
+\frac{61709721913247 e^{16}}{3277416038400000}  
\notag \\&
-\frac{10585818010370879 e^{18}}{59465436600729600000}
-\frac{6093295182537785141 e^{20}}{17839630980218880000000}  + \cdots \bigg) .
\end{align}
The 8.5PN non-log term can then be separated in an orderly fashion.  It features an appearance of 
$\mathcal{U}_{17/2L}$ itself and there is also a term with a second previously identified function, $\Xi_1(e)$ 
\cite{MunnEvan19a},
\be
\Xi_1(e) =\sum_{n=1}^{\infty} \left(\frac{n}{2} \right)^3   \log\Big(\frac{n}{2}\Big) \, g(n,e) ,
\ee
which is another analog of the function $\chi(e)$ and also depends exclusively upon the Newtonian mass quadrupole.  
This second term soaks up all the remaining appearances of transcendental numbers and leaves a final series with 
rational number coefficients.  The result is
\begin{align}
\label{eqn:U8p5}
\mathcal{U}_{17/2}  =& \, \, \pi (1 - e^2)^{3/2} \bigg( -\frac{2207224641326123}{1048863816000}
-\frac{71647706604932467 e^2}{1048863816000}-\frac{810445553448114013 e^4}{2581818624000}   \notag \\&
-\frac{18435323045231443783 e^6}{56638646064000}   
-\frac{4277286582396672996821 e^8}{51553754284032000}  
-\frac{403116895474561377313 e^{10}}{47734957670400000}    \notag \\&
-\frac{2243648120491392755469179  e^{12}}{477240468229324800000}
-\frac{15908263768163409948152909 e^{14}}{5456449353421946880000}   \notag \\&
-\frac{17132229781087995700006572689 e^{16}}{7982005911291533721600000}
-\frac{2510618005651588728917156811889 e^{18}}{1508599117234099873382400000}   \notag \\&
-\frac{194103765865113110422717940935393 e^{20}}{144825515254473587844710400000} + \cdots \bigg)
- 2 \left[\g_E +\log \left(\frac{8 \left(1-e^2\right)^{3/2}}{1+\sqrt{1-e^2}}\right) 
- \left(\frac{35}{107} \right) \pi^2 \right] 
\mathcal{U}_{17/2L}  
\notag \\&
+ \pi \left( \frac{23447552}{55125} \right) (1-e^2)^{19/2} \Xi_1(e).
\end{align}

\subsection{Redshift invariant as an expansion in $y$}

We can now use the easily derived PN expansion of $1/p$ in terms of the compactness parameter $y$ to give an 
alternative expansion of the redshift invariant.  The general form of that expansion is 
\begin{align}
\notag 
 \left<u_t\right>_\tau^1 =& \, \, \left( \frac{\mu}{M} \right) y \bigg[ \mathcal{V}_0 +  \mathcal{V}_1 y
+  \mathcal{V}_2 y^2 + \mathcal{V}_{3} y^3
+ \Big( \mathcal{V}_{4} + \mathcal{V}_{4L} \log y \Big) y^4
+ \Big( \mathcal{V}_{5} + \mathcal{V}_{5L} \log y \Big)y^5
+  \mathcal{V}_{11/2} y^{11/2}
\notag \\&
+ \Big( \mathcal{V}_{6} + \mathcal{V}_{6L} \log y  \Big) y^6    
+  \mathcal{V}_{13/2} y^{13/2}   
+ \Big( \mathcal{V}_{7} + \mathcal{V}_{7L} \log y 
+ \mathcal{V}_{7L2} \log^2 y \Big) y^7
+ \mathcal{V}_{15/2}  y^{15/2}  
\notag \\&
+ \Big( \mathcal{V}_{8} + \mathcal{V}_{8L} \log p 
+ \mathcal{V}_{8L2} \log^2 y \Big) y^8 
+ \Big( \mathcal{V}_{17/2} + \mathcal{V}_{17/2L} \log y \Big) y^{17/2}  
+ \Big( \mathcal{V}_{9} + \mathcal{V}_{9L} \log p 
+ \mathcal{V}_{9L2} \log^2 y \Big) y^9 
\notag \\&
+ \Big( \mathcal{V}_{19/2} + \mathcal{V}_{19/2L} \log y \Big) y^{19/2}  
+ \Big( \mathcal{V}_{10} + \mathcal{V}_{10L} \log p 
+ \mathcal{V}_{10L2} \log^2 y + \mathcal{V}_{10L3} \log^3 y \Big) y^{10} + \cdots \bigg] .
\end{align}
The $\mathcal{V}_k(e)$ functions in the $y$ expansion exhibit most of the same structure as the $\mathcal{U}_k(e)$ 
in the $1/p$ expansion, so as we list our findings we will primarily refer back to the previous section 
for comments found there.  As with the $1/p$ results, the first few terms in the $y$ series all yield closed 
forms, with the first two having been found previously \cite{HoppKavaOtte16}
\begin{align}
\mathcal{V}_0 =& -1,  \\
\mathcal{V}_1 =& - \frac{2 - 4 e^2}{1 - e^2},  \\
\mathcal{V}_2 =& \frac{1}{(1 - e^2)^2} \left(14-\frac{15 e^2}{2}-6 e^4+\sqrt{1-e^2} (-19+14 e^2)\right),  \\
\mathcal{V}_3 =& \frac{1}{(1 - e^2)^3}\bigg[58-132 e^2+76 e^4-\frac{8 e^6}{3} +\sqrt{1-e^2} \left(-\frac{295}{3}
+\frac{41 \pi ^2}{32}+e^2 \left(\frac{713}{6}+\frac{41 \pi^2}{64}\right)-\frac{171 e^4}{2}\right) \bigg] .
\end{align}

The 4PN terms share properties with $\mathcal{U}_4$ and $\mathcal{U}_{4L}$ as discussed near \eqref{eqn:U4} and 
\eqref{eqn:U4L} (with the added change of a sign flip between $\log(p)$ and $\log(y)$ terms)
\begin{align}
\mathcal{V}_4 =& \frac{1}{(1 - e^2)^4} \bigg[ -\frac{66827}{180}+\frac{117799 e^2}{180}
-\frac{442291 e^4}{1440}-\frac{65 e^6}{12}+\frac{85 e^8}{3}+\sqrt{1-e^2} \bigg(\frac{52943}{180}
-\frac{84751 e^2}{360}   \notag \\& \qq
+\frac{5881 e^4}{72}+\frac{3327 e^6}{16} \bigg) \bigg]  
+ 2 \left[\g_E +\log \left(\frac{8 \left(1-e^2\right)}{1+\sqrt{1-e^2}}\right) \right] \mathcal{V}_{4L}  \notag \\&\qq
+ \frac{\pi ^2}{(1 - e^2)^{7/2}} \bigg(\frac{677}{512} +\frac{2135 \pi ^2}{1536}e^2 
- \frac{33311 \pi ^2}{12288} e^4\bigg)  - \frac{128}{5}  \La_0(e) , \\
\mathcal{V}_{4L} =& -\frac{64}{5} \frac{1}{(1 - e^2)^{7/2}} \left( 1+\frac{73 e^2}{24}+\frac{37 e^4}{96} \right) 
= -\left(1 - e^2\right)^{-5} \mathcal{U}_{4L} = -\frac{64}{5} \, \mathcal{L}_0(e) .
\end{align}

The remaining higher-order terms are similar to their counterparts in the $1/p$ expansion.  In what follows, we give 
the closed-form parts and many of the infinite series in $e^2$ out to order $e^{20}$ (with a few exceptions).  For 
brevity we have omitted listing the $\chi$-like portions of the $y$-expansion terms (e.g., $\mathcal{V}^{\chi}_5$, 
$\mathcal{V}^{\chi}_6$, etc), relegating those along with the more complicated series expansions to the posting 
at \cite{BHPTK18}.  The 5PN terms have a structure that mirrors that discussed near \eqref{eqn:U5L} and \eqref{eqn:U5}
\begin{align}
\mathcal{V}_{5} =& \frac{1}{(1 - e^2)^5} \bigg( \frac{1606877}{3150}+\frac{7523659 e^2}{3150}
-\frac{733791 e^4}{280}+\frac{838289 e^6}{5040}-\frac{4985231 e^8}{8064} +\frac{35782139 e^{10}}{268800}
 \notag \\&
+\frac{11177111 e^{12}}{201600} +\frac{19535179 e^{14}}{806400}+\frac{70069067 e^{16}}{5734400}
+\frac{38935589 e^{18}}{5898240} + \frac{752638217 e^{20}}{206438400} + \cdots \bigg)  \notag \\&
+ 2 \left[\g_E +\log \left(\frac{8 \left(1-e^2\right)}{1+\sqrt{1-e^2}}\right) \right] \mathcal{V}_{5L}
+ \frac{\pi ^2}{(1 - e^2)^{9/2}} \bigg[  -\frac{84451}{768}-\frac{68375 e^2}{384}-\frac{87881 e^4}{6144}  
+\frac{4243 e^6}{2048}    \notag \\&
+\sqrt{1-e^2} \left(\frac{2009}{64}-\frac{2009 e^2}{128}-\frac{2009 e^4}{128}\right) \bigg]
+ \mathcal{V}^{\chi}_{5}    , \\
\mathcal{V}_{5L} =& \frac{1}{(1 - e^2)^{9/2}} \bigg(\frac{956}{105}+\frac{4714 e^2}{21}+\frac{12313 e^4}{30}
+\frac{6109 e^6}{140} \bigg)  .
\end{align}

As the first appearance of a term of its type, the 5.5PN redshift $\mathcal{V}_{11/2}(e)$ is identical to 
$\mathcal{U}_{11/2}(e)$ \eqref{eqn:U5p5} up to a power of $1 - e^2$ and is likewise proportional to the 1.5PN 
energy flux tail
\begin{align}
\mathcal{V}_{11/2} = - \frac{13696}{525} \pi \, \vp(e) = \left(1 - e^2\right)^{-13/2} \mathcal{U}_{11/2} .
\end{align}

An understanding of the structure of the 6PN terms follows from the discussion surrounding \eqref{eqn:U6} and 
\eqref{eqn:U6L}
\begin{align}
\mathcal{V}_{6} =& \frac{1}{(1 - e^2)^6} \bigg(\frac{17083661}{4050}+\frac{532849892 e^2}{19845}
-\frac{13121598937 e^4}{793800}+\frac{22207024181 e^6}{4762800} -\frac{171292853309 e^8}{33868800}
\notag \\&
-\frac{31878137383 e^{10}}{67737600} -\frac{2080903905791 e^{12}}{1219276800}
-\frac{99502106951 e^{14}}{81285120} -\frac{7814620751519 e^{16}}{8670412800}   \notag \\&
-\frac{109440738879557 e^{18}}{156067430400}  
-\frac{14748616145089 e^{20}}{26011238400} + \cdots \bigg)  
+ 2 \left[\g_E +\log \left(\frac{8 (1-e^2)}{1+\sqrt{1-e^2}}\right) \right] \mathcal{V}_{6L}  \notag \\&
+ \frac{\pi ^4}{(1 - e^2)^{11/2}} \bigg(\frac{2800873}{262144}+\frac{27872821 e^2}{524288}  
+\frac{41197641 e^4}{2097152}-\frac{135909 e^6}{4194304} \bigg)    \notag \\&
+ \frac{\pi ^2}{(1 - e^2)^{11/2}} \bigg[ -\frac{1380825359}{1769472}-\frac{3533232287 e^2}{884736}  
-\frac{3298849937 e^4}{2359296}-\frac{106812205 e^6}{262144}+\frac{384008563 e^8}{25165824}     \notag \\&
+\left(1-e^2\right)^{3/2} \left(\frac{77991}{1024}+\frac{110221 e^2}{3072}-\frac{2256613 e^4}{24576}\right) \bigg]
+ \mathcal{V}^{\chi}_{6} , \\
\mathcal{V}_{6L} =& \frac{1}{(1 - e^2)^{11/2}} \bigg( \frac{1326776}{2835}+\frac{5364808 e^2}{2835}
-\frac{1087216 e^4}{945}-\frac{141367 e^6}{63}-\frac{1010039 e^8}{5040} \bigg) 
+ \frac{59}{2 \sqrt{1 - e^2}}  \mathcal{V}_{4L} . 
\end{align}

The 6.5PN term is proportional to an infinite series with rational number coefficients, similar to the 
$\mathcal{U}_{13/2}$ term \eqref{eqn:U6p5}
\begin{align}
\mathcal{V}_{13/2} =& \frac{\pi}{(1 - e^2)^{6}} \bigg( \frac{81077}{3675}+\frac{10821932 e^2}{11025}
+\frac{518653529 e^4}{141120}+\frac{465892081 e^6}{235200}+\frac{10086118841 e^8}{116121600} \notag \\&
+\frac{3522792971 e^{10}}{10160640000}-\frac{489338360093 e^{12}}{7803371520000}
+\frac{8260641257293 e^{14}}{1147095613440000} -\frac{45676604096923 e^{16}}{117462590816256000}
\notag \\&
+\frac{1182988156073 e^{18}}{70792186429440000}
+\frac{1182698258651584849 e^{20}}{76115758848933888000000} + \cdots  \bigg) .
\end{align}

The redshift invariant at 7PN features the first appearance of a $\log^2(y)$ term, which is closed in form.  
As the discussion surrounding the corresponding terms, \eqref{eqn:U7}, \eqref{eqn:U7L}, and \eqref{eqn:U7L2}, in 
the $1/p$ expansion showed, the 7PN $y$-expansion terms separate similarly 
\begin{align}
\mathcal{V}_{7} =& \frac{1}{(1 - e^2)^{7}} \bigg[ \frac{12624956532163}{382016250}
-\frac{10327445038 \g_E}{5457375}+\frac{109568 \g_E^2}{525} 
-\frac{9041721471697 \pi ^2}{2477260800}-\frac{23851025 \pi ^4}{16777216}  \notag \\&  
-\frac{16983588526 \log (2)}{5457375}+\frac{438272}{525} \g_E \log (2)
+\frac{438272 \log ^2(2)}{525}-\frac{2873961 \log (3)}{24640}-\frac{1953125 \log (5)}{19008}    \notag \\&
-\frac{2048 \zeta (3)}{5}   
+ \bigg(  \frac{95954904706559}{312558750}-\frac{60219626092 \g_E}{1819125}+
\frac{4492288 \g_E^2}{1575}-\frac{29974613135297 \pi ^2}{1238630400}  \notag \\&
-\frac{99857911927 \pi ^4}{503316480}-\frac{357143036 \log (2)}{11025}
+\frac{219136}{315} \g_E \log (2)-\frac{7341056 \log ^2(2)}{1575}-\frac{153791486811 \log (3)}{2156000}
\notag \\&
+\frac{1872072}{175} \g_E \log (3)+\frac{1872072}{175} \log (2) \log (3)+\frac{936036 \log ^2(3)}{175}
+\frac{611328125 \log (5)}{28512}-\frac{83968 \zeta (3)}{15} \bigg) e^2  \notag \\&
+ \bigg( -\frac{408859577737}{22325625}-\frac{103798581437 \g_E}{1039500} +\frac{926192 \g_E^2}{225}
+\frac{77870248904627 \pi ^2}{2831155200}  -\frac{104270589937 \pi ^4}{402653184}  \notag \\&
-\frac{20802851168513 \log (2)}{21829500} +\frac{130800224}{525} \g_E \log (2)
+\frac{106275824 \log ^2(2)}{225}  +\frac{347735485526229 \log (3)}{275968000}   \notag \\&
-\frac{21996846}{175} \g_E \log (3)  -\frac{21996846}{175} \log (2) \log (3)-\frac{10998423 \log ^2(3)}{175}
-\frac{1417441796875 \log(5)}{2838528}
\notag \\&
-\frac{678223072849 \log (7)}{6082560}-\frac{121184 \zeta (3)}{15}  \bigg) e^4 
+ \cdots \bigg] , \\
\mathcal{V}_{7L} =& \frac{1}{(1 - e^2)^{13/2}}  \bigg( -\frac{5163722519}{5457375}
-\frac{36822832351 e^2}{2182950}-\frac{136448109487 e^4}{2425500} - \frac{505462816847 e^6}{14553000}
 \notag \\&
+\frac{2286191870693 e^8}{232848000} +\frac{114737952487 e^{10}}{66528000}+\frac{8196389 e^{12}}{54000} 
-\frac{545737 e^{14}}{60480} -\frac{4065182087 e^{16}}{77414400}    \\&
-\frac{70788229 e^{18}}{1105920} -\frac{14350431011 e^{20}}{221184000}  + \cdots  \bigg)   
+ 4 \left[\g_E +\log \left(\frac{8 (1-e^2)}{1+\sqrt{1-e^2}}\right) \right] \mathcal{V}_{7L2}  
+ \frac{109568}{525} \chi(e)
, \notag \\
\mathcal{V}_{7L2} =& \frac{1}{(1 - e^2)^{13/2}} \left(\frac{27392}{525} \right) 
\left(1 +\frac{85}{6} e^2 + \frac{5171}{192} e^4 + \frac{1751}{192} e^6 + \frac{297}{1024} e^8 \right) 
= \left(\frac{27392}{525}\right) F(e) .
\end{align}

The 7.5PN term is proportional to another infinite series in $e^2$ with rational coefficients
\begin{align}
\mathcal{V}_{15/2} =& \frac{\pi}{(1 - e^2)^{7}} \bigg(\frac{82561159}{467775}
+\frac{91675684453 e^2}{26195400}-\frac{16022780959 e^4}{69854400}
-\frac{691420819648151 e^6}{22632825600}    \notag \\&
-\frac{4122993297854251 e^8}{241416806400}  
-\frac{67223507326794097 e^{10}}{57940033536000}
-\frac{4573218369555807463 e^{12}}{20858412072960000}   \notag \\&
-\frac{10499957605818206893 e^{14}}{90849972584448000}
-\frac{36738832192598535471139 e^{16}}{523295842086420480000}  \notag \\&
-\frac{35781609047672532119149 e^{18}}{768924502657597440000}
-\frac{740782396859898339667508393 e^{20}}{22606380378133364736000000}+ \cdots  \bigg) .
\end{align}

As the discussion near \eqref{eqn:U8}, \eqref{eqn:U8L}, and \eqref{eqn:U8L2} would suggest, the 8PN term is 
quite complex but does allow a separation into closed-form and infinite series parts of different type.  We find
\begin{align}
\mathcal{V}_{8} =& \frac{1}{(1 - e^2)^{8}} \bigg[  -\frac{7516581717416867}{34763478750}
-\frac{1526970297506 \g_E}{496621125}-\frac{108064 \g_E^2}{2205}
-\frac{246847155756529 \pi ^2}{18496880640}    \notag \\&
+\frac{22759807747673 \pi ^4}{6442450944}-\frac{1363551923554 \log (2)}{496621125}
-\frac{3574208 \g_E \log (2)}{3675}  -\frac{2201898578589 \log (3)}{392392000}  \notag \\&
- \frac{2143328 \log ^2(2)}{1575} + \frac{37908}{49} \g_E \log (3)+\frac{37908}{49} \log (2) \log (3)
+\frac{18954 \log ^2(3)}{49}  +\frac{798828125 \log (5)}{741312}   \notag \\&
-\frac{41408 \zeta (3)}{105}
+ \bigg( -\frac{899118394693923097}{491654913750}+\frac{64711377904 \g_E}{1447875}   
-\frac{2962688 \g_E^2}{225}  
-\frac{1648884680043103 \pi^2}{26424115200}    \notag \\&
+\frac{879870714987551 \pi ^4}{32212254720}-\frac{2812127235664 \log (2)}{12733875}  
+\frac{398923264 \g_E \log (2)}{11025}+\frac{245903104 \log ^2(2)}{2205}   \notag \\&
+\frac{58970211975621 \log (3)}{196196000}-\frac{71372016 \g_E \log (3)}{1225}   
-\frac{71372016 \log (2) \log (3)}{1225}-\frac{35686008 \log ^2(3)}{1225}  \notag \\&
-\frac{1546498046875 \log (5)}{18162144}-\frac{678223072849 \log (7)}{46332000}   
+\frac{238208 \zeta (3)}{15}  \bigg) e^2  + \bigg( -\frac{1850695307993262971}{367102335600}   \notag \\&
+\frac{474991573730237 \g_E}{662161500}  +\frac{85298687127989143 \pi ^2}{184968806400} 
-\frac{108590956 \g_E^2}{1225} +\frac{318549856963953 \pi ^4}{85899345920}  +  \notag \\&
\frac{1927184488471871 \log (2)}{220720500} -\frac{28549712792 \g_E \log (2)}{11025}
-\frac{54378998732 \log ^2(2)}{11025}   +\frac{1575274959 \log ^2(3)}{5600}   \notag \\&
-\frac{3185922772104363 \log (3)}{717516800}+\frac{1575274959 \g_E \log (3)}{2800}   
+\frac{1575274959 \log (2) \log (3)}{2800}   \notag \\&
-\frac{33310535689609375 \log (5)}{16273281024}+\frac{3173828125 \g_E \log (5)}{7056}  
+\frac{3173828125 \log (2) \log (5)}{7056}   \notag \\&
+\frac{3173828125 \log ^2(5)}{14112}  
+\frac{11290185715696489 \log (7)}{5930496000}+\frac{1023368 \zeta (3)}{7}  \bigg) e^4   + \cdots \bigg]   , \\
\mathcal{V}_{8L} =& \frac{1}{(1 - e^2)^{8}} \bigg(  -\frac{769841899153}{496621125}
+\frac{1569199431248 e^2}{70945875}+\frac{461479855229957 e^4}{1324323000}  
+\frac{6380037660991 e^6}{9459450}  \notag \\&
+\frac{2850581679359 e^8}{454053600}-\frac{80964330820213 e^{10}}{216216000}  
-\frac{4478618712015803 e^{12}}{36324288000}-\frac{197888011746329 e^{14}}{2889432000}  \notag \\&
-\frac{96496373295233 e^{16}}{2012774400}   
-\frac{2126293356495073 e^{18}}{59175567360}-\frac{47751587053936639 e^{20}}{1690730496000} +\cdots\bigg)
 \notag \\&
+ 4 \left[\g_E +\log \left(\frac{8 (1-e^2)}{1+\sqrt{1-e^2}}\right) \right] \mathcal{V}_{8L2} 
+ \mathcal{V}^{\chi}_{8L}   ,  \\
\mathcal{V}_{8L2} =& \frac{-1}{(1 - e^2)^{15/2}} \bigg( \frac{27016}{2205} + \frac{4040052 e^2}{1225}
+ \frac{52505354 e^4}{2205} + \frac{349164251 e^6}{11025} + \frac{1500201151 e^8}{176400}
+ \frac{1800063 e^{10}}{7840}  \bigg) . 
\end{align}

The 8.5PN redshift term in the $y$-expansion is similar to \eqref{eqn:U8p5} and \eqref{eqn:U8p5L}, and separates 
into distinct infinite series with the appearance of the special functions $\Th_1(e)$ and $\Xi_1(e)$
\begin{align}
\mathcal{V}_{17/2}  =& \,\, \frac{\pi}{(1 - e^2)^{8}} \bigg(-\frac{2207224641326123}{1048863816000}
-\frac{2650649701927561 e^2}{38846808000}-\frac{323918330025102751 e^4}{860606208000}  \notag \\&
-\frac{26475630227556761683 e^6}{56638646064000}
+\frac{30869646375667808864131 e^8}{463983788556288000} 
+\frac{412364732751629164661647 e^{10}}{3866531571302400000}   \notag \\&
+\frac{4278649689952925300925061  e^{12}}{477240468229324800000}
+\frac{21539433195525668512769497  e^{14}}{16369348060265840640000}   \notag \\&
+\frac{974913490747601744257312111 e^{16}}{7982005911291533721600000}  
-\frac{414022013332738056999952875889 e^{18}}{1508599117234099873382400000}    
\notag \\&
-\frac{59927587971876739009859734907233 e^{20}}{144825515254473587844710400000} + \cdots \bigg)  +
2 \left[\g_E +\log \left(\frac{8 \left(1-e^2\right)}{1+\sqrt{1-e^2}}\right) 
- \left(\frac{35}{107} \right) \pi^2 \right] 
\mathcal{V}_{17/2L}  \notag \\&
+ \pi \left( \frac{23447552}{55125} \right) \Xi_1(e)
,  \notag \\
\mathcal{V}_{17/2L} =& \,\,  \pi \left(\frac{11723776}{55125} \right) 
\sum_{n=1}^{\infty} \left(\frac{n}{2} \right)^3 \, g(n,e)  \notag \\
=& \,\, \frac{\pi}{(1 - e^2)^{8}}\bigg(\frac{11723776}{55125}+\frac{179108156 e^2}{33075}
+\frac{3476454503 e^4}{165375}+\frac{30371758363 e^6}{1587600}+\frac{29489429729 e^8}{7620480} \notag \\&
+\frac{6377226117523 e^{10}}{76204800000}-\frac{51655953119 e^{12}}{130636800000}
+\frac{56934823428673 e^{14}}{477956505600000}-\frac{61709721913247 e^{16}}{3277416038400000}  \notag \\&
+\frac{10585818010370879 e^{18}}{59465436600729600000}
+\frac{6093295182537785141 e^{20}}{17839630980218880000000}+ \cdots \bigg)  .
\end{align}

\end{widetext}

\subsection{Discussion}
\label{sec:ResultDiss}

A primary focus of this paper is on taking the high-order analytic PN series of the redshift invariant to a 
deeper expansion in eccentricity, with the goal of finding as many closed-form terms or fully-determinable 
infinite series expansions in $e$ as possible.  By pushing our analytic self-force calculation to $e^{20}$, we 
found a surprising amount of this fully-explicable structure in the eccentricity dependence, including a series 
of connections between the (conservative-sector) redshift invariant and the (dissipative-sector) energy flux to 
infinity.  We summarize the findings here.

The 0PN and 1PN redshift functions, $\mathcal{V}_0(e)$ and $\mathcal{V}_1(e)$, were found previously through 
self-force calculations and known to have closed form \cite{HoppKavaOtte16}.  (This effectively is also 
true of $\mathcal{U}_0(e)$ and $\mathcal{U}_1(e)$, and everything we say in this section about the analytic 
determinability of $\mathcal{U}_k(e)$ functions in the $1/p$ expansion pertains equally well to $\mathcal{V}_k(e)$ 
functions in the $y$ expansion.)  In taking the self-force calculations further, we are able to show that the 2PN 
and 3PN redshift terms can be condensed into closed expressions of a particular form, with a dominant-subdominant 
eccentricity singular factor structure, though these two terms were previously found \cite{AkcaETC15}, in slightly 
different form, using a full PN theory calculation (see their Eq.~(4.51)).  The 2PN term, which contains only 
rational number coefficients, is reminiscent of the 2PN energy flux, $\mathcal{L}_2(e)$ \cite{ArunETC08a}.  The 
3PN redshift term contains both rational numbers and appearances of coefficients with $\pi^2$, which is unlike any 
PN term in the flux expansion prior to the advent of tail-squared terms.  The appearance of $\pi^2$ in this 
early term in the redshift is traceable to the sum over infinite $l$ that occurs in conservative sector self-force 
calculations.  (Recall that the sum of inverse square integers is $\zeta(2)$ and totals to $\pi^2/6$.)  

It is at 4PN and beyond in the redshift expansion that our calculation reveals new results.  Indeed, at 4PN order 
itself there emerges a more profound connection between the redshift expansion and the energy flux at infinity.  We 
found the $\mathcal{U}_{4L}(e)$ term in the redshift invariant to be \emph{exactly proportional} to the Peters-Mathews 
energy flux $\mathcal{L}_0(e)$ \cite{PeteMath63}.  (This result was implicitly present in the 4PN log term, 
$\mathcal{V}_{4L}$, found in \cite{HoppKavaOtte16} but its closed form and connection to the energy flux was missed 
by resumming on a different eccentricity singular factor.)  Then, within the 4PN non-log term, $\mathcal{U}_{4}(e)$, 
we found the Peters-Mathews flux function reappearing, as well as a second function, $\La_0(e)$, that also depends 
exclusively on a sum over the Fourier power spectrum $g(n,e)$ \cite{MunnEvan19a} of the Newtonian mass quadrupole 
moment.  The function $\La_0(e)$ is an infinite series in $e^2$ but its coefficients are fully determinable at all 
orders.  Other than $\La_0(e)$, the rest of the $\mathcal{U}_4(e)$ term has a closed form.  The structure of 
$\mathcal{U}_4(e)$ is similar to the 3PN non-log energy flux \cite{ArunETC08a}, except that where the two 
Newtonian-quadrupole-derived functions, $F(e)$ and $\chi(e) = \La_1(e)$, appear in $\mathcal{L}_3(e)$, it is 
$\mathcal{L}_0(e)$ and $\La_0(e)$ that appear in $\mathcal{U}_4(e)$.  

Following the discussion in \cite{MunnEvan19a}, we can recognize the connection between $\mathcal{U}_{4L}(e)$ and 
$\mathcal{U}_{4}(e)$ is similar to that occurring between leading and subleading logarithmic terms in the PN 
expansion of the flux at infinity.  As a reminder, we showed previously \cite{MunnEvan19a} that a pair of infinite 
sequences of leading logarithms exists in the energy flux PN expansion, at integer orders $p^{-3k} \, \log^k(p)$ 
and at half-integer orders $p^{-(3k+3/2)} \, \log^k(p)$ (for $k \ge 0$).  The terms in these sequences are 
completely determined by sets of special functions, $T_k(e)$ and $\Th_k(e)$, that are defined by infinite sums 
over $n$ of even and odd integer powers of $n/2$ multiplying the Newtonian mass quadrupole spectrum $g(n,e)$.  A new 
term in each sequence arises every 3PN orders as a new power of $\log(p)$ appears.  Every integer-order leading log 
(proportional to $T_k(e)$) can be shown to have a closed-form expression.  Every half-integer-order leading log 
(proportional to $\Th_k(e)$) remains an infinite series, but with exactly known rational number coefficients.  

The subleading-log terms formed a second pair of infinite sequences in the energy flux at integer orders 
$p^{-3k} \, \log^{k-1}(p)$ and at half-integer orders $p^{-(3k+3/2)} \, \log^{k-1}(p)$.  At the same PN order as 
a leading-log term, these are terms with one lower power of $\log(p)$.  In another context \cite{MunnEvan20a}, 
these terms were referred to as 3PN-corrected leading logs (or 3PN-log terms).  
At integer PN order these subleading-log terms were shown to feature a return appearance of the functions $T_k(e)$ 
and another special function, $\La_k(e)$, with the remaining analytic dependence requiring self-force calculation 
to determine.  At half-integer PN order these subleading-log terms were shown to feature a return appearance of the 
functions $\Th_k(e)$ and yet another special function, $\Xi_k(e)$.  The positions of these leading and subleading-log 
terms in the energy flux PN expansion are graphically depicted in Fig.~1 in \cite{MunnEvan20a} by the four red and 
green sequences.

If we now carry these ideas over to the PN expansion of the redshift invariant, we can take the 4PN log term, 
$\mathcal{U}_{4L}(e)$, as the start of the integer-order redshift leading-log sequence.  The next terms in that 
sequence would be $\mathcal{U}_{7L2}(e)$, $\mathcal{U}_{10L3}(e)$, etc.  If the observed pattern holds to higher 
order, each term in the sequence will be fully determined by the Newtonian quadrupole and proportional to the 
closed-form functions $T_k(e)$, starting with $k=0$.  Since the leading logs in the energy flux actually begin with 
the non-log $\mathcal{L}_0(e)$ Peters-Mathews term, in principle it is an open question whether we should consider 
the non-log term $\mathcal{U}_1(e)$ as the start of the redshift leading logs.  The conjectures regarding the 
primacy of the $T_k(e)$ functions and $\mathcal{U}_1(e)$ will ultimately be settled by a formal PN theory calculation.

The first half-integer-order term in the redshift is $\mathcal{U}_{11/2}(e)$ at 5.5PN order.  This first appearance 
of a half-integer order term was found in high precision numerical work \cite{ShahFrieWhit14} and its connection to 
the tail field was discussed in \cite{BlanFayeWhit14a}.  In our self-force calculations this term emerged as exactly 
proportional to the 1.5PN energy flux tail enhancement function $\vp(e)$ (see also \cite{BiniDamoGera20a}), which is 
the first function $\Th_0(e)$ in the $\Th_k(e)$ function sequence \cite{MunnEvan19a}.  It is reasonable to regard 
$\mathcal{U}_{11/2}(e)$ as the first element in the redshift half-integer leading-log sequence, which is lagged 
by four PN orders relative to the corresponding sequence in the energy flux.  The next elements in this sequence would 
be $\mathcal{U}_{17/2L}(e)$, $\mathcal{U}_{23/2L2}(e)$, etc.  As we showed in the previous subsections, 
$\mathcal{U}_{17/2L}(e)$ is directly proportional to $\Th_1(e)$, which supports a conjecture that the entire run of
half-integer-order leading logs will be determined by the $\Th_k(e)$ functions.  The next element in that sequence, 
$\mathcal{U}_{23/2L}(e)$, is beyond where we have taken our present calculations.

If the redshift leading logs begin with $\mathcal{U}_{4L}(e)$, then the redshift integer-order subleading logs start 
with $\mathcal{U}_{4}(e)$.  This sequence continues with $\mathcal{U}_{7L}(e)$, $\mathcal{U}_{10L2}(e)$, etc.  Our 
present calculations overlap the first three elements in this sequence.  Like in the corresponding sequence in 
the energy flux, we find that the redshift subleading logs feature a return appearance of the closed-form 
leading-log function (proportional to the relevant $T_k(e)$) and functions from the $\La_k(e)$ ($\chi$-like 
\cite{ArunETC08a}) sequence, starting with $k=0$).

The half-integer-order subleading logs in the redshift would begin with $\mathcal{U}_{17/2}(e)$ and continue 
with terms with $k \ge 1$ that have PN dependence $p^{-3k-17/2} \, \log^{k-1}(p)$.  Our present calculations only 
overlap the first term in this sequence, but we do see the expected behavior that $\mathcal{U}_{17/2}(e)$ 
depends in part on $\mathcal{U}_{17/2L}(e)$ ($\propto \Th_1(e)$) and on $\Xi_1(e)$.

It seems reasonable to conjecture that, like in the energy flux \cite{MunnEvan20a}, there will be 1PN-log sequences 
in the redshift.  At integer PN orders, this sequence would be $\mathcal{U}_{5L}(e)$, $\mathcal{U}_{8L2}(e)$, 
$\mathcal{U}_{11L3}(e)$, etc.  Our results reveal the first two terms in this sequence and find that they are both 
closed-form expressions.  Following the logic, the half-integer-order 1PN-logs would be the sequence that starts with
$\mathcal{U}_{13/2}(e)$, $\mathcal{U}_{19/2L}(e)$, $\mathcal{U}_{25/2L2}(e)$, etc.  The first two functions follow 
the expectation of being rational-number coefficient infinite series (times an overall factor of $\pi$).  Like with 
the 1PN-logs in the energy flux expansion, these terms may derive from what we called earlier \cite{MunnEvan20a} 
the 1PN (source) multipole moments--the Newtonian current quadrupole, the Newtonian mass octupole, and the 1PN 
correction to the mass quadrupole.  This idea deserves further study.

Finally, the analogues in the redshift of the 4PN-logs in the energy flux \cite{MunnEvan20a} would be the 
integer-order sequence $\mathcal{U}_{5}(e)$, $\mathcal{U}_{8L}(e)$, $\mathcal{U}_{11L2}(e)$, etc and 
the half-integer-order sequence $\mathcal{U}_{19/2}(e)$, $\mathcal{U}_{25/2L}(e)$, etc.  If the analysis of the 
corresponding terms in the energy flux are a guide, these too might be partly determined through application of 
known levels of PN theory.  Note that once a derivation of $\mathcal{U}_5^\chi$ is found, all PN terms through 
5.5PN in the redshift invariant will be completely known functions of eccentricity.

\subsection{Comparison to numerical data}

We can assess the validity of these expansions by comparing them to the numerical redshift data given in Table II of
\cite{AkcaETC15}.  Results from larger orbits in that data set are well fit by our PN expansions, which leads us to 
single out smaller orbits ($p=10$ and $p=20$) for comparison, where convergence is expected to be slower.  We compare 
the numerical self-force results to both the $y$ and $1/p$ versions of our PN expansions, and we try a few added 
resummation methods to check for improved convergence.  Two such methods are logarithmic resummation (in which the 
log of the series is taken, the new series is evaluated numerically, and then the result is exponentiated) and 
reciprocal resummation \cite{IsoyETC13, JohnMcDa14}.  

\begin{figure*}
\hspace{-1.5em}\includegraphics[scale=.71]{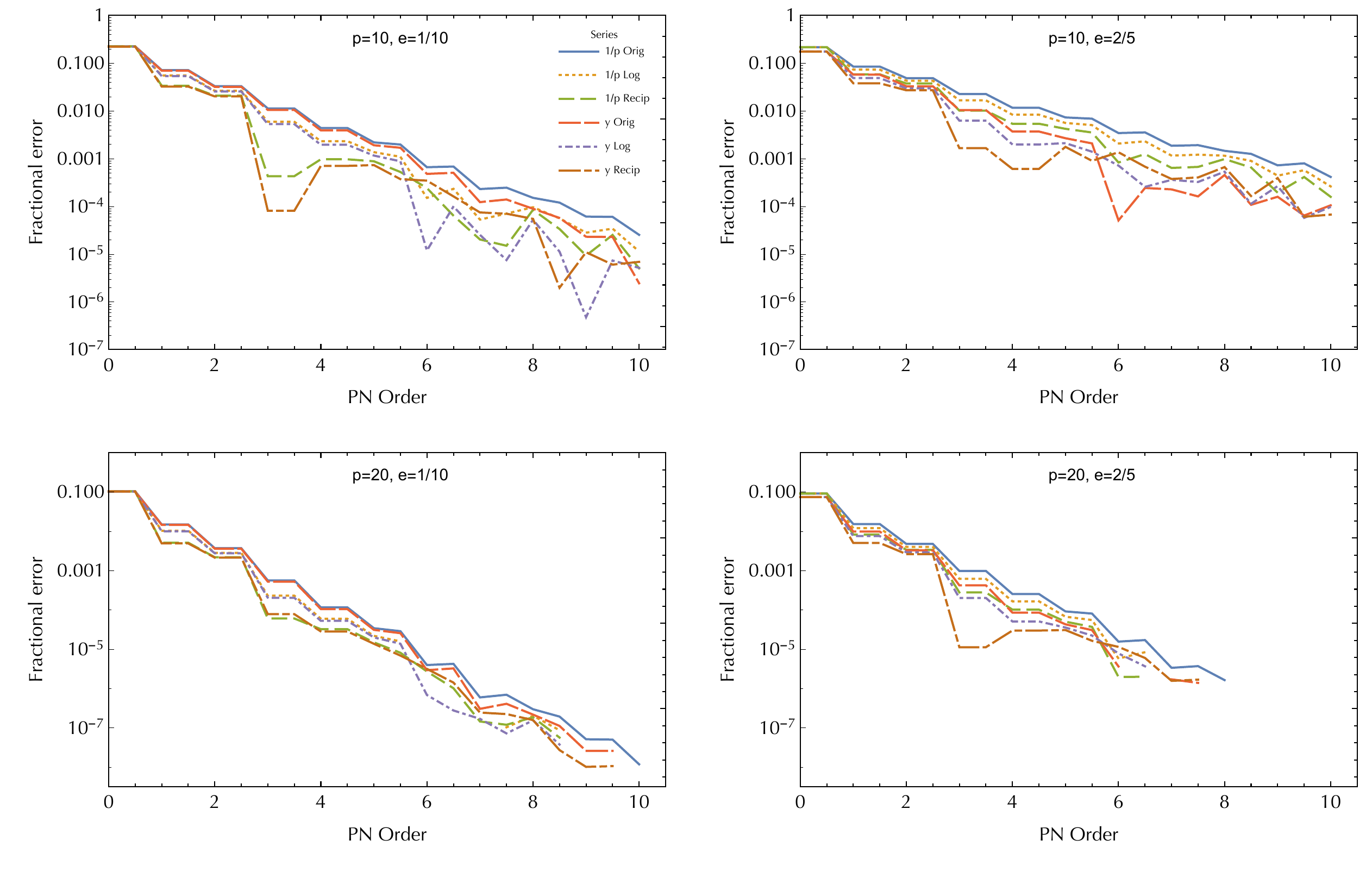}
\caption{Accuracy of the redshift invariant PN expansion and its resummations for several individual orbits.  The
numerical values of our redshift expansion are plotted against data from \cite{AkcaETC15} for the orbits 
$(p=10, e=1/10), (p=10, e=2/5), (p=20, e=1/10), (p=20, e=2/5)$.  Within each plot comparisons are made for 
both the $1/p$ and $y$ expansions, both with and without the use of logarithmic and reciprocal summations. 
Note the changes in vertical scaling in the bottom two plots.  Lines in the bottom right plot vanish where the 
expansions reproduce all numerical digits given in \cite{AkcaETC15}.
\label{fig:p1020ords}}
\end{figure*}

\begin{figure*}
\hspace{-1.5em}\includegraphics[scale=.71]{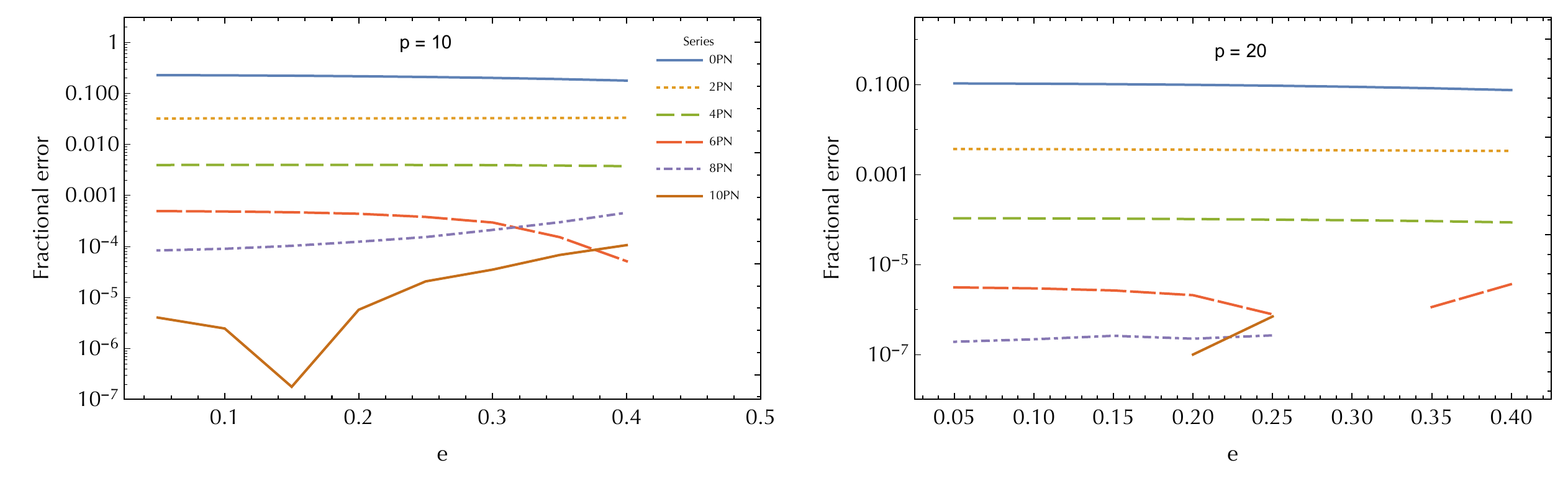}
\caption{Accuracy of the redshift PN expansion with increasing $e$.  The (simple) $y$ expansion is compared to 
numerical data for the $e$ values $0.05$ to $0.40$ at $0.05$ intervals (plots are made continuous for clarity) for both 
$p = 10$ and $p = 20$.  Decreasing residuals are observed with increasing PN order, though with some unexpected 
variations at high order.  Lines in the right-hand plot vanish where the expansions reproduce all numerical digits given 
in \cite{AkcaETC15}.
\label{fig:eAll}}
\end{figure*}

Results from making the comparisons at two orbital sizes ($p=10$ and $p=20$) are shown in 
Figs.~\ref{fig:p1020ords} and \ref{fig:eAll}.  Fig.~\ref{fig:p1020ords} considers four distinct orbits (with the 
two separations and with two eccentricities, $e=0.1$ and $e=0.4$).  The wider orbit ($p=20$) with low ($e=0.1$) 
eccentricity converges rapidly and uniformly with increasing PN terms.  At the other extreme, the orbit with 
$p=10$ and $e=0.4$ is decidedly slower to converge but still reaches a relative error of order $10^{-4}$ when 
using the $y$ expansion and its resummations.  The energy flux required an expansion to 19PN to attain an error 
of $10^{-5}$ for the $p=10$, $e=0.5$ orbit \cite{Munn20}, which suggests that the redshift invariant has better 
convergence properties.  Furthermore, it is worth remembering that in EMRI calculations the contributions to the 
orbital phase evolution from conservative terms in the flux are suppressed by the mass ratio relative to flux 
contributions \cite{HindFlan08}.  This suggests that even a slow-to-converge PN expansion of the conservative part of 
the self-force may be useful in close orbits.  

While the rate of convergence varies with orbital parameters, we do observe at least continued monotonic approach 
to the numerical self-force results as we add PN terms.  Despite the fact that the PN expansion is expected to be 
an asymptotic expansion, there is still no evidence even at 10PN order in the redshift at $p=10$ of added terms 
becoming detrimental to accuracy of the approximation.  Interestingly, the $y$ expansion appears consistently better 
than the $1/p$ expansion in the orbits we have considered.  Finally, it is notable that for the $p=20$, $e=0.4$ orbit 
we reproduce the numerical self-force data with our expansions taken to 8PN order or less.  Hence, the accuracy of 
the full 10PN order expansion with that orbit remains unknown.  

Fig.~\ref{fig:eAll} shows how our expansion offers generally consistently useful accuracy with increasing eccentricity 
$e$.  At higher PN order (6-10PN) the eccentricity dependence is not known exactly but instead contains infinite 
series in $e^2$ that are truncated at $e^{20}$ in the present work.  Factoring out the right eccentricity singular 
factor at each PN order helps, but the truncated series lead to variations in accuracy with $e$.  These effects can
be seen in the rise and fall, and local minimum in the 6PN, 8PN, and 10PN comparisons in the $p=10$ orbit in 
Fig.~\ref{fig:eAll}.  Nevertheless, we expect that the residuals would generally continue to fall if the 
PN series were extended marginally further.  In the $p=20$ orbit in Fig.~\ref{fig:eAll}, the curves of residuals are 
incomplete or missing at 6-10PN orders because of limits on the accuracy of the numerical self-force results to 
which we are making comparison.

\section{Conclusions}
\label{sec:ConsConc}

We have presented the PN and eccentricity expansion of the gravitational redshift invariant, for a point mass in 
eccentric bound motion about a Schwarzschild black hole, to a higher order than has been achieved previously.  
We determine the redshift analytically to 10 PN order and, importantly, to $e^{20}$ in eccentricity.  We present 
results in this paper to 8.5PN, while relegating 9PN to 10PN terms to a posting at the Black Hole Perturbation 
Toolkit \cite{BHPTK18} website.  The depth of the eccentricity expansion allows us to resum on expected singular 
factors and simplify the remaining eccentricity dependence at each PN order.  In many cases we find closed-form 
expressions for the eccentricity dependence.  Some of these closed-form functions are identifiable as terms that 
appear in the PN expansion of the energy flux at infinity, associated with leading-logarithm and subleading-logarithm 
sequences in the energy flux.  The leading-logarithm terms in the energy flux all depend solely upon the Newtonian 
quadrupole moment power spectrum $g(n,e)$ (over eccentric motion harmonics $n$).  Once the presence of these terms 
from the dissipative sector was identified as showing up in the (conservative) redshift invariant, it was possible 
to find known infinite series terms and, using techniques developed in \cite{MunnEvan19a} and \cite{MunnETC20}, to 
uncover added terms in the redshift whose eccentricity dependence follows merely from $g(n,e)$.  A full summary of 
these findings and their significance is found in Sec.~\ref{sec:ResultDiss}.

We also compared the high-order expansions to published close-orbit numerical results to examine the accuracy of 
the PN expansion.  We found the PN expansion to still be converging at 10PN for orbits with semi-latus $p=10$.
It is conceivable that the series might be extended further and still improve accuracy.  The present calculation 
taken to 10PN and $e^{20}$ required about 7 days on the UNC Longleaf cluster.  

All of the machinery presented here is readily extendible to calculating the spin-precession invariant $\psi$ and 
other higher-order invariants.  We will present results on $\psi$ in a subsequent paper.  Additionally, now that 
we are calculating PN expansions in the conservative sector, we may be able to make connection with the EOB 
formalism.  The redshift invariant can be transcribed to yield portions of the EOB $Q(1/r,p_r;\nu)$ potential 
by extending a procedure described in \cite{Leti15}.  However, the process is difficult, with each new order in $e^2$ 
requiring the derivation of an additional transformation.  It is not presently possible to transform closed-form
eccentricity functions in $\langle u^t \rangle_\tau$ to find closed functions in $Q(1/r,p_r;\nu)$.  A similar fact 
is true of the spin-precession invariant, whose (complicated) transformation to the EOB gyromagnetic ratio 
$g_{S*}(1/r;p_r;p_\vp)$ is mapped out in \cite{KavaETC17}.  The derivation of a procedure to transform all powers 
of $e$ would be highly beneficial in the context of this work on closed forms.  Otherwise, it may be possible to 
perform the two transformations to high finite order in $e^2$ and then use factorizations and resummations to extract 
closed forms.  These possibilities will be explored in future work.

\acknowledgements

This work was supported by NSF Grant Nos.~PHY-1806447 and PHY-2110335 to the University of North Carolina--Chapel 
Hill.  C.M.M.~acknowledges additional support from NASA ATP Grant 80NSSC18K1091 to MIT.

\bibliography{redshift}

\begin{thebibliography}{70}%
\makeatletter
\providecommand \@ifxundefined [1]{%
 \@ifx{#1\undefined}
}%
\providecommand \@ifnum [1]{%
 \ifnum #1\expandafter \@firstoftwo
 \else \expandafter \@secondoftwo
 \fi
}%
\providecommand \@ifx [1]{%
 \ifx #1\expandafter \@firstoftwo
 \else \expandafter \@secondoftwo
 \fi
}%
\providecommand \natexlab [1]{#1}%
\providecommand \enquote  [1]{``#1''}%
\providecommand \bibnamefont  [1]{#1}%
\providecommand \bibfnamefont [1]{#1}%
\providecommand \citenamefont [1]{#1}%
\providecommand \href@noop [0]{\@secondoftwo}%
\providecommand \href [0]{\begingroup \@sanitize@url \@href}%
\providecommand \@href[1]{\@@startlink{#1}\@@href}%
\providecommand \@@href[1]{\endgroup#1\@@endlink}%
\providecommand \@sanitize@url [0]{\catcode `\\12\catcode `\$12\catcode
  `\&12\catcode `\#12\catcode `\^12\catcode `\_12\catcode `\%12\relax}%
\providecommand \@@startlink[1]{}%
\providecommand \@@endlink[0]{}%
\providecommand \url  [0]{\begingroup\@sanitize@url \@url }%
\providecommand \@url [1]{\endgroup\@href {#1}{\urlprefix }}%
\providecommand \urlprefix  [0]{URL }%
\providecommand \Eprint [0]{\href }%
\providecommand \doibase [0]{http://dx.doi.org/}%
\providecommand \selectlanguage [0]{\@gobble}%
\providecommand \bibinfo  [0]{\@secondoftwo}%
\providecommand \bibfield  [0]{\@secondoftwo}%
\providecommand \translation [1]{[#1]}%
\providecommand \BibitemOpen [0]{}%
\providecommand \bibitemStop [0]{}%
\providecommand \bibitemNoStop [0]{.\EOS\space}%
\providecommand \EOS [0]{\spacefactor3000\relax}%
\providecommand \BibitemShut  [1]{\csname bibitem#1\endcsname}%
\let\auto@bib@innerbib\@empty
\bibitem [{\citenamefont {{Munna}}(2020{\natexlab{a}})}]{Munn20}%
  \BibitemOpen
  \bibfield  {author} {\bibinfo {author} {\bibfnamefont {C.}~\bibnamefont
  {{Munna}}},\ }\href {\doibase 10.1103/PhysRevD.102.124001} {\bibfield
  {journal} {\bibinfo  {journal} {Phys. Rev. D}\ }\textbf {\bibinfo {volume}
  {102}},\ \bibinfo {eid} {124001} (\bibinfo {year} {2020}{\natexlab{a}})},\
  \Eprint {http://arxiv.org/abs/2008.10622} {arXiv:2008.10622 [gr-qc]}
  \BibitemShut {NoStop}%
\bibitem [{\citenamefont {{Munna}}(2020{\natexlab{b}})}]{Munn20c}%
  \BibitemOpen
  \bibfield  {author} {\bibinfo {author} {\bibfnamefont {C.}~\bibnamefont
  {{Munna}}},\ }\emph {\bibinfo {title} {{Eccentric-orbit binary black hole
  inspirals: informing the post-Newtonian expansion through black hole
  perturbation theory and multipole moment analysis}}},\ \href@noop {} {Ph.D.
  thesis},\ \bibinfo  {school} {University of North Carolina at Chapel Hill}
  (\bibinfo {year} {2020}{\natexlab{b}})\BibitemShut {NoStop}%
\bibitem [{\citenamefont {{Munna}}\ \emph {et~al.}(2020)\citenamefont
  {{Munna}}, \citenamefont {{Evans}}, \citenamefont {{Hopper}},\ and\
  \citenamefont {{Forseth}}}]{MunnETC20}%
  \BibitemOpen
  \bibfield  {author} {\bibinfo {author} {\bibfnamefont {C.}~\bibnamefont
  {{Munna}}}, \bibinfo {author} {\bibfnamefont {C.~R.}\ \bibnamefont
  {{Evans}}}, \bibinfo {author} {\bibfnamefont {S.}~\bibnamefont {{Hopper}}}, \
  and\ \bibinfo {author} {\bibfnamefont {E.}~\bibnamefont {{Forseth}}},\ }\href
  {\doibase 10.1103/PhysRevD.102.024047} {\bibfield  {journal} {\bibinfo
  {journal} {Phys. Rev. D}\ }\textbf {\bibinfo {volume} {102}},\ \bibinfo {eid}
  {024047} (\bibinfo {year} {2020})},\ \Eprint
  {http://arxiv.org/abs/2005.03044} {arXiv:2005.03044 [gr-qc]} \BibitemShut
  {NoStop}%
\bibitem [{\citenamefont {{Munna}}\ and\ \citenamefont
  {{Evans}}(2019)}]{MunnEvan19a}%
  \BibitemOpen
  \bibfield  {author} {\bibinfo {author} {\bibfnamefont {C.}~\bibnamefont
  {{Munna}}}\ and\ \bibinfo {author} {\bibfnamefont {C.~R.}\ \bibnamefont
  {{Evans}}},\ }\href {\doibase 10.1103/PhysRevD.100.104060} {\bibfield
  {journal} {\bibinfo  {journal} {Phys. Rev. D}\ }\textbf {\bibinfo {volume}
  {100}},\ \bibinfo {eid} {104060} (\bibinfo {year} {2019})},\ \Eprint
  {http://arxiv.org/abs/1909.05877} {arXiv:1909.05877 [gr-qc]} \BibitemShut
  {NoStop}%
\bibitem [{\citenamefont {Munna}\ and\ \citenamefont
  {Evans}(2020{\natexlab{a}})}]{MunnEvan20a}%
  \BibitemOpen
  \bibfield  {author} {\bibinfo {author} {\bibfnamefont {C.}~\bibnamefont
  {Munna}}\ and\ \bibinfo {author} {\bibfnamefont {C.~R.}\ \bibnamefont
  {Evans}},\ }\href {\doibase 10.1103/PhysRevD.102.104006} {\bibfield
  {journal} {\bibinfo  {journal} {Phys. Rev. D}\ }\textbf {\bibinfo {volume}
  {102}},\ \bibinfo {eid} {104006} (\bibinfo {year} {2020}{\natexlab{a}})},\
  \Eprint {http://arxiv.org/abs/2009.01254} {arXiv:2009.01254 [gr-qc]}
  \BibitemShut {NoStop}%
\bibitem [{\citenamefont {Munna}\ and\ \citenamefont
  {Evans}(2020{\natexlab{b}})}]{MunnEvan20b}%
  \BibitemOpen
  \bibfield  {author} {\bibinfo {author} {\bibfnamefont {C.}~\bibnamefont
  {Munna}}\ and\ \bibinfo {author} {\bibfnamefont {C.~R.}\ \bibnamefont
  {Evans}},\ }\href@noop {} {\bibfield  {journal} {\bibinfo  {journal} {to be
  submitted to Phys. Rev. D}\ } (\bibinfo {year}
  {2020}{\natexlab{b}})}\BibitemShut {NoStop}%
\bibitem [{\citenamefont {Hinderer}\ and\ \citenamefont
  {Flanagan}(2008)}]{HindFlan08}%
  \BibitemOpen
  \bibfield  {author} {\bibinfo {author} {\bibfnamefont {T.}~\bibnamefont
  {Hinderer}}\ and\ \bibinfo {author} {\bibfnamefont {E.~E.}\ \bibnamefont
  {Flanagan}},\ }\href {\doibase 10.1103/PhysRevD.78.064028} {\bibfield
  {journal} {\bibinfo  {journal} {Phys. Rev. D}\ }\textbf {\bibinfo {volume}
  {78}},\ \bibinfo {pages} {064028} (\bibinfo {year} {2008})},\ \Eprint
  {http://arxiv.org/abs/0805.3337} {arXiv:0805.3337 [gr-qc]} \BibitemShut
  {NoStop}%
\bibitem [{\citenamefont {{Katz}}\ \emph {et~al.}(2021)\citenamefont {{Katz}},
  \citenamefont {{Chua}}, \citenamefont {{Speri}}, \citenamefont
  {{Warburton}},\ and\ \citenamefont {{Hughes}}}]{KatzETC21}%
  \BibitemOpen
  \bibfield  {author} {\bibinfo {author} {\bibfnamefont {M.~L.}\ \bibnamefont
  {{Katz}}}, \bibinfo {author} {\bibfnamefont {A.~J.~K.}\ \bibnamefont
  {{Chua}}}, \bibinfo {author} {\bibfnamefont {L.}~\bibnamefont {{Speri}}},
  \bibinfo {author} {\bibfnamefont {N.}~\bibnamefont {{Warburton}}}, \ and\
  \bibinfo {author} {\bibfnamefont {S.~A.}\ \bibnamefont {{Hughes}}},\ }\href
  {\doibase 10.1103/PhysRevD.104.064047} {\bibfield  {journal} {\bibinfo
  {journal} {Phys. Rev. D}\ }\textbf {\bibinfo {volume} {104}},\ \bibinfo {eid}
  {064047} (\bibinfo {year} {2021})},\ \Eprint
  {http://arxiv.org/abs/2104.04582} {arXiv:2104.04582 [gr-qc]} \BibitemShut
  {NoStop}%
\bibitem [{\citenamefont {Detweiler}(2005)}]{Detw05}%
  \BibitemOpen
  \bibfield  {author} {\bibinfo {author} {\bibfnamefont {S.~L.}\ \bibnamefont
  {Detweiler}},\ }\href {\doibase 10.1088/0264-9381/22/15/006} {\bibfield
  {journal} {\bibinfo  {journal} {Class. Quant. Grav.}\ }\textbf {\bibinfo
  {volume} {22}},\ \bibinfo {pages} {S681} (\bibinfo {year} {2005})},\ \Eprint
  {http://arxiv.org/abs/gr-qc/0501004} {arXiv:gr-qc/0501004 [gr-qc]}
  \BibitemShut {NoStop}%
\bibitem [{\citenamefont {Detweiler}(2008)}]{Detw08}%
  \BibitemOpen
  \bibfield  {author} {\bibinfo {author} {\bibfnamefont {S.}~\bibnamefont
  {Detweiler}},\ }\href {\doibase 10.1103/PhysRevD.77.124026} {\bibfield
  {journal} {\bibinfo  {journal} {Phys. Rev. D}\ }\textbf {\bibinfo {volume}
  {77}},\ \bibinfo {pages} {124026} (\bibinfo {year} {2008})},\ \Eprint
  {http://arxiv.org/abs/0804.3529} {arXiv:0804.3529 [gr-qc]} \BibitemShut
  {NoStop}%
\bibitem [{\citenamefont {Kavanagh}\ \emph {et~al.}(2015)\citenamefont
  {Kavanagh}, \citenamefont {Ottewill},\ and\ \citenamefont
  {Wardell}}]{KavaOtteWard15}%
  \BibitemOpen
  \bibfield  {author} {\bibinfo {author} {\bibfnamefont {C.}~\bibnamefont
  {Kavanagh}}, \bibinfo {author} {\bibfnamefont {A.~C.}\ \bibnamefont
  {Ottewill}}, \ and\ \bibinfo {author} {\bibfnamefont {B.}~\bibnamefont
  {Wardell}},\ }\href {\doibase 10.1103/PhysRevD.92.084025} {\bibfield
  {journal} {\bibinfo  {journal} {Phys. Rev. D}\ }\textbf {\bibinfo {volume}
  {92}},\ \bibinfo {pages} {084025} (\bibinfo {year} {2015})},\ \Eprint
  {http://arxiv.org/abs/1503.02334} {arXiv:1503.02334 [gr-qc]} \BibitemShut
  {NoStop}%
\bibitem [{\citenamefont {{Barack}}\ and\ \citenamefont
  {{Sago}}(2011)}]{BaraSago11}%
  \BibitemOpen
  \bibfield  {author} {\bibinfo {author} {\bibfnamefont {L.}~\bibnamefont
  {{Barack}}}\ and\ \bibinfo {author} {\bibfnamefont {N.}~\bibnamefont
  {{Sago}}},\ }\href {\doibase 10.1103/PhysRevD.83.084023} {\bibfield
  {journal} {\bibinfo  {journal} {Phys. Rev. D}\ }\textbf {\bibinfo {volume}
  {83}},\ \bibinfo {pages} {084023} (\bibinfo {year} {2011})},\ \Eprint
  {http://arxiv.org/abs/1101.3331} {arXiv:1101.3331 [gr-qc]} \BibitemShut
  {NoStop}%
\bibitem [{\citenamefont {{Akcay}}\ \emph {et~al.}(2015)\citenamefont
  {{Akcay}}, \citenamefont {{Le Tiec}}, \citenamefont {{Barack}}, \citenamefont
  {{Sago}},\ and\ \citenamefont {{Warburton}}}]{AkcaETC15}%
  \BibitemOpen
  \bibfield  {author} {\bibinfo {author} {\bibfnamefont {S.}~\bibnamefont
  {{Akcay}}}, \bibinfo {author} {\bibfnamefont {A.}~\bibnamefont {{Le Tiec}}},
  \bibinfo {author} {\bibfnamefont {L.}~\bibnamefont {{Barack}}}, \bibinfo
  {author} {\bibfnamefont {N.}~\bibnamefont {{Sago}}}, \ and\ \bibinfo {author}
  {\bibfnamefont {N.}~\bibnamefont {{Warburton}}},\ }\href {\doibase
  10.1103/PhysRevD.91.124014} {\bibfield  {journal} {\bibinfo  {journal} {Phys.
  Rev. D}\ }\textbf {\bibinfo {volume} {91}},\ \bibinfo {eid} {124014}
  (\bibinfo {year} {2015})},\ \Eprint {http://arxiv.org/abs/1503.01374}
  {arXiv:1503.01374 [gr-qc]} \BibitemShut {NoStop}%
\bibitem [{\citenamefont {{Blanchet}}(2014)}]{Blan14}%
  \BibitemOpen
  \bibfield  {author} {\bibinfo {author} {\bibfnamefont {L.}~\bibnamefont
  {{Blanchet}}},\ }\href {\doibase 10.12942/lrr-2014-2} {\bibfield  {journal}
  {\bibinfo  {journal} {Living Reviews in Relativity}\ }\textbf {\bibinfo
  {volume} {17}},\ \bibinfo {pages} {2} (\bibinfo {year} {2014})},\ \Eprint
  {http://arxiv.org/abs/1310.1528} {arXiv:1310.1528 [gr-qc]} \BibitemShut
  {NoStop}%
\bibitem [{\citenamefont {Barack}\ and\ \citenamefont
  {Sago}(2009)}]{BaraSago09}%
  \BibitemOpen
  \bibfield  {author} {\bibinfo {author} {\bibfnamefont {L.}~\bibnamefont
  {Barack}}\ and\ \bibinfo {author} {\bibfnamefont {N.}~\bibnamefont {Sago}},\
  }\href {\doibase 10.1103/PhysRevLett.102.191101} {\bibfield  {journal}
  {\bibinfo  {journal} {Physical Review Letters}\ }\textbf {\bibinfo {volume}
  {102}},\ \bibinfo {pages} {191101} (\bibinfo {year} {2009})},\ \Eprint
  {http://arxiv.org/abs/0902.0573} {arXiv:0902.0573 [gr-qc]} \BibitemShut
  {NoStop}%
\bibitem [{\citenamefont {{Dolan}}\ \emph {et~al.}(2014)\citenamefont
  {{Dolan}}, \citenamefont {{Warburton}}, \citenamefont {{Harte}},
  \citenamefont {{Le Tiec}}, \citenamefont {{Wardell}},\ and\ \citenamefont
  {{Barack}}}]{DolaETC14a}%
  \BibitemOpen
  \bibfield  {author} {\bibinfo {author} {\bibfnamefont {S.~R.}\ \bibnamefont
  {{Dolan}}}, \bibinfo {author} {\bibfnamefont {N.}~\bibnamefont
  {{Warburton}}}, \bibinfo {author} {\bibfnamefont {A.~I.}\ \bibnamefont
  {{Harte}}}, \bibinfo {author} {\bibfnamefont {A.}~\bibnamefont {{Le Tiec}}},
  \bibinfo {author} {\bibfnamefont {B.}~\bibnamefont {{Wardell}}}, \ and\
  \bibinfo {author} {\bibfnamefont {L.}~\bibnamefont {{Barack}}},\ }\href
  {\doibase 10.1103/PhysRevD.89.064011} {\bibfield  {journal} {\bibinfo
  {journal} {Phys. Rev. D}\ }\textbf {\bibinfo {volume} {89}},\ \bibinfo {eid}
  {064011} (\bibinfo {year} {2014})},\ \Eprint {http://arxiv.org/abs/1312.0775}
  {arXiv:1312.0775 [gr-qc]} \BibitemShut {NoStop}%
\bibitem [{\citenamefont {Bini}\ and\ \citenamefont
  {Damour}(2014{\natexlab{a}})}]{BiniDamo14b}%
  \BibitemOpen
  \bibfield  {author} {\bibinfo {author} {\bibfnamefont {D.}~\bibnamefont
  {Bini}}\ and\ \bibinfo {author} {\bibfnamefont {T.}~\bibnamefont {Damour}},\
  }\href {\doibase 10.1103/PhysRevD.90.024039} {\bibfield  {journal} {\bibinfo
  {journal} {Phys. Rev.}\ }\textbf {\bibinfo {volume} {D90}},\ \bibinfo {pages}
  {024039} (\bibinfo {year} {2014}{\natexlab{a}})},\ \Eprint
  {http://arxiv.org/abs/1404.2747} {arXiv:1404.2747 [gr-qc]} \BibitemShut
  {NoStop}%
\bibitem [{\citenamefont {{Dolan}}\ \emph {et~al.}(2015)\citenamefont
  {{Dolan}}, \citenamefont {{Nolan}}, \citenamefont {{Ottewill}}, \citenamefont
  {{Warburton}},\ and\ \citenamefont {{Wardell}}}]{DolaETC14b}%
  \BibitemOpen
  \bibfield  {author} {\bibinfo {author} {\bibfnamefont {S.~R.}\ \bibnamefont
  {{Dolan}}}, \bibinfo {author} {\bibfnamefont {P.}~\bibnamefont {{Nolan}}},
  \bibinfo {author} {\bibfnamefont {A.~C.}\ \bibnamefont {{Ottewill}}},
  \bibinfo {author} {\bibfnamefont {N.}~\bibnamefont {{Warburton}}}, \ and\
  \bibinfo {author} {\bibfnamefont {B.}~\bibnamefont {{Wardell}}},\ }\href
  {\doibase 10.1103/PhysRevD.91.023009} {\bibfield  {journal} {\bibinfo
  {journal} {Phys. Rev. D}\ }\textbf {\bibinfo {volume} {91}},\ \bibinfo {eid}
  {023009} (\bibinfo {year} {2015})},\ \Eprint {http://arxiv.org/abs/1406.4890}
  {arXiv:1406.4890 [gr-qc]} \BibitemShut {NoStop}%
\bibitem [{\citenamefont {{Nolan}}\ \emph {et~al.}(2015)\citenamefont
  {{Nolan}}, \citenamefont {{Kavanagh}}, \citenamefont {{Dolan}}, \citenamefont
  {{Ottewill}}, \citenamefont {{Warburton}},\ and\ \citenamefont
  {{Wardell}}}]{NolaETC15}%
  \BibitemOpen
  \bibfield  {author} {\bibinfo {author} {\bibfnamefont {P.}~\bibnamefont
  {{Nolan}}}, \bibinfo {author} {\bibfnamefont {C.}~\bibnamefont {{Kavanagh}}},
  \bibinfo {author} {\bibfnamefont {S.~R.}\ \bibnamefont {{Dolan}}}, \bibinfo
  {author} {\bibfnamefont {A.~C.}\ \bibnamefont {{Ottewill}}}, \bibinfo
  {author} {\bibfnamefont {N.}~\bibnamefont {{Warburton}}}, \ and\ \bibinfo
  {author} {\bibfnamefont {B.}~\bibnamefont {{Wardell}}},\ }\href {\doibase
  10.1103/PhysRevD.92.123008} {\bibfield  {journal} {\bibinfo  {journal} {Phys.
  Rev. D}\ }\textbf {\bibinfo {volume} {92}},\ \bibinfo {eid} {123008}
  (\bibinfo {year} {2015})},\ \Eprint {http://arxiv.org/abs/1505.04447}
  {arXiv:1505.04447 [gr-qc]} \BibitemShut {NoStop}%
\bibitem [{\citenamefont {{Barack}}\ \emph {et~al.}(2010)\citenamefont
  {{Barack}}, \citenamefont {{Damour}},\ and\ \citenamefont
  {{Sago}}}]{BaraDamoSago10}%
  \BibitemOpen
  \bibfield  {author} {\bibinfo {author} {\bibfnamefont {L.}~\bibnamefont
  {{Barack}}}, \bibinfo {author} {\bibfnamefont {T.}~\bibnamefont {{Damour}}},
  \ and\ \bibinfo {author} {\bibfnamefont {N.}~\bibnamefont {{Sago}}},\ }\href
  {\doibase 10.1103/PhysRevD.82.084036} {\bibfield  {journal} {\bibinfo
  {journal} {Phys. Rev. D}\ }\textbf {\bibinfo {volume} {82}},\ \bibinfo {eid}
  {084036} (\bibinfo {year} {2010})},\ \Eprint {http://arxiv.org/abs/1008.0935}
  {arXiv:1008.0935 [gr-qc]} \BibitemShut {NoStop}%
\bibitem [{\citenamefont {{Le Tiec}}\ \emph {et~al.}(2012)\citenamefont {{Le
  Tiec}}, \citenamefont {{Blanchet}},\ and\ \citenamefont
  {{Whiting}}}]{LetiBlanWhit12}%
  \BibitemOpen
  \bibfield  {author} {\bibinfo {author} {\bibfnamefont {A.}~\bibnamefont {{Le
  Tiec}}}, \bibinfo {author} {\bibfnamefont {L.}~\bibnamefont {{Blanchet}}}, \
  and\ \bibinfo {author} {\bibfnamefont {B.}~\bibnamefont {{Whiting}}},\ }\href
  {\doibase 10.1103/PhysRevD.85.064039} {\bibfield  {journal} {\bibinfo
  {journal} {Phys. Rev. D}\ }\textbf {\bibinfo {volume} {85}},\ \bibinfo {eid}
  {064039} (\bibinfo {year} {2012})},\ \Eprint {http://arxiv.org/abs/1111.5378}
  {arXiv:1111.5378} \BibitemShut {NoStop}%
\bibitem [{\citenamefont {Bini}\ and\ \citenamefont
  {Damour}(2014{\natexlab{b}})}]{BiniDamo14c}%
  \BibitemOpen
  \bibfield  {author} {\bibinfo {author} {\bibfnamefont {D.}~\bibnamefont
  {Bini}}\ and\ \bibinfo {author} {\bibfnamefont {T.}~\bibnamefont {Damour}},\
  }\href {\doibase 10.1103/PhysRevD.90.124037} {\bibfield  {journal} {\bibinfo
  {journal} {Phys. Rev. D}\ }\textbf {\bibinfo {volume} {90}},\ \bibinfo
  {pages} {124037} (\bibinfo {year} {2014}{\natexlab{b}})},\ \Eprint
  {http://arxiv.org/abs/1409.6933} {arXiv:1409.6933 [gr-qc]} \BibitemShut
  {NoStop}%
\bibitem [{\citenamefont {Bini}\ \emph {et~al.}(2016)\citenamefont {Bini},
  \citenamefont {Damour},\ and\ \citenamefont {Geralico}}]{BiniDamoGera15}%
  \BibitemOpen
  \bibfield  {author} {\bibinfo {author} {\bibfnamefont {D.}~\bibnamefont
  {Bini}}, \bibinfo {author} {\bibfnamefont {T.}~\bibnamefont {Damour}}, \ and\
  \bibinfo {author} {\bibfnamefont {A.}~\bibnamefont {Geralico}},\ }\href
  {\doibase 10.1103/PhysRevD.90.124037} {\bibfield  {journal} {\bibinfo
  {journal} {Phys. Rev. D}\ }\textbf {\bibinfo {volume} {93}},\ \bibinfo
  {pages} {064023} (\bibinfo {year} {2016})},\ \Eprint
  {http://arxiv.org/abs/1511.04533} {arXiv:1511.04533 [gr-qc]} \BibitemShut
  {NoStop}%
\bibitem [{\citenamefont {Le~Tiec}(2015)}]{Leti15}%
  \BibitemOpen
  \bibfield  {author} {\bibinfo {author} {\bibfnamefont {A.}~\bibnamefont
  {Le~Tiec}},\ }\href {\doibase 10.1103/PhysRevD.92.084021} {\bibfield
  {journal} {\bibinfo  {journal} {Phys. Rev.}\ }\textbf {\bibinfo {volume}
  {D92}},\ \bibinfo {pages} {084021} (\bibinfo {year} {2015})},\ \Eprint
  {http://arxiv.org/abs/1506.05648} {arXiv:1506.05648 [gr-qc]} \BibitemShut
  {NoStop}%
\bibitem [{\citenamefont {{Hopper}}\ \emph {et~al.}(2016)\citenamefont
  {{Hopper}}, \citenamefont {{Kavanagh}},\ and\ \citenamefont
  {{Ottewill}}}]{HoppKavaOtte16}%
  \BibitemOpen
  \bibfield  {author} {\bibinfo {author} {\bibfnamefont {S.}~\bibnamefont
  {{Hopper}}}, \bibinfo {author} {\bibfnamefont {C.}~\bibnamefont
  {{Kavanagh}}}, \ and\ \bibinfo {author} {\bibfnamefont {A.~C.}\ \bibnamefont
  {{Ottewill}}},\ }\href {\doibase 10.1103/PhysRevD.93.044010} {\bibfield
  {journal} {\bibinfo  {journal} {Phys. Rev. D}\ }\textbf {\bibinfo {volume}
  {93}},\ \bibinfo {eid} {044010} (\bibinfo {year} {2016})},\ \Eprint
  {http://arxiv.org/abs/1512.01556} {arXiv:1512.01556 [gr-qc]} \BibitemShut
  {NoStop}%
\bibitem [{\citenamefont {Kavanagh}\ \emph {et~al.}(2017)\citenamefont
  {Kavanagh}, \citenamefont {Bini}, \citenamefont {Damour}, \citenamefont
  {Hopper}, \citenamefont {Ottewil},\ and\ \citenamefont
  {Wardell}}]{KavaETC17}%
  \BibitemOpen
  \bibfield  {author} {\bibinfo {author} {\bibfnamefont {C.}~\bibnamefont
  {Kavanagh}}, \bibinfo {author} {\bibfnamefont {D.}~\bibnamefont {Bini}},
  \bibinfo {author} {\bibfnamefont {T.}~\bibnamefont {Damour}}, \bibinfo
  {author} {\bibfnamefont {S.}~\bibnamefont {Hopper}}, \bibinfo {author}
  {\bibfnamefont {A.}~\bibnamefont {Ottewil}}, \ and\ \bibinfo {author}
  {\bibfnamefont {B.}~\bibnamefont {Wardell}},\ }\href {\doibase
  10.1103/PhysRevD.96.064012} {\bibfield  {journal} {\bibinfo  {journal} {Phys.
  Rev. D}\ }\textbf {\bibinfo {volume} {96}},\ \bibinfo {eid} {064012}
  (\bibinfo {year} {2017}),\ 10.1103/PhysRevD.96.064012},\ \Eprint
  {http://arxiv.org/abs/1706.00459} {arXiv:1706.00459 [gr-qc]} \BibitemShut
  {NoStop}%
\bibitem [{\citenamefont {{Bini}}\ \emph {et~al.}(2018)\citenamefont {{Bini}},
  \citenamefont {{Damour}},\ and\ \citenamefont {{Geralico}}}]{BiniDamoGera18}%
  \BibitemOpen
  \bibfield  {author} {\bibinfo {author} {\bibfnamefont {D.}~\bibnamefont
  {{Bini}}}, \bibinfo {author} {\bibfnamefont {T.}~\bibnamefont {{Damour}}}, \
  and\ \bibinfo {author} {\bibfnamefont {A.}~\bibnamefont {{Geralico}}},\
  }\href {\doibase 10.1103/PhysRevD.97.104046} {\bibfield  {journal} {\bibinfo
  {journal} {Phys. Rev. D}\ }\textbf {\bibinfo {volume} {97}},\ \bibinfo {eid}
  {104046} (\bibinfo {year} {2018})},\ \Eprint
  {http://arxiv.org/abs/1801.03704} {arXiv:1801.03704 [gr-qc]} \BibitemShut
  {NoStop}%
\bibitem [{\citenamefont {{Bini}}\ \emph {et~al.}(2019)\citenamefont {{Bini}},
  \citenamefont {{Damour}},\ and\ \citenamefont {{Geralico}}}]{BiniDamoGera19}%
  \BibitemOpen
  \bibfield  {author} {\bibinfo {author} {\bibfnamefont {D.}~\bibnamefont
  {{Bini}}}, \bibinfo {author} {\bibfnamefont {T.}~\bibnamefont {{Damour}}}, \
  and\ \bibinfo {author} {\bibfnamefont {A.}~\bibnamefont {{Geralico}}},\
  }\href {\doibase 10.1103/PhysRevLett.123.231104} {\bibfield  {journal}
  {\bibinfo  {journal} {Phys. Rev. Lett.}\ }\textbf {\bibinfo {volume} {123}},\
  \bibinfo {eid} {231104} (\bibinfo {year} {2019})},\ \Eprint
  {http://arxiv.org/abs/1909.02375} {arXiv:1909.02375 [gr-qc]} \BibitemShut
  {NoStop}%
\bibitem [{\citenamefont {{Bini}}\ \emph
  {et~al.}(2020{\natexlab{a}})\citenamefont {{Bini}}, \citenamefont
  {{Damour}},\ and\ \citenamefont {{Geralico}}}]{BiniDamoGera20a}%
  \BibitemOpen
  \bibfield  {author} {\bibinfo {author} {\bibfnamefont {D.}~\bibnamefont
  {{Bini}}}, \bibinfo {author} {\bibfnamefont {T.}~\bibnamefont {{Damour}}}, \
  and\ \bibinfo {author} {\bibfnamefont {A.}~\bibnamefont {{Geralico}}},\
  }\href@noop {} {\  (\bibinfo {year} {2020}{\natexlab{a}})},\ \Eprint
  {http://arxiv.org/abs/2003.11891} {arXiv:2003.11891 [gr-qc]} \BibitemShut
  {NoStop}%
\bibitem [{\citenamefont {{Bini}}\ \emph
  {et~al.}(2020{\natexlab{b}})\citenamefont {{Bini}}, \citenamefont
  {{Damour}},\ and\ \citenamefont {{Geralico}}}]{BiniDamoGera20b}%
  \BibitemOpen
  \bibfield  {author} {\bibinfo {author} {\bibfnamefont {D.}~\bibnamefont
  {{Bini}}}, \bibinfo {author} {\bibfnamefont {T.}~\bibnamefont {{Damour}}}, \
  and\ \bibinfo {author} {\bibfnamefont {A.}~\bibnamefont {{Geralico}}},\
  }\href@noop {} {\  (\bibinfo {year} {2020}{\natexlab{b}})},\ \Eprint
  {http://arxiv.org/abs/2004.05407} {arXiv:2004.05407 [gr-qc]} \BibitemShut
  {NoStop}%
\bibitem [{\citenamefont {{Bini}}\ \emph
  {et~al.}(2016{\natexlab{a}})\citenamefont {{Bini}}, \citenamefont
  {{Damour}},\ and\ \citenamefont {{Geralico}}}]{BiniDamoGera16a}%
  \BibitemOpen
  \bibfield  {author} {\bibinfo {author} {\bibfnamefont {D.}~\bibnamefont
  {{Bini}}}, \bibinfo {author} {\bibfnamefont {T.}~\bibnamefont {{Damour}}}, \
  and\ \bibinfo {author} {\bibfnamefont {A.}~\bibnamefont {{Geralico}}},\
  }\href {\doibase 10.1103/PhysRevD.93.064023} {\bibfield  {journal} {\bibinfo
  {journal} {Physical Review D}\ }\textbf {\bibinfo {volume} {93}},\ \bibinfo
  {eid} {064023} (\bibinfo {year} {2016}{\natexlab{a}})},\ \Eprint
  {http://arxiv.org/abs/1511.04533} {arXiv:1511.04533 [gr-qc]} \BibitemShut
  {NoStop}%
\bibitem [{\citenamefont {{Bini}}\ \emph
  {et~al.}(2016{\natexlab{b}})\citenamefont {{Bini}}, \citenamefont
  {{Damour}},\ and\ \citenamefont {{Geralico}}}]{BiniDamoGera16c}%
  \BibitemOpen
  \bibfield  {author} {\bibinfo {author} {\bibfnamefont {D.}~\bibnamefont
  {{Bini}}}, \bibinfo {author} {\bibfnamefont {T.}~\bibnamefont {{Damour}}}, \
  and\ \bibinfo {author} {\bibfnamefont {A.}~\bibnamefont {{Geralico}}},\
  }\href {\doibase 10.1103/PhysRevD.93.124058} {\bibfield  {journal} {\bibinfo
  {journal} {Physical Review D}\ }\textbf {\bibinfo {volume} {93}},\ \bibinfo
  {eid} {124058} (\bibinfo {year} {2016}{\natexlab{b}})},\ \Eprint
  {http://arxiv.org/abs/1602.08282} {arXiv:1602.08282 [gr-qc]} \BibitemShut
  {NoStop}%
\bibitem [{\citenamefont {Regge}\ and\ \citenamefont
  {Wheeler}(1957)}]{ReggWhee57}%
  \BibitemOpen
  \bibfield  {author} {\bibinfo {author} {\bibfnamefont {T.}~\bibnamefont
  {Regge}}\ and\ \bibinfo {author} {\bibfnamefont {J.}~\bibnamefont
  {Wheeler}},\ }\href@noop {} {\bibfield  {journal} {\bibinfo  {journal} {Phys.
  Rev.}\ }\textbf {\bibinfo {volume} {108}},\ \bibinfo {pages} {1063} (\bibinfo
  {year} {1957})}\BibitemShut {NoStop}%
\bibitem [{\citenamefont {Zerilli}(1970)}]{Zeri70}%
  \BibitemOpen
  \bibfield  {author} {\bibinfo {author} {\bibfnamefont {F.}~\bibnamefont
  {Zerilli}},\ }\href@noop {} {\bibfield  {journal} {\bibinfo  {journal} {Phys.
  Rev. D}\ }\textbf {\bibinfo {volume} {2}},\ \bibinfo {pages} {2141} (\bibinfo
  {year} {1970})}\BibitemShut {NoStop}%
\bibitem [{\citenamefont {{Mano}}\ \emph {et~al.}(1996)\citenamefont {{Mano}},
  \citenamefont {{Suzuki}},\ and\ \citenamefont
  {{Takasugi}}}]{ManoSuzuTaka96a}%
  \BibitemOpen
  \bibfield  {author} {\bibinfo {author} {\bibfnamefont {S.}~\bibnamefont
  {{Mano}}}, \bibinfo {author} {\bibfnamefont {H.}~\bibnamefont {{Suzuki}}}, \
  and\ \bibinfo {author} {\bibfnamefont {E.}~\bibnamefont {{Takasugi}}},\
  }\href {\doibase 10.1143/PTP.96.549} {\bibfield  {journal} {\bibinfo
  {journal} {Progress of Theoretical Physics}\ }\textbf {\bibinfo {volume}
  {96}},\ \bibinfo {pages} {549} (\bibinfo {year} {1996})},\ \Eprint
  {http://arxiv.org/abs/gr-qc/9605057} {gr-qc/9605057} \BibitemShut {NoStop}%
\bibitem [{\citenamefont {Bini}\ and\ \citenamefont
  {Damour}(2013)}]{BiniDamo13}%
  \BibitemOpen
  \bibfield  {author} {\bibinfo {author} {\bibfnamefont {D.}~\bibnamefont
  {Bini}}\ and\ \bibinfo {author} {\bibfnamefont {T.}~\bibnamefont {Damour}},\
  }\href {\doibase 10.1103/PhysRevD.87.121501} {\bibfield  {journal} {\bibinfo
  {journal} {Phys. Rev.}\ }\textbf {\bibinfo {volume} {D87}},\ \bibinfo {pages}
  {121501} (\bibinfo {year} {2013})},\ \Eprint {http://arxiv.org/abs/1305.4884}
  {arXiv:1305.4884 [gr-qc]} \BibitemShut {NoStop}%
\bibitem [{\citenamefont {{Bini}}\ and\ \citenamefont
  {{Damour}}(2014)}]{BiniDamo14a}%
  \BibitemOpen
  \bibfield  {author} {\bibinfo {author} {\bibfnamefont {D.}~\bibnamefont
  {{Bini}}}\ and\ \bibinfo {author} {\bibfnamefont {T.}~\bibnamefont
  {{Damour}}},\ }\href {\doibase 10.1103/PhysRevD.89.064063} {\bibfield
  {journal} {\bibinfo  {journal} {Phys. Rev. D}\ }\textbf {\bibinfo {volume}
  {89}},\ \bibinfo {eid} {064063} (\bibinfo {year} {2014})},\ \Eprint
  {http://arxiv.org/abs/1312.2503} {arXiv:1312.2503 [gr-qc]} \BibitemShut
  {NoStop}%
\bibitem [{BHP()}]{BHPTK18}%
  \BibitemOpen
  \href@noop {} {\enquote {\bibinfo {title} {{Black Hole Perturbation
  Toolkit}},}\ }\bibinfo {note} {\url{bhptoolkit.org}}\BibitemShut {NoStop}%
\bibitem [{\citenamefont {Forseth}\ \emph {et~al.}(2016)\citenamefont
  {Forseth}, \citenamefont {Evans},\ and\ \citenamefont
  {Hopper}}]{ForsEvanHopp16}%
  \BibitemOpen
  \bibfield  {author} {\bibinfo {author} {\bibfnamefont {E.}~\bibnamefont
  {Forseth}}, \bibinfo {author} {\bibfnamefont {C.~R.}\ \bibnamefont {Evans}},
  \ and\ \bibinfo {author} {\bibfnamefont {S.}~\bibnamefont {Hopper}},\ }\href
  {\doibase 10.1103/PhysRevD.93.064058} {\bibfield  {journal} {\bibinfo
  {journal} {Phys. Rev. D}\ }\textbf {\bibinfo {volume} {93}},\ \bibinfo
  {pages} {064058} (\bibinfo {year} {2016})}\BibitemShut {NoStop}%
\bibitem [{\citenamefont {{Martel}}\ and\ \citenamefont
  {{Poisson}}(2005)}]{MartPois05}%
  \BibitemOpen
  \bibfield  {author} {\bibinfo {author} {\bibfnamefont {K.}~\bibnamefont
  {{Martel}}}\ and\ \bibinfo {author} {\bibfnamefont {E.}~\bibnamefont
  {{Poisson}}},\ }\href {\doibase 10.1103/PhysRevD.71.104003} {\bibfield
  {journal} {\bibinfo  {journal} {Phys. Rev. D}\ }\textbf {\bibinfo {volume}
  {71}},\ \bibinfo {pages} {104003} (\bibinfo {year} {2005})},\ \Eprint
  {http://arxiv.org/abs/arXiv:gr-qc/0502028} {arXiv:gr-qc/0502028} \BibitemShut
  {NoStop}%
\bibitem [{\citenamefont {{Sasaki}}\ and\ \citenamefont
  {{Tagoshi}}(2003)}]{SasaTago03}%
  \BibitemOpen
  \bibfield  {author} {\bibinfo {author} {\bibfnamefont {M.}~\bibnamefont
  {{Sasaki}}}\ and\ \bibinfo {author} {\bibfnamefont {H.}~\bibnamefont
  {{Tagoshi}}},\ }\href {\doibase 10.12942/lrr-2003-6} {\bibfield  {journal}
  {\bibinfo  {journal} {Living Reviews in Relativity}\ }\textbf {\bibinfo
  {volume} {6}},\ \bibinfo {pages} {6} (\bibinfo {year} {2003})},\ \Eprint
  {http://arxiv.org/abs/gr-qc/0306120} {gr-qc/0306120} \BibitemShut {NoStop}%
\bibitem [{\citenamefont {{Darwin}}(1959)}]{Darw59}%
  \BibitemOpen
  \bibfield  {author} {\bibinfo {author} {\bibfnamefont {C.}~\bibnamefont
  {{Darwin}}},\ }\href {\doibase 10.1098/rspa.1959.0015} {\bibfield  {journal}
  {\bibinfo  {journal} {Proc. R. Soc. Lond. A}\ }\textbf {\bibinfo {volume}
  {249}},\ \bibinfo {pages} {180} (\bibinfo {year} {1959})}\BibitemShut
  {NoStop}%
\bibitem [{\citenamefont {Cutler}\ \emph {et~al.}(1994)\citenamefont {Cutler},
  \citenamefont {Kennefick},\ and\ \citenamefont {Poisson}}]{CutlKennPois94}%
  \BibitemOpen
  \bibfield  {author} {\bibinfo {author} {\bibfnamefont {C.}~\bibnamefont
  {Cutler}}, \bibinfo {author} {\bibfnamefont {D.}~\bibnamefont {Kennefick}}, \
  and\ \bibinfo {author} {\bibfnamefont {E.}~\bibnamefont {Poisson}},\ }\href
  {\doibase 10.1103/PhysRevD.50.3816} {\bibfield  {journal} {\bibinfo
  {journal} {Phys. Rev. D}\ }\textbf {\bibinfo {volume} {50}},\ \bibinfo
  {pages} {3816} (\bibinfo {year} {1994})}\BibitemShut {NoStop}%
\bibitem [{\citenamefont {{Barack}}\ and\ \citenamefont
  {{Sago}}(2010)}]{BaraSago10}%
  \BibitemOpen
  \bibfield  {author} {\bibinfo {author} {\bibfnamefont {L.}~\bibnamefont
  {{Barack}}}\ and\ \bibinfo {author} {\bibfnamefont {N.}~\bibnamefont
  {{Sago}}},\ }\href {\doibase 10.1103/PhysRevD.81.084021} {\bibfield
  {journal} {\bibinfo  {journal} {Phys. Rev. D}\ }\textbf {\bibinfo {volume}
  {81}},\ \bibinfo {eid} {084021} (\bibinfo {year} {2010})},\ \Eprint
  {http://arxiv.org/abs/1002.2386} {arXiv:1002.2386 [gr-qc]} \BibitemShut
  {NoStop}%
\bibitem [{\citenamefont {Hopper}\ \emph {et~al.}(2015)\citenamefont {Hopper},
  \citenamefont {Forseth}, \citenamefont {Osburn},\ and\ \citenamefont
  {Evans}}]{HoppETC15}%
  \BibitemOpen
  \bibfield  {author} {\bibinfo {author} {\bibfnamefont {S.}~\bibnamefont
  {Hopper}}, \bibinfo {author} {\bibfnamefont {E.}~\bibnamefont {Forseth}},
  \bibinfo {author} {\bibfnamefont {T.}~\bibnamefont {Osburn}}, \ and\ \bibinfo
  {author} {\bibfnamefont {C.~R.}\ \bibnamefont {Evans}},\ }\href {\doibase
  10.1103/PhysRevD.92.044048} {\bibfield  {journal} {\bibinfo  {journal} {Phys.
  Rev. D}\ }\textbf {\bibinfo {volume} {92}},\ \bibinfo {eid} {044048}
  (\bibinfo {year} {2015})},\ \Eprint {http://arxiv.org/abs/1506.04742}
  {arXiv:1506.04742 [gr-qc]} \BibitemShut {NoStop}%
\bibitem [{\citenamefont {{Gradshteyn}}\ \emph {et~al.}(2007)\citenamefont
  {{Gradshteyn}}, \citenamefont {{Ryzhik}}, \citenamefont {{Jeffrey}},\ and\
  \citenamefont {{Zwillinger}}}]{GradETC07}%
  \BibitemOpen
  \bibfield  {author} {\bibinfo {author} {\bibfnamefont {I.~S.}\ \bibnamefont
  {{Gradshteyn}}}, \bibinfo {author} {\bibfnamefont {I.~M.}\ \bibnamefont
  {{Ryzhik}}}, \bibinfo {author} {\bibfnamefont {A.}~\bibnamefont {{Jeffrey}}},
  \ and\ \bibinfo {author} {\bibfnamefont {D.}~\bibnamefont {{Zwillinger}}},\
  }\href@noop {} {\emph {\bibinfo {title} {Table of Integrals, Series, and
  Products, Seventh Edition~Elsevier Academic Press, 2007.~ISBN
  012-373637-4}}}\ (\bibinfo {year} {2007})\BibitemShut {NoStop}%
\bibitem [{\citenamefont {Hopper}\ and\ \citenamefont
  {Evans}(2010)}]{HoppEvan10}%
  \BibitemOpen
  \bibfield  {author} {\bibinfo {author} {\bibfnamefont {S.}~\bibnamefont
  {Hopper}}\ and\ \bibinfo {author} {\bibfnamefont {C.~R.}\ \bibnamefont
  {Evans}},\ }\href {\doibase 10.1103/PhysRevD.82.084010} {\bibfield  {journal}
  {\bibinfo  {journal} {Phys. Rev. D}\ }\textbf {\bibinfo {volume} {82}},\
  \bibinfo {pages} {084010} (\bibinfo {year} {2010})}\BibitemShut {NoStop}%
\bibitem [{\citenamefont {{Chandrasekhar}}(1975)}]{Chan75}%
  \BibitemOpen
  \bibfield  {author} {\bibinfo {author} {\bibfnamefont {S.}~\bibnamefont
  {{Chandrasekhar}}},\ }\href {\doibase 10.1098/rspa.1975.0066} {\bibfield
  {journal} {\bibinfo  {journal} {Royal Society of London Proceedings Series
  A}\ }\textbf {\bibinfo {volume} {343}},\ \bibinfo {pages} {289} (\bibinfo
  {year} {1975})}\BibitemShut {NoStop}%
\bibitem [{\citenamefont {{Chandrasekhar}}\ and\ \citenamefont
  {{Detweiler}}(1975)}]{ChanDetw75}%
  \BibitemOpen
  \bibfield  {author} {\bibinfo {author} {\bibfnamefont {S.}~\bibnamefont
  {{Chandrasekhar}}}\ and\ \bibinfo {author} {\bibfnamefont {S.}~\bibnamefont
  {{Detweiler}}},\ }\href {\doibase 10.1098/rspa.1975.0130} {\bibfield
  {journal} {\bibinfo  {journal} {Royal Society of London Proceedings Series
  A}\ }\textbf {\bibinfo {volume} {345}},\ \bibinfo {pages} {145} (\bibinfo
  {year} {1975})}\BibitemShut {NoStop}%
\bibitem [{\citenamefont {Chandrasekhar}(1983)}]{Chan83}%
  \BibitemOpen
  \bibfield  {author} {\bibinfo {author} {\bibfnamefont {S.}~\bibnamefont
  {Chandrasekhar}},\ }\href@noop {} {\emph {\bibinfo {title} {{The Mathematical
  Theory of Black Holes}}}},\ \bibinfo {series} {The International Series of
  Monographs on Physics}, Vol.~\bibinfo {volume} {69}\ (\bibinfo  {publisher}
  {Clarendon},\ \bibinfo {address} {Oxford},\ \bibinfo {year}
  {1983})\BibitemShut {NoStop}%
\bibitem [{\citenamefont {Berndston}(2007)}]{Bern07}%
  \BibitemOpen
  \bibfield  {author} {\bibinfo {author} {\bibfnamefont {M.}~\bibnamefont
  {Berndston}},\ }\emph {\bibinfo {title} {Harmonic Gauge Perturbations of the
  Schwarzschild Metric}},\ \href@noop {} {Ph.D. thesis},\ \bibinfo  {school}
  {University of Colorado} (\bibinfo {year} {2007}),\ \Eprint
  {http://arxiv.org/abs/0904.0033v1} {arXiv:0904.0033v1} \BibitemShut {NoStop}%
\bibitem [{\citenamefont {Nakano}\ \emph {et~al.}(2003)\citenamefont {Nakano},
  \citenamefont {Sago},\ and\ \citenamefont {Sasaki}}]{NakaSagoSasa03}%
  \BibitemOpen
  \bibfield  {author} {\bibinfo {author} {\bibfnamefont {H.}~\bibnamefont
  {Nakano}}, \bibinfo {author} {\bibfnamefont {N.}~\bibnamefont {Sago}}, \ and\
  \bibinfo {author} {\bibfnamefont {M.}~\bibnamefont {Sasaki}},\ }\href
  {\doibase 10.1103/PhysRevD.68.124003} {\bibfield  {journal} {\bibinfo
  {journal} {Phys. Rev. D}\ }\textbf {\bibinfo {volume} {68}},\ \bibinfo
  {pages} {124003} (\bibinfo {year} {2003})},\ \Eprint
  {http://arxiv.org/abs/gr-qc/0308027} {arXiv:gr-qc/0308027} \BibitemShut
  {NoStop}%
\bibitem [{\citenamefont {Detweiler}\ and\ \citenamefont
  {Poisson}(2004)}]{DetwPois04}%
  \BibitemOpen
  \bibfield  {author} {\bibinfo {author} {\bibfnamefont {S.~L.}\ \bibnamefont
  {Detweiler}}\ and\ \bibinfo {author} {\bibfnamefont {E.}~\bibnamefont
  {Poisson}},\ }\href {\doibase 10.1103/PhysRevD.69.084019} {\bibfield
  {journal} {\bibinfo  {journal} {Phys. Rev. D}\ }\textbf {\bibinfo {volume}
  {69}},\ \bibinfo {pages} {084019} (\bibinfo {year} {2004})},\ \Eprint
  {http://arxiv.org/abs/gr-qc/0312010} {arXiv:gr-qc/0312010} \BibitemShut
  {NoStop}%
\bibitem [{\citenamefont {Barack}\ and\ \citenamefont
  {Lousto}(2005)}]{BaraLous05}%
  \BibitemOpen
  \bibfield  {author} {\bibinfo {author} {\bibfnamefont {L.}~\bibnamefont
  {Barack}}\ and\ \bibinfo {author} {\bibfnamefont {C.~O.}\ \bibnamefont
  {Lousto}},\ }\href {\doibase 10.1103/PhysRevD.72.104026} {\bibfield
  {journal} {\bibinfo  {journal} {Phys. Rev. D}\ }\textbf {\bibinfo {volume}
  {72}},\ \bibinfo {pages} {104026} (\bibinfo {year} {2005})},\ \Eprint
  {http://arxiv.org/abs/gr-qc/0510019} {arXiv:gr-qc/0510019} \BibitemShut
  {NoStop}%
\bibitem [{\citenamefont {Barack}\ and\ \citenamefont
  {Sago}(2007)}]{BaraSago07}%
  \BibitemOpen
  \bibfield  {author} {\bibinfo {author} {\bibfnamefont {L.}~\bibnamefont
  {Barack}}\ and\ \bibinfo {author} {\bibfnamefont {N.}~\bibnamefont {Sago}},\
  }\href {\doibase 10.1103/PhysRevD.75.064021} {\bibfield  {journal} {\bibinfo
  {journal} {Phys. Rev. D}\ }\textbf {\bibinfo {volume} {75}},\ \bibinfo
  {pages} {064021} (\bibinfo {year} {2007})},\ \Eprint
  {http://arxiv.org/abs/gr-qc/0701069} {arXiv:gr-qc/0701069} \BibitemShut
  {NoStop}%
\bibitem [{\citenamefont {Sago}\ \emph {et~al.}(2008)\citenamefont {Sago},
  \citenamefont {Barack},\ and\ \citenamefont {Detweiler}}]{SagoBaraDetw08}%
  \BibitemOpen
  \bibfield  {author} {\bibinfo {author} {\bibfnamefont {N.}~\bibnamefont
  {Sago}}, \bibinfo {author} {\bibfnamefont {L.}~\bibnamefont {Barack}}, \ and\
  \bibinfo {author} {\bibfnamefont {S.~L.}\ \bibnamefont {Detweiler}},\ }\href
  {\doibase 10.1103/PhysRevD.78.124024} {\bibfield  {journal} {\bibinfo
  {journal} {Phys. Rev. D}\ }\textbf {\bibinfo {volume} {78}},\ \bibinfo
  {pages} {124024} (\bibinfo {year} {2008})},\ \Eprint
  {http://arxiv.org/abs/0810.2530} {arXiv:0810.2530 [gr-qc]} \BibitemShut
  {NoStop}%
\bibitem [{\citenamefont {Detweiler}\ and\ \citenamefont
  {Whiting}(2003)}]{DetwWhit03}%
  \BibitemOpen
  \bibfield  {author} {\bibinfo {author} {\bibfnamefont {S.~L.}\ \bibnamefont
  {Detweiler}}\ and\ \bibinfo {author} {\bibfnamefont {B.~F.}\ \bibnamefont
  {Whiting}},\ }\href {\doibase 10.1103/PhysRevD.67.024025} {\bibfield
  {journal} {\bibinfo  {journal} {Phys. Rev. D}\ }\textbf {\bibinfo {volume}
  {67}},\ \bibinfo {pages} {024025} (\bibinfo {year} {2003})},\ \Eprint
  {http://arxiv.org/abs/gr-qc/0202086} {arXiv:gr-qc/0202086} \BibitemShut
  {NoStop}%
\bibitem [{\citenamefont {Barack}(2001)}]{Bara01}%
  \BibitemOpen
  \bibfield  {author} {\bibinfo {author} {\bibfnamefont {L.}~\bibnamefont
  {Barack}},\ }\href {\doibase 10.1103/PhysRevD.64.084021} {\bibfield
  {journal} {\bibinfo  {journal} {Phys. Rev. D}\ }\textbf {\bibinfo {volume}
  {64}},\ \bibinfo {pages} {084021} (\bibinfo {year} {2001})}\BibitemShut
  {NoStop}%
\bibitem [{\citenamefont {{Barack}}\ and\ \citenamefont
  {{Ori}}(2003)}]{BaraOri03b}%
  \BibitemOpen
  \bibfield  {author} {\bibinfo {author} {\bibfnamefont {L.}~\bibnamefont
  {{Barack}}}\ and\ \bibinfo {author} {\bibfnamefont {A.}~\bibnamefont
  {{Ori}}},\ }\href {\doibase 10.1103/PhysRevD.67.024029} {\bibfield  {journal}
  {\bibinfo  {journal} {Phys. Rev. D}\ }\textbf {\bibinfo {volume} {67}},\
  \bibinfo {eid} {024029} (\bibinfo {year} {2003})},\ \Eprint
  {http://arxiv.org/abs/arXiv:gr-qc/0209072} {arXiv:gr-qc/0209072} \BibitemShut
  {NoStop}%
\bibitem [{\citenamefont {Heffernan}\ \emph {et~al.}(2012)\citenamefont
  {Heffernan}, \citenamefont {Ottewill},\ and\ \citenamefont
  {Wardell}}]{HeffOtteWard12a}%
  \BibitemOpen
  \bibfield  {author} {\bibinfo {author} {\bibfnamefont {A.}~\bibnamefont
  {Heffernan}}, \bibinfo {author} {\bibfnamefont {A.}~\bibnamefont {Ottewill}},
  \ and\ \bibinfo {author} {\bibfnamefont {B.}~\bibnamefont {Wardell}},\ }\href
  {\doibase 10.1103/PhysRevD.86.104023} {\bibfield  {journal} {\bibinfo
  {journal} {Phys. Rev. D}\ }\textbf {\bibinfo {volume} {86}},\ \bibinfo
  {pages} {104023} (\bibinfo {year} {2012})},\ \Eprint
  {http://arxiv.org/abs/1204.0794} {arXiv:1204.0794 [gr-qc]} \BibitemShut
  {NoStop}%
\bibitem [{\citenamefont {{Wardell}}\ and\ \citenamefont
  {{Warburton}}(2015)}]{WardWarb15}%
  \BibitemOpen
  \bibfield  {author} {\bibinfo {author} {\bibfnamefont {B.}~\bibnamefont
  {{Wardell}}}\ and\ \bibinfo {author} {\bibfnamefont {N.}~\bibnamefont
  {{Warburton}}},\ }\href {\doibase 10.1103/PhysRevD.92.084019} {\bibfield
  {journal} {\bibinfo  {journal} {Phys. Rev. D}\ }\textbf {\bibinfo {volume}
  {92}},\ \bibinfo {eid} {084019} (\bibinfo {year} {2015})},\ \Eprint
  {http://arxiv.org/abs/1505.07841} {arXiv:1505.07841 [gr-qc]} \BibitemShut
  {NoStop}%
\bibitem [{\citenamefont {{Blanchet}}\ \emph {et~al.}(2010)\citenamefont
  {{Blanchet}}, \citenamefont {{Detweiler}}, \citenamefont {{Le Tiec}},\ and\
  \citenamefont {{Whiting}}}]{BlanETC10}%
  \BibitemOpen
  \bibfield  {author} {\bibinfo {author} {\bibfnamefont {L.}~\bibnamefont
  {{Blanchet}}}, \bibinfo {author} {\bibfnamefont {S.}~\bibnamefont
  {{Detweiler}}}, \bibinfo {author} {\bibfnamefont {A.}~\bibnamefont {{Le
  Tiec}}}, \ and\ \bibinfo {author} {\bibfnamefont {B.~F.}\ \bibnamefont
  {{Whiting}}},\ }\href {\doibase 10.1103/PhysRevD.81.084033} {\bibfield
  {journal} {\bibinfo  {journal} {Phys. Rev. D}\ }\textbf {\bibinfo {volume}
  {81}},\ \bibinfo {eid} {084033} (\bibinfo {year} {2010})},\ \Eprint
  {http://arxiv.org/abs/1002.0726} {arXiv:1002.0726 [gr-qc]} \BibitemShut
  {NoStop}%
\bibitem [{\citenamefont {Damour}\ \emph {et~al.}(2015)\citenamefont {Damour},
  \citenamefont {Jaranowski},\ and\ \citenamefont
  {Sch\"afer}}]{DamoJaraScha15}%
  \BibitemOpen
  \bibfield  {author} {\bibinfo {author} {\bibfnamefont {T.}~\bibnamefont
  {Damour}}, \bibinfo {author} {\bibfnamefont {P.}~\bibnamefont {Jaranowski}},
  \ and\ \bibinfo {author} {\bibfnamefont {G.}~\bibnamefont {Sch\"afer}},\
  }\href {\doibase 10.1103/PhysRevD.91.084024} {\bibfield  {journal} {\bibinfo
  {journal} {Phys. Rev. D}\ }\textbf {\bibinfo {volume} {91}},\ \bibinfo
  {pages} {084024} (\bibinfo {year} {2015})},\ \Eprint
  {http://arxiv.org/abs/1502.07245} {arXiv:1502.07245 [gr-qc]} \BibitemShut
  {NoStop}%
\bibitem [{\citenamefont {{Peters}}\ and\ \citenamefont
  {{Mathews}}(1963)}]{PeteMath63}%
  \BibitemOpen
  \bibfield  {author} {\bibinfo {author} {\bibfnamefont {P.~C.}\ \bibnamefont
  {{Peters}}}\ and\ \bibinfo {author} {\bibfnamefont {J.}~\bibnamefont
  {{Mathews}}},\ }\href {\doibase 10.1103/PhysRev.131.435} {\bibfield
  {journal} {\bibinfo  {journal} {Physical Review}\ }\textbf {\bibinfo {volume}
  {131}},\ \bibinfo {pages} {435} (\bibinfo {year} {1963})}\BibitemShut
  {NoStop}%
\bibitem [{\citenamefont {{Blanchet}}\ and\ \citenamefont
  {{Sch{\"a}fer}}(1993)}]{BlanScha93}%
  \BibitemOpen
  \bibfield  {author} {\bibinfo {author} {\bibfnamefont {L.}~\bibnamefont
  {{Blanchet}}}\ and\ \bibinfo {author} {\bibfnamefont {G.}~\bibnamefont
  {{Sch{\"a}fer}}},\ }\href {\doibase 10.1088/0264-9381/10/12/026} {\bibfield
  {journal} {\bibinfo  {journal} {Classical and Quantum Gravity}\ }\textbf
  {\bibinfo {volume} {10}},\ \bibinfo {pages} {2699} (\bibinfo {year}
  {1993})}\BibitemShut {NoStop}%
\bibitem [{\citenamefont {{Arun}}\ \emph {et~al.}(2008)\citenamefont {{Arun}},
  \citenamefont {{Blanchet}}, \citenamefont {{Iyer}},\ and\ \citenamefont
  {{Qusailah}}}]{ArunETC08a}%
  \BibitemOpen
  \bibfield  {author} {\bibinfo {author} {\bibfnamefont {K.~G.}\ \bibnamefont
  {{Arun}}}, \bibinfo {author} {\bibfnamefont {L.}~\bibnamefont {{Blanchet}}},
  \bibinfo {author} {\bibfnamefont {B.~R.}\ \bibnamefont {{Iyer}}}, \ and\
  \bibinfo {author} {\bibfnamefont {M.~S.~S.}\ \bibnamefont {{Qusailah}}},\
  }\href {\doibase 10.1103/PhysRevD.77.064034} {\bibfield  {journal} {\bibinfo
  {journal} {Phys. Rev. D}\ }\textbf {\bibinfo {volume} {77}},\ \bibinfo {eid}
  {064034} (\bibinfo {year} {2008})},\ \Eprint {http://arxiv.org/abs/0711.0250}
  {arXiv:0711.0250 [gr-qc]} \BibitemShut {NoStop}%
\bibitem [{\citenamefont {{Shah}}\ \emph {et~al.}(2014)\citenamefont {{Shah}},
  \citenamefont {{Friedman}},\ and\ \citenamefont
  {{Whiting}}}]{ShahFrieWhit14}%
  \BibitemOpen
  \bibfield  {author} {\bibinfo {author} {\bibfnamefont {A.~G.}\ \bibnamefont
  {{Shah}}}, \bibinfo {author} {\bibfnamefont {J.~L.}\ \bibnamefont
  {{Friedman}}}, \ and\ \bibinfo {author} {\bibfnamefont {B.~F.}\ \bibnamefont
  {{Whiting}}},\ }\href {\doibase 10.1103/PhysRevD.89.064042} {\bibfield
  {journal} {\bibinfo  {journal} {Phys. Rev. D}\ }\textbf {\bibinfo {volume}
  {89}},\ \bibinfo {eid} {064042} (\bibinfo {year} {2014})},\ \Eprint
  {http://arxiv.org/abs/1312.1952} {arXiv:1312.1952 [gr-qc]} \BibitemShut
  {NoStop}%
\bibitem [{\citenamefont {{Blanchet}}\ \emph {et~al.}(2014)\citenamefont
  {{Blanchet}}, \citenamefont {{Faye}},\ and\ \citenamefont
  {{Whiting}}}]{BlanFayeWhit14a}%
  \BibitemOpen
  \bibfield  {author} {\bibinfo {author} {\bibfnamefont {L.}~\bibnamefont
  {{Blanchet}}}, \bibinfo {author} {\bibfnamefont {G.}~\bibnamefont {{Faye}}},
  \ and\ \bibinfo {author} {\bibfnamefont {B.~F.}\ \bibnamefont {{Whiting}}},\
  }\href {\doibase 10.1103/PhysRevD.89.064026} {\bibfield  {journal} {\bibinfo
  {journal} {Phys. Rev. D}\ }\textbf {\bibinfo {volume} {89}},\ \bibinfo {eid}
  {064026} (\bibinfo {year} {2014})},\ \Eprint {http://arxiv.org/abs/1312.2975}
  {arXiv:1312.2975 [gr-qc]} \BibitemShut {NoStop}%
\bibitem [{\citenamefont {{Isoyama}}\ \emph {et~al.}(2013)\citenamefont
  {{Isoyama}}, \citenamefont {{Fujita}}, \citenamefont {{Nakano}},
  \citenamefont {{Sago}},\ and\ \citenamefont {{Tanaka}}}]{IsoyETC13}%
  \BibitemOpen
  \bibfield  {author} {\bibinfo {author} {\bibfnamefont {S.}~\bibnamefont
  {{Isoyama}}}, \bibinfo {author} {\bibfnamefont {R.}~\bibnamefont {{Fujita}}},
  \bibinfo {author} {\bibfnamefont {H.}~\bibnamefont {{Nakano}}}, \bibinfo
  {author} {\bibfnamefont {N.}~\bibnamefont {{Sago}}}, \ and\ \bibinfo {author}
  {\bibfnamefont {T.}~\bibnamefont {{Tanaka}}},\ }\href {\doibase
  10.1093/ptep/ptt034} {\bibfield  {journal} {\bibinfo  {journal} {Progress of
  Theoretical and Experimental Physics}\ }\textbf {\bibinfo {volume} {2013}},\
  \bibinfo {eid} {063E01} (\bibinfo {year} {2013})},\ \Eprint
  {http://arxiv.org/abs/1302.4035} {arXiv:1302.4035 [gr-qc]} \BibitemShut
  {NoStop}%
\bibitem [{\citenamefont {{Johnson-McDaniel}}(2014)}]{JohnMcDa14}%
  \BibitemOpen
  \bibfield  {author} {\bibinfo {author} {\bibfnamefont {N.~K.}\ \bibnamefont
  {{Johnson-McDaniel}}},\ }\href {\doibase 10.1103/PhysRevD.90.024043}
  {\bibfield  {journal} {\bibinfo  {journal} {Phys. Rev. D}\ }\textbf {\bibinfo
  {volume} {90}},\ \bibinfo {eid} {024043} (\bibinfo {year} {2014})},\ \Eprint
  {http://arxiv.org/abs/1405.1572} {arXiv:1405.1572 [gr-qc]} \BibitemShut
  {NoStop}%
\end{thebibliography}%

\end{document}